\documentclass{LMCS}
\usepackage{graphicx}
\usepackage{color}
\usepackage{umlaut}
\usepackage{amsmath}
\usepackage{amssymb}

\def\PDFVARS{%
  pdftitle={Linear Encodings of Bounded LTL Model Checking},%
  pdfsubject={Bounded Model Checking},%
  pdfauthor={Armin Biere, Keijo Heljanko, Tommi Junttila, Timo Latvala, and Viktor Schuppan},%
  pdfkeywords={Bounded Model Checking, LTL, PLTL, Liveness to Safety, NuSMV},%
  pdfpagemode=None,%
  pdfborder={0 0 0}}

\usepackage[dvips,pagebackref=true,\PDFVARS]{hyperref}
\def\texpic{pstex_t}

\newcommand{\Until}[2]{\ensuremath{#1\mathbin{\mathbf{U}}#2}}
\newcommand{\Release}[2]{\ensuremath{#1\mathbin{\mathbf{R}}#2}}
\newcommand{\Since}[2]{\ensuremath{#1\mathbin{\mathbf{S}}#2}}
\newcommand{\Trigger}[2]{\ensuremath{#1\mathbin{\mathbf{T}}#2}}
\newcommand{\Globally}[1]{\ensuremath{\mathbf{G}\,#1}}
\newcommand{\Historically}[1]{\ensuremath{\mathbf{H}\,#1}}
\newcommand{\Finally}[1]{\ensuremath{\mathbf{F}\,#1}}
\newcommand{\Once}[1]{\ensuremath{\mathbf{O}\,#1}}
\newcommand{\Next}[1]{\ensuremath{\mathbf{X}\,#1}}
\newcommand{\Prev}[1]{\ensuremath{\mathbf{Y}\,#1}}
\newcommand{\PrevZ}[1]{\ensuremath{\mathbf{Z}\,#1}}
\newcommand{\Trans}[1]{\ensuremath{{\vert [#1]\vert}}}
\newcommand{\ATrans}[1]{\ensuremath{{\langle\langle #1 \rangle\rangle}}}
\newcommand{\CTrans}[1]{\ensuremath{{\langle\langle {#1} \rangle\rangle}}}
\newcommand{\lvar}[1]{\ensuremath{l_{#1}}}
\newcommand{\InLoop}[1]{\ensuremath{\textup{InLoop}_{#1}}}
\newcommand{\LoopExists}{\ensuremath{\textup{LoopExists}}}
\newcommand{\Implies}{\ensuremath{\Rightarrow}}
\newcommand{\Equiv}{\ensuremath{\Leftrightarrow}}
\newcommand{\TLF}{\psi}
\newcommand{\TSF}{\varphi}
\newcommand{\ASF}{s_\varphi}
\newcommand{\SF}{\psi_1}
\newcommand{\LSF}{\psi_1}
\newcommand{\RSF}{\psi_2}

\newcommand{\Nat}{{\mathbb N}}

\newcommand{\Bool}{{\mathbb B}}
\newcommand{\Lang}[1]{\ensuremath{\textup{Lang}(#1)}}
\newcommand{\Succs}[1]{\ensuremath{{#1}'}} 
\newcommand{\ltoscopy}[1]{\ensuremath{\hat{#1}}} 
\newcommand{\ltos}[1]{\ensuremath{{#1}^{\mathbf{S}}}} 
\newcommand{\LoopClosed}{\ensuremath{\textup{LoopClosed}}}
\newcommand{\lstart}{\mathit{l_s}}
\newcommand{\lend}{\mathit{le}}

\newcommand{\modification}[2]{\ensuremath{\textcolor{blue}{\setlength{\fboxsep}{1.5pt} \fbox{$#2{#1}$}}}}
\newcommand{\modelsnl}[0]{\ensuremath{\models_{\mathrm{nl}}}}

\theoremstyle{plain}\newtheorem{definition}[thm]{Definition}
\theoremstyle{plain}\newtheorem{proposition}[thm]{Proposition}
\theoremstyle{plain}\newtheorem{theorem}[thm]{Theorem}
\theoremstyle{plain}\newtheorem{lemma}[thm]{Lemma}
\theoremstyle{plain}\newtheorem{corollary}[thm]{Corollary}

\def\encsize{\smallskip\small}
\def\tightencsize{\smallskip\footnotesize}
\def\tighterencsize{\smallskip\scriptsize}

\def\doi{2 (5:5) 2006}
\lmcsheading%
{\doi}
{1--64}
{}
{}
{Feb.~16, 2006}
{Nov.~15, 2006}
{}   

\begin{document}
  \title{Linear Encodings of Bounded LTL Model Checking} 
  \author[A.~Biere]{Armin Biere\rsuper a}
  \address{{\lsuper a}Institute for Formal Models and Verification,
    Johannes Kepler University,
    Altenbergerstrasse 69, A-4040 Linz,
    Austria
  }
  \email{biere@jku.at}

  \author[K.~Heljanko]{Keijo Heljanko\rsuper b}
  \address{{\lsuper{b,c}}Laboratory for Theoretical Computer Science,
    Helsinki University of Technology,
    P.O.\ Box 5400, FI-02015 TKK, Finland}
  \email{\{Keijo.Heljanko,Tommi.Junttila\}@tkk.fi}

  \author[T.~Junttila]{Tommi Junttila\rsuper c}
  \address{\vskip-8 pt}
  \email{}

  \author[T.Latvala]{Timo Latvala\rsuper d}
  \address{{\lsuper d}Department of Computer Science,
    University of Illinois at Urbana-Champaign,
    201 Goodwin Ave.,
    Urbana, IL 61801-2302,
    USA}
  \email{tlatvala@uiuc.edu}

  \author[V.~Schuppan]{Viktor Schuppan\rsuper e}
  \address{{\lsuper e}Computer Systems Institute,
    ETH Zentrum, CH-8092 Z{\"u}rich, Switzerland
  }
  \email{vschuppan@acm.org}


  \titlecomment{}

  \begin{abstract}
    We consider the problem of bounded model checking (BMC) for linear
    temporal logic (LTL). We present several efficient encodings that
    have size linear in the bound. Furthermore, we show how the
    encodings can be extended to LTL with past operators (PLTL). The
    generalised encoding is still of linear size, but cannot detect
    minimal length counterexamples. By using the virtual unrolling
    technique minimal length counterexamples can be captured, however,
    the size of the encoding is quadratic in the specification. We
    also extend virtual unrolling to B{\"u}chi automata, enabling them to
    accept minimal length counterexamples.

    Our BMC encodings can be made incremental in order to benefit from
    incremental SAT technology. With fairly small modifications the
    incremental encoding can be further enhanced with a termination
    check, allowing us to prove properties with BMC\@.

    An analysis of the liveness-to-safety transformation reveals many
    similarities to the BMC encodings in this paper. We conduct
    experiments to determine the advantage of employing dedicated BMC
    encodings for PLTL over combining more general but potentially
    less efficient approaches with BMC: the liveness-to-safety
    transformation with invariant checking and B{\"u}chi automata with
    fair cycle detection.

    Experiments clearly show that our new encodings improve
    performance of BMC considerably, particularly in the case of the
    incremental encoding, and that they are very competitive for
    finding bugs. Dedicated encodings seem to have an advantage over
    using more general methods with BMC. Using the liveness-to-safety
    translation with BDD-based invariant checking results in an
    efficient method to find shortest counterexamples that complements
    the BMC-based approach. For proving complex properties BDD-based
    methods still tend to perform better.
  \end{abstract}

\maketitle

\section*{Introduction}\label{intro}
Bounded model checking~\cite{BCCZ99} was introduced as an alternative
to binary decisions diagrams (BDDs) to implement symbolic model checking. This paper describes some 
of the key results of~\cite{HUT-TCS-A95,Schuppan06} on bounded model checking, and some extensions.
The main results have been published in~\cite{LBHJ04,LBHJ05,HelJunLat:CAV05,SB04,SB05}.

The basic idea behind bounded model checking (BMC) is to restrict the general 
model checking problem to a 
\emph{boun\-ded} problem. Instead of asking whether the system $M$ violates the property $\TLF$,
we ask whether the system $M$ has any counterexample of length $k$ to $\TLF$.
This bounded problem is \emph{encoded} into SAT, the propositional satisfiability problem, in order to obtain 
the benefits of symbolic representations of states. In other words, a Boolean formula 
$\Trans{M,\neg\TLF,k}$ is generated which is satisfiable iff $M$ has a 
counterexample to $\TLF$ of length $k$. The satisfiability of this formula can then
be checked with a SAT solver.

The key insight behind BMC for linear-time formalisms such as 
linear temporal logic (LTL)    
is that a witness for LTL given as an \emph{infinite} execution path of the system
can be captured by a \emph{finite} path in two ways: either the finite path represents all its 
infinite extensions or the finite path loops and in fact captures
the behaviour of an infinite path. Let $\pi=s_0s_1s_2\ldots$ be an
infinite path of a system. We say that $\pi$ is a $(k,l)$-loop if 
$\pi=(s_0s_1\ldots s_{l-1})(s_l\dots s_k)^\omega$ such that $0 < l \leq k$ and $s_{l-1} = s_k$.

In BMC the transition relation $T(s,s^\prime)$ of a system $M$ is represented symbolically 
as a Boolean formula, where the states $s,s^\prime$ are modelled as bit vectors. 
To capture the finite paths of length $k$, we unroll the transition relation $k$ times and 
obtain the following Boolean formula:
{\encsize
\begin{equation*}
\Trans{M}_k \Leftrightarrow I(s_0)\wedge\bigwedge_{i=1}^k T(s_{i-1},s_i).
\end{equation*}
}\smallskip%

\noindent
Here $I(s)$ is the initial state predicate and $T(s,s^\prime)$ a total
transition relation predicate. Since only counterexamples to the given LTL formula $\TLF$ should be
accepted, additional constraints must be generated to restrict the 
models of the Boolean formula. If we denote the formula constraints by $\Trans{\neg\TLF}_k$,
the Boolean formula $\Trans{M,\neg\TLF,k} \Leftrightarrow \Trans{M}_k\wedge\Trans{\neg\TLF}_k$ is
satisfiable iff $M$ has a counterexample of length $k$ to $\TLF$. 

Compared with using BDDs to implement symbolic model checking, BMC has 
a few advantages. BMC can leverage the impressive gains that 
have been achieved in SAT solver technology in recent years \cite{SATcompetition04}.
The increase in efficiency of the solvers can directly be translated to more effective BMC\@.
The use of SAT procedures as a practical implementation technique 
to search for bounded length executions of systems has also been used in the context of
SAT-based artificial intelligence (AI) planning~\cite{DBLP:conf/ecai/KautzS92,aaai96-2*1194} and 
in sequential ATPG \cite{konuk93explorations}.
In practice, SAT solvers seem to be able to solve certain problems that are not 
feasible for BDDs.


An important advantage of BMC is that the counterexamples produced by most BMC encodings
are minimal and that the counterexample is immediately available.
Producing short counterexamples using BDDs is a fairly involved process~\cite{DBLP:conf/dac/ClarkeGMZ95} 
and minimality is seldom guaranteed. In many cases producing the counterexample consumes
more resources than answering the model checking query~\cite{DBLP:conf/dac/ClarkeGMZ95}.
However, recently a BDD model checking procedure~\cite{SB05} based on the BMC encoding of~\cite{LBHJ05}
was presented that provably produces minimal counterexamples. The method appears to consume
more memory than standard BDD model checkers, but can in some cases be faster.

Boolean formulas, or more specifically circuits, are a more compact encoding than BDDs 
for many Boolean functions: 
there are Boolean functions whose BDDs are exponential in the number of propositional 
variables~\cite{Bry86} that still have polynomial circuits. 
However, since BMC represents the length of the paths explicitly it is not always 
more space efficient than using BDDs~\cite{CKOS04}. For instance, for a simple binary counter system an exponential 
number of unrollings of the transition relation is required before the system loops and 
we can be sure that the whole behaviour of the system has been covered.

Although BMC has been very successful in practice~\cite{BCRZ99,CFFGKTV01,STR04}, improving BMC remains a 
high priority. Increasing the efficiency of BMC can be done in several ways. 
Two important approaches are developing smarter encodings of the problem to SAT and utilising improvements 
in solver technology. 
Better encodings of the problem boil down to finding 
new representations of the formula $\Trans{M,\neg\TLF,k}$,
which are easier for the SAT solver. As a rule of thumb,
good BMC encodings are compact but still
propagate information efficiently, thus minimising the non-deterministic choices the
solver has to make.


LTL with temporal operators that can reference the past is exponentially
more succinct than LTL~\cite{LMS02}.
In many cases the future fragment of LTL, which is the only fragment
usually supported, is not expressive enough in practice.
The main argument for adding support for 
past operators is motivated by practice: LTL with past operators (PLTL) allows more succinct 
and natural specifications. Especially compositional reasoning benefits from 
the added succinctness~\cite{LPZ85}. Efficient encodings for LTL with past operators
is therefore one way to increase the usability, efficiency, and the scope of BMC\@.
Utilising new solver technology such as incremental SAT solvers can result
in huge benefits for BMC~\cite{WKS01,Str01}. When solving a sequence of similar SAT problems, as is the
case in BMC, an incremental solver can retain much of the learned clauses obtained while solving
earlier related instances. This can result in large time savings for solving the whole sequence 
of problems. The benefits of incremental SAT technology can be maximised by adapting 
BMC encodings to suit the incremental framework.

In this paper we will introduce several efficient BMC encodings for LTL that 
all have linear size encodings in the bound $k$. Efficient 
encodings can make a big difference when the specification is complex~\cite{LBHJ04}.
We will present several encodings that take a slightly different view of the 
problem.  In particular we highlight the relation of BMC encodings to the automata-theoretic
approach to model checking~\cite{Kur94,VW86b}. We also show how our encodings can
be efficiently generalised to PLTL\@. The generalised encoding is still of linear size in
the bound and in the size of the PLTL formula
but does not detect minimal length counterexamples. By increasing the size
of the encoding to quadratic in the size of the PLTL formula, minimal length counterexamples can be guaranteed.
Our technique is based on \emph{virtual unrolling}~\cite{BC03}.
We also show how virtual unrolling enables
symbolic B\"uchi automata to detect minimal length counterexamples. 

Furthermore, with some modifications,
our new more efficient
encoding for PLTL can be adapted to utilise incremental SAT technology.
We try to maximise the number of learnt clauses which can be kept when the
solver moves from one problem instance to the next (i.e.,{\ }when the bound $k$ is increased).
Experiments show that the increase in efficiency can be quite dramatic.
 
Model checking $\omega$-regular properties depends on finding fair loops in the
system. Using the \emph{liveness-to-safety} model transformation~\cite{SB04}, fair loop detection 
can be integrated in the system model. This effectively reduces the general unbounded model
checking problem to reachability of bad states. We present the technique and discuss
similarities with the BMC approaches introduced in this paper. 
In experiments we compare the performance of invariant checking, after translating liveness to safety,
with dedicated BMC encodings for PLTL.

From its inception BMC has been predominantly seen as an efficient method for
finding bugs. BDD-based methods have had the advantage of being complete and thus being
able to prove that no counterexample exists.  However, several methods have been developed
in the recent years which can be used to achieve completeness with BMC (see for instance
the recent survey \cite{PBG05}).
Our incremental encoding can also be extended with a termination check. The approach 
naturally integrates with our incremental approach and can prove properties 
for full PLTL\@.

We implemented the BMC encodings and the liveness-to-safety
transformation on top of the NuSMV system \cite{NuSMV}, version 2.2.3. Starting
with version 2.4.0, the BMC encoding variant published in~\cite{HelJunLat:CAV05}
and discussed in more detail in this work has recently become a part of the standard
distribution of NuSMV \cite{NuSMVURL}. Based on the former, we have
experimentally evaluated the encodings using a large set of models
with complex specifications. Compared to the original
encoding~\cite{BCCZ99} and its newer versions~\cite{CPRS02,BC03}, our
new linear encodings are clearly superior.  We observed additional
impressive performance gains for the incremental versions.
Alternative linear sized encodings to do BMC based either on the
liveness-to-safety transformation and invariant checking or on B{\"u}chi
automata and fair loop detection did not prove quite as effective as
the dedicated BMC encodings, although they were clearly more efficient
than the original encoding and its relatives. With the termination
check activated our linear BMC encoding did not perform quite as well
as without it, but still better than old encodings. For proving
properties BDD-based methods perform better.  It is clear that the
termination check must developed further in order for BMC to be
competitive also for proving properties. Combining the
liveness-to-safety transformation with BDD-based invariant checking
results in an efficient BDD-based method to find shortest
counterexamples. It significantly reduces the length of
counterexamples in comparison to the standard BDD-based algorithm. It
performs competitively with SAT-based methods for this purpose and
complements them with respect to solved examples.  Using virtual
unrolling for B{\"u}chi automata with the standard BDD-based algorithm
significantly increases running time and gives mixed results at best
in terms of counterexample length.

In the next section we will introduce basic notation and recall fundamental
definitions that will be used throughout the paper. In Sect.~\ref{sec:BMC}
the basics of bounded model checking are described and the results of the 
original BMC-paper~\cite{BCCZ99} are discussed. Section~\ref{sec:ltl-bmc}
presents our efficient BMC encoding for LTL published in~\cite{LBHJ04}.
The section also considers alternative encodings of the BMC problem and
contrasts the encodings to model checking based on symbolic B\"uchi automata. 
Section~\ref{sec:l2s} presents the liveness-to-safety transformation
and discusses its connection to the presented BMC encodings. 
To extend BMC to full PLTL, we use the technique of ``virtual unrolling''. 
We present our generalised BMC encoding that encompasses full PLTL in Sect.~\ref{sec:pltl-bmc}. 
We also show that virtual unrolling also can be applied to symbolic B\"uchi automata. 
Section~\ref{sec:incbmc} shows how our encodings can be adapted to the incremental setting~\cite{HelJunLat:CAV05}. 
The adapted encodings are developed to maximise the information learnt between the SAT solver 
invocations. In Sect.~\ref{sec:completeness} we discuss how BMC can be made complete. Specifically
we show how our encodings can be extended with a termination check to achieve completeness.
Section~\ref{sec:experiments} experimentally compares the different encodings
presented in the paper. We discuss conclusions and directions of future work in
Sect.~\ref{sec:conclusions}.

\section{Preliminaries}

\subsection{Linear Temporal Logic with Past}
Linear temporal logic with past (PLTL) is a commonly used specification logic.
Although all PLTL properties are definable using only two basic temporal operators ($\Until{}{}$
and $\Next{}$), it has been argued that especially compositional reasoning benefits
from the use of past operators~\cite{LPZ85}. Using only the basic operators results in 
a logic that is exponentially less succinct than PLTL~\cite{LMS02}.

The \emph{syntax} of PLTL is defined over a set of atomic propositions $\mathit{AP}$. 
Boolean operators we use are negation, disjunction and conjunction.
The temporal operators we will use are ``next time'' ($\Next{}$) and its two
past-time counterparts, the ``previous time'' past temporal operators 
($\Prev{}$, $\PrevZ{}$); the future temporal connectives ``until'' ($\Until{}{}$)  
and ``release'' ($\Release{}{}$) and their past-time counterparts ``since'' 
$(\Since{}{})$ and ``trigger'' $(\Trigger{}{})$.
We will call the commonly used subset of PLTL that does not contain any past temporal
operators linear temporal logic (LTL).

The \emph{semantics} of a PLTL formula is defined along infinite paths
$\pi = s_0 s_1 \ldots$\footnote{We use commas between elements of a
tuple (such as a state consisting of the valuations of several state
variables) and no separator between elements of a sequence (such as a
path). While we generally follow the latter convention also for
composition of sequences, we sometimes prefer to emphasise composition
using $\circ$, e.g., if the entire sequence spans multiple lines.} of
states $s_i$ where we assume a mapping $L$ from each state to
the set of atomic propositions true in that state.
Let
$\pi^i$ denote the path $\pi$ with a designated formula evaluation
position $i$. The semantics can then be defined inductively as follows:
{\encsize
\[\begin{array}{lcl}
\pi^i\models p & \Leftrightarrow & p\in L(s_i) \text{ for }p\in\mathit{AP}. \\
 \pi^i\models \neg p & \Leftrightarrow & \pi^i \not\models p. \\
 \pi^i\models \LSF\vee\RSF & \Leftrightarrow &\pi^i\models\LSF \text{ or } \pi^i\models\RSF. \\
 \pi^i\models \LSF\wedge\RSF & \Leftrightarrow &\pi^i\models\LSF \text{ and } \pi^i\models\RSF. \\
 \pi^i\models \Next{\SF} & \Leftrightarrow & \pi^{i+1}\models\SF. \\
\pi^i\models \Until{\LSF}{\RSF} & \Leftrightarrow &
   \exists j\geq i \text{ such that }
\pi^j\models\RSF \text{ and } \pi^n\models\LSF \text{ for all } i\leq n<j. \\
\pi^i\models \Release{\LSF}{\RSF} & \Leftrightarrow & \text{for all } j\geq i:
\pi^j\models\RSF \text{ or } \pi^n\models\LSF \text{ for some } i\leq n < j. \\
\pi^i\models \Prev{\SF} & \Leftrightarrow & i> 0 \text{ and } \pi^{i-1}\models\SF. \\
\pi^i\models \PrevZ{\SF} & \Leftrightarrow & i=0 \text{ or } \pi^{i-1}\models\SF. \\
\pi^i\models \Since{\LSF}{\RSF} & \Leftrightarrow &
   \exists \ 0\leq j\leq i \text{ such that } \pi^{j}\models\RSF
 \text{ and } \pi^{n}\models\LSF \text{ for all } j< n\leq i. \\ 
\pi^i\models \Trigger{\LSF}{\RSF} & \Leftrightarrow & \text{for all } 0\leq j\leq i: \pi^{j}\models\RSF
\text{ or } \pi^{n}\models\LSF \text{ for some } j<n\leq i. \\ 
\end{array}\]
}\smallskip

Commonly used \emph{abbreviations} for PLTL formulas are the standard Boolean shorthands 
$\top\equiv p\vee\neg p$ for some $p\in\mathit{AP}$, $\bot\equiv\neg\top$, 
$p\Rightarrow q\equiv \neg p\vee q$, 
$p\Leftrightarrow q\equiv \left(p\Rightarrow q\right)\wedge \left( q\Rightarrow p\right)$, 
and the derived temporal operators $\Finally{\SF}\equiv\Until{\top}{\SF}$ ('finally'), 
$\Globally{\SF}\equiv\neg\Finally{\neg\SF}$ ('globally'),
$\Once{\SF}\equiv\Since{\top}{\SF}$ ('once'), and
$\Historically{\SF}\equiv\Trigger{\bot}{\SF}$ ('historically').

\begin{sloppypar}
It is always possible to rewrite any formula to \emph{positive normal form}, where all
negations only appear in front of atomic propositions. This can be accomplished
by using the dualities $\neg\left(\Until{\LSF}{\RSF}\right)\equiv\Release{\neg\LSF}{\neg\RSF}$, 
$\neg\left(\Release{\LSF}{\RSF}\right)\equiv\Until{\neg\LSF}{\neg\RSF}$, 
$\neg\Next{\SF}\equiv\Next{\neg\SF}$, 
$\neg\Prev{\SF}\equiv\PrevZ{\neg\SF}$, 
$\neg\PrevZ{\SF}\equiv\Prev{\neg\SF}$, 
$\neg\left(\Since{\LSF}{\RSF}\right)\equiv\Trigger{\neg\LSF}{\neg\RSF}$, 
$\neg\left(\Trigger{\LSF}{\RSF}\right)\equiv\Since{\neg\LSF}{\neg\RSF}$, 
and DeMorgan's rules for propositional logic.
In this paper we assume all formulas are in positive normal form unless otherwise
explicitly stated.
\end{sloppypar}

The maximum number of nested past operators of a PLTL formula is called the \emph{past operator depth}.
\begin{definition}
The past operator depth~\cite{LMS02,BC03} for a PLTL formula $\TLF$ is denoted by $\delta(\TLF)$ and is inductively
defined as:
\begin{displaymath}
\begin{array}{lll}
\delta(p)  & = 0 & \text{\rm for\ \ } p\in\mathit{AP}, \\
\delta(\circ\, \SF) & = \delta(\SF) & \text{\rm for\ \ } \circ\in\left\{\neg,\Next{}\right\}, \\
\delta(\LSF\,\circ\,\RSF) & = \mathit{max}\left(\delta(\LSF),\delta(\RSF)\right) & 
\text{\rm for\ \ } \circ\in\left\{\vee, \wedge, \Until{}{},\Release{}{}\right\}, \\
\delta(\circ\, \SF) & = 1+\delta(\SF) & 
\text{\rm for\ \ }  \circ\in\left\{\Prev{},\PrevZ{}\right\}, \text{\rm and }\\
\delta(\LSF\,\circ\,\RSF) & = 1+\mathit{max}\left(\delta(\LSF),\delta(\RSF)\right) & 
\text{\rm for\ \ } \circ\in\left\{\Since{}{},\Trigger{}{}\right\}. \\
\end{array}
\end{displaymath}
\end{definition}

The \emph{set of subformulas} of a PLTL formula $\TLF$ is denoted by $\mathit{cl}(\TLF)$
and is defined as the smallest set satisfying the following conditions:
\begin{displaymath}
\begin{array}{lll}
\TLF\in\mathit{cl}(\TLF),\\
\text{if } \circ\, \SF\in\mathit{cl}(\TLF) & 
\text{ for } \circ\in\left\{\neg,\Next{},\Prev{},\PrevZ{}\right\}  &\text{ then } \SF\in\mathit{cl}(\TLF), \text{ and }\\
\text{if } \LSF\,\circ\,\RSF\in \mathit{cl}(\TLF) & 
\text{ for } \circ\in\left\{\vee, \wedge, \Until{}{},\Release{}{}, \Since{}{}, \Trigger{}{}\right\} & 
\text{ then } \LSF,\RSF\in\mathit{cl}(\TLF).\\
\end{array}
\end{displaymath}

\subsection{Kripke Structures}

The states of a path are members of the finite set of states $S$ of a
model (a Kripke structure) $M=(S,T,I,L)$ with a total transition
relation $T$, a set of initial states $I$, and a mapping $L: S \mapsto
2^\mathit{AP}$ indicating the set of atomic propositions that are true
in a state. $L$ is extended to sequences of states (paths) in the
natural way. A path is initialised iff its first state belongs to
$I$. The set of initialised infinite paths is denoted $\Pi$. The
language of a Kripke structure can then be defined as $\Lang{M} =
\{\alpha \mid \exists \pi \in \Pi \;.\; L(\pi) = \alpha\}$.

Sometimes we equip a Kripke structure with a number of
\emph{acceptance sets} (or fairness constraints)
$F_0,\ldots,F_f$, where each $F_m$, $0 \le m
\le f$ is a subset of $S$. $M=(S,T,I,L,F=\{F_0,\ldots,F_f\})$ is then
called a fair Kripke structure. A path in $M$ is fair iff it contains
infinitely many occurrences of states from each acceptance set. $\Pi$
and $\Lang{M}$ are then restricted to fair paths.

We usually construct a Kripke structure \emph{symbolically} over a set of
variables $V$. In that case the set of states $S$ is given by the set
of valuations of $V$, possibly constrained by a set of state
invariants. Similarly, $I$, $T$, and $F_0,\ldots,F_f$ are the largest
subsets of $S$ or $S \times S$ fulfilling certain constraints. The
valuation of a variable $v$ in a state $s$ is denoted $v(s)$.

For a Kripke structure $M$ we say that a \emph{formula $\SF$ holds} in
$M$ if for every infinite initialised path $\pi$ of $M$ we have that
$\pi\models\SF$. This is denoted $M\models\SF$.  For a formula to hold
in a fair Kripke structure it is required to hold only along all fair
paths.

\subsection{B{\"u}chi Automata}

B{\"u}chi automata are frequently used as an operational model of the more
descriptive PLTL formulae~\cite{VW86b}. In this paper a B{\"u}chi
automaton is simply a fair Kripke structure. However, if we speak of a
``model'' we refer to a Kripke structure that is to be verified (it is
used as a language generator). When we say ``B{\"u}chi automaton'' we intend
a Kripke structure to serve as a specification (it is used as a
language acceptor).

A B{\"u}chi automaton $B$ has a \emph{run} $\pi$ on an infinite
sequence $\alpha$ over $2^\mathit{AP}$ iff $\pi$ is an initialised
path in $B$ with $L(\pi) = \alpha$. The run is accepting iff it is
fair. Hence, $B$ has an accepting run on $\alpha$ iff $\alpha \in
\Lang{B}$.

Typically a B{\"u}chi automaton $B$ specifies undesirable behaviour. The
question whether a model $M$ conforms to the specification then
reduces to the question whether there is an initialised fair path in the product $M
\times B$~\cite{VW86b}. As both $M$ and $B$ are finite state the
search for such a path can be restricted to lasso-shaped paths,
i.e., paths which are of form $\beta \gamma^\omega$, where
$\beta$ and $\gamma$ are finite paths.

If a witness to the violation of the specification is to be extracted
from an initialised fair path in the product of $M$ and $B$ for debugging, it is
desirable that this path is short. A B{\"u}chi automaton is
\emph{tight} iff for every $\alpha = \beta \gamma^\omega \in \Lang{B}$
it has an accepting run $\rho = \sigma \tau^\omega$ such that $\alpha$
and $\rho$ have the same shape: $|\beta| = |\sigma|$ and $|\gamma| =
|\tau|$~\cite{SB05,KV01}. Hence, the B{\"u}chi automaton can adapt as
a chameleon to the shape of any potential lasso-shaped witness.

\section{Bounded Model Checking}\label{sec:BMC}

\begin{figure}\centering\includegraphics[scale=1.0]{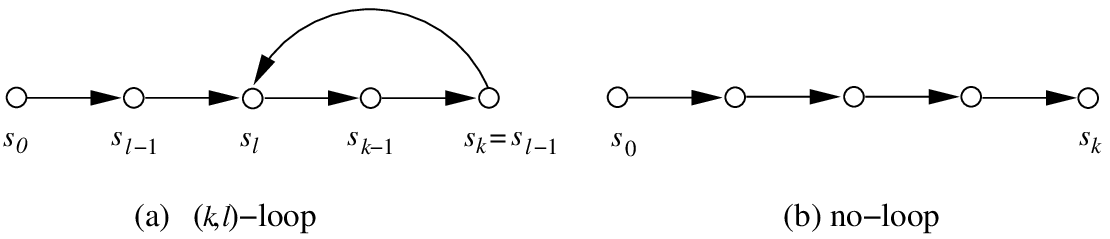}
\caption{\label{fig-loop}The two possible cases for a bounded path}
\end{figure}

The main idea of bounded model checking~\cite{BCCZ99} is to search for
\emph{bounded witnesses} for a temporal property.  A bounded witness is an
initialised infinite path in which the property holds, and which can be represented by a
finite path of length $k$.  A finite path can represent infinite behaviour,
in the following sense. In (a) the $(k,l)$-loop case the finite path forms a \emph{loop}
and contains all infinite behaviour, or (b) the no-loop case when the finite path represents 
all its infinite extensions.
More formally, an infinite path $\pi =
s_0s_1s_2\ldots$ of states contains a $(k,l)$-loop, or just a $k$-loop, if
$\pi = (s_0s_1\ldots s_{l-1})(s_l \ldots s_k)^\omega$ such that $0 < l \leq k$
and $s_{l-1} = s_k$.  The two cases we
consider are depicted in Fig.~\ref{fig-loop}.

In BMC all possible $k$-length bounded witnesses of the \emph{negation} of the
specification are encoded as a SAT problem. The bound 
$k$ is increased until either a witness is found (the instance is satisfiable) 
or a sufficiently high value of $k$ to guarantee completeness is reached. 

Note that as in~\cite{FSW02,BC03,LBHJ04,LBHJ05,HelJunLat:CAV05} the shape of the loop and accordingly the
meaning of the bound $k$ is slightly different from~\cite{BCCZ99}. In
this paper a finite path of length $k$ always has $k$ transitions, and 
an infinite path with a loop contains the \emph{looping state} twice, 
at position $l-1$ and at position $k$.

Bounded model checking uses a \emph{bounded semantics of PLTL} which safely under-approxi\-mates
the normal semantics. It allows us to use a bounded prefix $\pi_k=s_0s_1\ldots s_k$ of an 
initialised infinite path $\pi$ to check the formula. 
The semantics is split into two cases.
If the infinite path $\pi$ is a $k$-loop a different semantics is used than in the case
where it is not a $k$-loop.
The definition below assumes the formula is in 
positive normal form.

\begin{definition}(See also~\cite{BCCZ99,FSW02}.)
\label{def:boundedsemantics}
Given an initialised infinite path $\pi$ and bound $k\in\mathbb{N}$, 
$\pi \models_k\TLF$ iff (a) $\pi$ is a $(k,l)$-loop
for some $0 < l \leq k$ and $\pi^0 \models \TLF$, or (b) $\pi^0 \modelsnl\TLF$,
where:
{\small
\begin{displaymath}
\begin{array}{l}
\begin{array}{lcl}
\pi^i \modelsnl p & \Leftrightarrow & \pi^i \models p. \\
\pi^i \modelsnl\neg p & \Leftrightarrow & \pi^i \models \neg p. \\
\pi^i \modelsnl\LSF\wedge\RSF & \Leftrightarrow & \pi^i \modelsnl\LSF \ \mathit{and} \ \pi^i \modelsnl\RSF. \\
\pi^i \modelsnl\LSF\vee\RSF & \Leftrightarrow & \pi^i \modelsnl\LSF \ \mathit{or} \ \pi^i \modelsnl\RSF. \\
\pi^i \modelsnl\Next{\SF} & \Leftrightarrow & i<k \text{ and } \pi^{i+1} \modelsnl \SF. \\
\pi^i \modelsnl\Until{\LSF}{\RSF} & \Leftrightarrow &
\exists i\le j\le k \text{ such that } \pi^j \modelsnl \RSF \text{ and } \pi^n \modelsnl\LSF \text{ for all } i\le n < j. \\
\pi^i \modelsnl\Release{\LSF}{\RSF} & \Leftrightarrow &
\exists i\le j\le k \text{ such that } \pi^j \modelsnl \LSF \text { and } \pi^n \modelsnl \RSF \text{ for all } i\le n\le j. \\
\pi^i\modelsnl \Prev{\SF} & \Leftrightarrow & i> 0 \text{ and } \pi^{i-1}\modelsnl\SF. \\
\pi^i\modelsnl \PrevZ{\SF} & \Leftrightarrow & i=0 \text{ or } \pi^{i-1}\modelsnl\SF. \\
\pi^i\modelsnl \Since{\LSF}{\RSF} & \Leftrightarrow & \exists 0\leq j\leq i \text{ such that } \pi^{j}\modelsnl\RSF
 \text{ and } \pi^{n}\modelsnl\LSF \text{ for all } j< n\leq i. \\ 
\pi^i\modelsnl \Trigger{\LSF}{\RSF} & \Leftrightarrow & \text{for all } 0\leq j\leq i: \pi^{j}\modelsnl\RSF
\text{ or } \pi^{n}\modelsnl\LSF \text{ for some } j<n\leq i. \\ 
\end{array}
\end{array}
\end{displaymath}
}
\end{definition}
Because the language defined by the models of a PLTL formula belong to 
the $\omega$-regular languages, we can restrict ourselves to searching 
for ultimately periodic witnesses in
our models. Notice that for every ultimately periodic infinite path $\pi$,
the bounded semantics becomes equivalent to the exact semantics when the
$k$ grows large enough to represent $\pi$ as a $(k,l)$-loop. Thus 
for a model $M$ and a PLTL property $\TLF$
there always exists some $k \in \mathbb{N}$ such that the bounded semantics
becomes exact, i.e., $M \models\TLF$ iff $M \models_k\TLF$.

\subsection{Original BMC Encoding for LTL}\label{sec:ltl-orig}

The original encoding~\cite{BCCZ99} is defined recursively over the
structure of the LTL formula $\psi$ and the current position $i$.
It is parameterised by the bound $k$, the start of the loop $l$ and
closely follows the bounded semantics of Def.~\ref{def:boundedsemantics}.
Therefore, for fixed $i$, $k$, and $l$, each subformula $\Finally{\psi_1}$
resp.~$\Globally{\psi_1}$ of $\psi$ requires constraints of size $O(k)$
using the encoding of $\psi_1$ at various positions.  The binary operators
$\Until{}{}$ and ~$\Release{}{}$ need constraints of size $O(k^2)$. Since
the encoding of a subformula $\psi_2$ is only dependent on $i$, $l$, and
$k$, and, in particular, multiple occurrences of the encoding of $\psi_2$ under
the same set of parameters can be shared, the overall size can be bounded by
$O(|\psi| \cdot k^4)$.  

Parts of the constraints can be shared for different $i$. This reduces the
overall complexity of the original encoding to $O(|\psi| \cdot k^3)$.  It
can be reduced even further to $O(|\psi|\cdot k)$, if only unary future temporal
operators occur in $\psi$.  As example consider the formula $\psi \equiv
\Finally{\Globally p}$.  As shown in~\cite{CPRS02,LBHJ04} a linear encoding of
$\psi$ can be obtained by optimising the original encoding using
associativity and sharing.  The encoding of, for instance, $\Globally{(r \to
(\Until{p}{q}))}$ is at least quadratic no matter what simplifications based
on sharing and associativity are used~\cite{LBHJ04}.

Even if more sophisticated circuit optimisations would allow to reduce the
cubic original encoding to linear size, it is much more natural to start with
a linear encoding in the first place.  Finally, the original encoding
translates \emph{looping} and \emph{non-looping} witnesses separately, while
more advanced encodings, as discussed in this article, combine both.

It is tempting to use the recursive one step identities of the (unbounded) semantics
of temporal operators $\Until{\LSF}{\RSF} \equiv \RSF \vee ( \LSF \wedge \Next{(\Until{\LSF}{\RSF})})$ and
$\Release{\LSF}{\RSF} \equiv \RSF \wedge ( \LSF \vee \Next{(\Release{\LSF}{\RSF})})$
without any notion of fairness
to encode LTL in a straightforward way, as for
instance suggested in~\cite{BCCSZ03}.  In order to represent all witnesses
for $\Globally{p}$ in a Kripke structure consisting of a single state with a self loop the
following propositional formula can be used:
{\encsize
\[
I(s_0)
\;\wedge\;
T(s_0,s_0)
\;\wedge\;
\Trans{\Globally{p}}_0
\;\wedge\;
(\Trans{\Globally{p}}_0 \Leftrightarrow p_0 \wedge \Trans{\Globally{p}}_0).
\]
}\smallskip%

\noindent 
Note the direct translation of the one step identity of the semantics of
$\Globally{}$ on the right side.  In this case, and in general for temporal
operators with greatest fix-point semantics, this construction is sound,
because the existence of an arbitrary fix-point implies the existence of the
greatest fix-point and the recursively defined variable, denoted $\Trans{\Globally{p}}_0$, has only positive
occurrences. 

If applied in a naïve way, the same construction is incorrect for temporal
operators with \emph{least} fix-point semantics, as the following
example shows.  Again we are interested in all witnesses consisting of a
single state with a self loop, but now for the LTL formula $\Finally{p}$.
Using the same construction as above, simply following the one step LTL identities of
$\Finally$ without any notion of fairness
the following propositional encoding is obtained:
{\encsize
\[
I(s_0)
\;\wedge\;
T(s_0,s_0)
\;\wedge\;
\Trans{\Finally{p}}_0
\;\wedge\;
(\Trans{\Finally{p}}_0 \Leftrightarrow p_0 \vee \Trans{\Finally{p}}_0).
\]
}

\noindent
This formula can always be satisfied as long the transition relation has a self loop
in an initial state by setting the boolean variable $\Trans{\Finally{p}}_0$ to $\top$.
Therefore it will hold even if $p$ is false in the initial state, 
and the encoding is therefore incorrect.

\section{Improved Encodings of Bounded Model Checking for LTL}\label{sec:ltl-bmc}
In this section several alternative bounded model checking encodings
for LTL (i.e., PLTL without past temporal operators) are presented.
How to extend the approaches to full PLTL containing also past temporal
formulas is the topic of Sect.~\ref{sec:pltl-bmc}.

\subsection{BMC for LTL with Fixpoint Evaluation}\label{sec:ltl-fixe}

One of the key factors affecting the efficiency of BMC is the size of the resulting SAT encoding. 
If the encoding produces unnecessarily large formulas
the solver can quickly be overwhelmed, and we may not be able to proceed
deep enough to find all violations to the specification in the design
under model checking.

In~\cite{LBHJ04} we presented a BMC encoding to SAT for LTL which is linear 
in $k$ that outperformed previous encodings. 
It consists of three types of constraints on the state variables representing
the possible paths of length $k$: model constraints, loop constraints 
and LTL constraints. \emph{Model constraints} $\Trans{M}_k$ encode legal initialised 
finite paths of the model $M$ of length $k$:
{\encsize
\[
\Trans{M}_k \Leftrightarrow I(s_0)\wedge\bigwedge_{i=1}^k T(s_{i-1},s_i),
\]
}\smallskip%

\noindent
where $I(s)$ is the initial state predicate and $T(s,s^\prime)$ is a total transition
relation predicate. The \emph{loop constraints} are used to non-deterministically select 
loops of paths encoded by the model constraints. 
We introduce $k+1$ fresh \emph{loop selector variables} $l_0,\ldots,l_k$ which determine 
where the path loops. At most one loop selector variable is allowed to be true. 
If $l_j$ is true then $s_{j-1}=s_k$, i.e., the bit vectors representing the state
$s_{j-1}$ and state $s_k$ have bitwise identical values.
In this case the LTL constraints treat the
bounded path as a $(k,j)$-loop. If no loop selector variable is true then the
LTL constraints treat the path as not having a loop  
(the \emph{no-loop} case). 
Some counterexamples can be detected at lower bounds with the no-loop case 
(informative safety counterexamples of~\cite{KV01} to be exact).
The loop constraints are encoded by conjuncting the constraints below.
Only the loop selector variables $\lvar{i}$ require fresh unconstrained
variables; everything else can be implemented as constraints, i.e., variables that
are constrained to be functionally dependent on the other variables of the formula.
We denote the constraints by $\Trans{\mathit{LoopConstraints}}_k$:
{\encsize
\[\begin{array}{l|rcl}
\hline
\mathrm{Base} & \lvar{0} & \Equiv & \bot\\[1ex]
&\InLoop{0} & \Leftrightarrow & \bot\\[1ex]
\hline
& \lvar{i} & \Rightarrow & (s_{i-1}=s_k) \\[1ex]
1\leq i\leq k&\InLoop{i} & \Leftrightarrow & \InLoop{i-1}\vee \lvar{i},\\[1ex]
&  \InLoop{i-1} & \Implies & \neg\lvar{i}\\[1ex]  
& \LoopExists & \Equiv &  \InLoop{k}\\[1ex]
\end{array}\]
}\smallskip%

\noindent
$\mathit{InLoop}_i$ is true if the position $i$ is in the loop part of the path.
The loop selector variables indicate where the bounded path loops and select either a $(k,j)$-loop
when $l_j$ holds\footnote{There is at most one index $j$ where $l_j$ holds, as 
otherwise $\Trans{\mathit{LoopConstraints}}_k$ would be unsatisfiable.}
or the no-loop case when no $l_j$ holds.
In the $(k,j)$-loop case the variable $\LoopExists$ will be true and
in the no-loop case it will be false.
Finally, the \emph{LTL constraints} check if the bounded path defined by the model
constraints and loop constraints is a model of the LTL formula.
The LTL encoding utilises the fact that for $(k,l)$-loops
the semantics of CTL and LTL coincide, see e.g., \cite{KV01,TauHel:STTT02}.
The intuitive reason is that if each state has exactly one successor (i.e., the path is lasso-shaped)
then the semantics of the path quantifiers $\mathbf{A}$ and $\mathbf{E}$ of CTL agree.
An LTL formula can therefore be evaluated in a lasso-shaped Kripke structure by a CTL model
checker in linear time by prefixing each temporal operator by an $\mathbf{E}$ path
quantifier~\cite{TauHel:STTT02}, which results in a CTL formula.\footnote{Naturally, we could
also use the $\mathbf{A}$ path quantifier.}
The encoding can be seen as a~CTL model checker for lasso-shaped
Kripke structures based on using the least and greatest fixpoint characterisations
of $\Until{}$ and $\Release{}{}$. 
In CTL the until operator $\mathbf{E}(\Until{\psi_1}{\psi_2})$ can be evaluated by computing the least
fixed point $\mathbf{E}(\Until{\psi_1}{\psi_2})=\mu Z . \psi_2 \vee (\psi_1\wedge\mathbf{E}\Next{Z})$
while the release operator $\mathbf{E}(\Release{\psi_1}{\psi_2})$ can be
evaluated by computing the
greatest fixpoint $\mathbf{E}(\Release{\psi_1}{\psi_2})=\nu Z . \psi_2 \wedge (\psi_1\vee\mathbf{E}
\Next{Z})$, see e.g., \cite{CGP99}. 
The encoding model checks lasso-shaped Kripke structures by computing 
the least and greatest fixpoints for $\Until{}{}$ and $\Release{}{}$.

Given a formula $\TSF$ we denote by $\Trans{\TSF}_i$ the Boolean formula  
for computing the truth value of $\TSF$ at position $i$. To evaluate whether
a formula $\TSF$ holds in the initial state we must generate the formula for 
$\Trans{\TSF}_0$.
The computation of the fixpoints
for $\Until{}{}$ and $\Release{}{}$ is done in two parts. The \emph{auxiliary translation}
$\ATrans{\cdot}$ computes an over-approximation for greatest fixpoints and
an under-approximation for least fixpoints. The approximations are refined to exact 
values by $\Trans{\cdot}$.
The auxiliary translation $\ATrans{\cdot}$ under-approximates $\Until{\LSF}{\RSF}$-formulas
by assuming that $\Until{\LSF}{\RSF}$ does \emph{not hold} in the successor of the end state $s_k$.
Conversely, $\Release{\LSF}{\RSF}$ is over-approximated by assuming that $\Release{\LSF}{\RSF}$
\emph{holds} in the successor of the end state $s_k$. Both of these approximations are exact at
the loop point $j$ where $l_j$ holds, because of the
simple looping structure of the models.

The encoding can be understood as a recursively defined function where there is a 
case for each logical or temporal connective.
For propositional LTL formulas the encoding is as follows:
{\encsize
\[
\begin{array}{c@{\quad}|@{\quad}c@{\quad}|}
\Trans{\TSF}_i & 0\leq i\leq k \\
\hline&\\[-2ex]
\Trans{p}_i & 
  p \in L(s_i)
\\[1ex]
\Trans{\neg p}_{i} &
  p \not \in L(s_i)
\\[1ex]
\Trans{\LSF\wedge\RSF}_i &
\Trans{\LSF}_i\wedge\Trans{\RSF}_i  
\\[1ex]
\Trans{\LSF\vee\RSF}_i &
\Trans{\LSF}_i\vee\Trans{\RSF}_i  
\end{array}
\]
}\smallskip%

The encoding for temporal LTL formulas is as follows:
{\tightencsize
\[
\begin{array}{c@{\quad}|@{\quad}c@{\quad}|@{\quad}c|}
\Trans{\TSF}_i & 0 \leq i < k  & i=k \\
\hline& & \\[-2ex]
\Trans{\Next{\LSF}}_{i} &
\Trans {\LSF}_{i+1} &
\bigvee_{j=1}^k \left( l_{j}\wedge \Trans{\LSF}_{j}\right) 
\\[1ex]
\Trans{\Until{\LSF}{\RSF}}_{i} &
\Trans{\RSF}_{i}\vee
\left(\Trans{\LSF}_i\wedge\Trans{\Until{\LSF}{\RSF}}_{i+1}\right) &
\Trans{\RSF}_{i}\vee\left(\Trans{\LSF}_i\wedge\left(\bigvee_{j=1}^k \left( l_{j}\wedge \ATrans{\Until{\LSF}{\RSF}}_j\right)\right)\right)
\\[1ex]
\Trans{\Release{\LSF}{\RSF}}_{i} &
\Trans{\RSF}_{i}\wedge
\left(\Trans{\LSF}_i\vee\Trans{\Release{\LSF}{\RSF}}_{i+1}\right) &
\Trans{\RSF}_{i}\wedge\left(\Trans{\LSF}_i\vee\left(\bigvee_{j=1}^k \left( l_{j}\wedge \ATrans{\Release{\LSF}{\RSF}}_j\right)\right)\right) 
\end{array}
\]
}\smallskip

\noindent The until (release) formulas at $k$ refer to an auxiliary translation
$\ATrans{\Until{\LSF}{\RSF}}_j$ ($\ATrans{\Release{\LSF}{\RSF}}_j$) at
the loop point $j$ where $l_j$ holds. It computes an approximation of
the semantics of until (release). If a loop exists this approximation
is, in fact, exact for $\Until{\LSF}{\RSF}$ ($\Release{\LSF}{\RSF}$)
at the loop index $j$ corresponding to the time point $k+1$ in the $(k,j)$-loop path.
In the no-loop case the effect of the encoding 
at index $k$ is the same
as if all subformulas at the index $k+1$ would be evaluated to $\bot$.

The auxiliary encoding for temporal LTL formulas is as follows:
{\encsize
\[
\begin{array}{c@{\quad}|@{\quad}c@{\quad}|@{\quad}c|}
\Trans{\TSF}_i &  1\leq i < k & i=k 
\\[1ex]
\hline & &\\[-2ex]
\ATrans{\Until{\LSF}{\RSF}}_i & 
\Trans{\RSF}_{i}\vee
\left(\Trans{\LSF}_i\wedge\ATrans{\Until{\LSF}{\RSF}}_{i+1} \right) & \Trans{{\RSF}}_{k} \\[1ex]
\ATrans{\Release{\LSF}{\RSF}}_i & 
\Trans{\RSF}_{i}\wedge
\left(\Trans{\LSF}_i\vee\ATrans{\Release{\LSF}{\RSF}}_{i+1} \right) & \Trans{{\RSF}}_{k}\\
\end{array}
\]
}\smallskip%

\noindent Because the semantics of until is a least fixpoint,
the encoding of $\ATrans{\Until{\LSF}{\RSF}}_{k}$ is just the simplified form
of the expression
$\Trans{\RSF}_{k} \vee \left(\Trans{\LSF}_{k} \wedge \bot\right)$,
where $\ATrans{\Until{\LSF}{\RSF}}_{k+1}$ has been replaced by $\bot$.
Similarly, because the semantics of release is a greatest fixpoint, we have
$\Trans{\RSF}_{k} \wedge \left(\Trans{\LSF}_{k} \vee \top\right)$ for
$\ATrans{\Release{\LSF}{\RSF}}_{k}$.

The conjunction of these three sets of constraints forms the full \emph{fixpoint evaluation encoding} of the
boun\-ded model checking problem into SAT:
{\encsize
\[
\Trans{M,\TLF,k} \Leftrightarrow \Trans{M}_k\wedge\Trans{\mathit{LoopConstraints}}_k\wedge\Trans{\TLF}_0.
\]
}

\noindent We have the following result:
\begin{theorem}\label{thm:ltl-fixe}
Given a Kripke structure $M$ and an LTL formula $\TLF$, 
$M$ has an initialised path $\pi$ such that $\pi\models\TLF$ iff there exists a $k\in\mathbb{N}$ such
that the fixpoint evaluation encoding $\Trans{M,\TLF,k}$ is satisfiable. In particular, if
$\pi\models_k\TLF$ then the fixpoint evaluation encoding $\Trans{M,\TLF,k}$ is satisfiable.
\footnote{As immediate corollary minimal length $(k,l)$-loop counterexamples for LTL can be detected.
The encoding also detects minimal length informative safety counterexamples for LTL\@.}
\end{theorem}
\proof
We first prove a stronger result than the second part of the theorem:
$M$ has an initialised path $\pi$ such that $\pi\models_k\TLF$ iff the fixpoint
evaluation encoding $\Trans{M,\TLF,k}$ is satisfiable.
The first part of the theorem follows from this together with the fact that
when the bound $k$ is increased large enough $M \models \TLF$ iff $M \models_k \TLF$.

It is easy to see that the model constraints $\Trans{M}_k$ encode all legal
initialised finite paths $\pi'$ of the model $M$ of length $k$.
Now consider the loop
constraints $\Trans{\mathit{LoopConstraints}}_k$.
As in the definition of
the semantics of $\models_k$, we have two cases: (a) $\pi'$ is a $(k,j)$-loop
for some $j$ inducing an infinite path $\pi$:
In this case by setting $l_j$ to true and all other $l_i$ to false
the truth values of all other variables in $\Trans{\mathit{LoopConstraints}}_k$
are uniquely determined, in particular
$\LoopExists$ will be true. Because $s_{j-1} = s_k$ we can satisfy all constraints in
$\Trans{\mathit{LoopConstraints}}_k$.
It is also easy to check that if more than one $l_i$ variable is true, these
constraints are unsatisfiable.
The second case is: (b) We are in the no-loop case: $\pi'$ is a finite prefix
of some initialised infinite path $\pi$ through the system.
The only remaining option is that all $l_i$ variables are false. Now 
the truth values of all other variables in $\Trans{\mathit{LoopConstraints}}_k$
are again uniquely determined,
in particular $\LoopExists$ will be false. Thus all constraints in 
$\Trans{\mathit{LoopConstraints}}_k$ are satisfied.

Consider a satisfying truth assignment of
$\Trans{M}_k\wedge\Trans{\mathit{LoopConstraints}}_k$
inducing an initialised infinite path $\pi$.
We want to check that it can be extended to a model of the
full encoding $\Trans{M,\TLF,k}$ iff $\pi \models_k \TLF$. Because the encoding
$\Trans{\TLF}_0$ is just a Boolean circuit, the truth value of each of its
nodes are uniquely determined by the other variables of the encoding,
and we will evaluate these values in what follows.

We will prove by induction on the structure
of the LTL formula $\TLF$ that for all  $\TSF \in \mathit{cl}(\TLF), 0 \leq i \leq k$:
$\pi^i\models_k \TSF$ iff $\Trans{\TSF}_i$ is true. In particular,
$\pi \models_k \TLF$ iff $\Trans{\TLF}_0$ is true.

The cases where $\TSF$ is an atomic proposition or its negation are trivial
in both cases (a) and (b). The same holds for all propositional cases, where
the claim holds for the subformulas by the induction hypothesis.

What remains to be proven are the cases where $\TSF$ is a temporal operator.
Because the encoding of $\Trans{\TSF}_i$ for all indices $0 \leq i < k$
just uses the one-step identities for LTL formulas, the claim
holds for all of them provided that for the last index $k$ it holds that
$\pi^k\models_k \TSF$ iff $\Trans{\TSF}_k$ is true.

First consider the easier no-loop case (b): By the above we have that none of the $l_i$
variables is true. In this case we can simplify the encoding by substituting
$\bot$ for every $l_i$ variable and simplifying the result. After doing this
it is easy to check that the encoding of $\Trans{\TSF}_k$ behaves as
if $\pi^{k+1} \not \models \LSF$ for all subformulas
$\LSF \in \mathit{cl}(\TLF)$. It is now easy to check that at index $k$ this matches
the definition of the no-loop semantics $\modelsnl$ for all temporal operators, and thus
the semantics matches $\modelsnl$ also for all indexes $0 \leq i < k$.

Now consider the $(k,j)$-loop case (a): 
Recall that
an LTL formula can be evaluated in a lasso-shaped Kripke structure by a CTL model
checker by prefixing each temporal operator by an $\mathbf{E}$ path
quantifier~\cite{TauHel:STTT02}, which results in a CTL state formula. Thus 
in a $(k,j)$-loop we need to
only consider the truth value of LTL formulas at indexes $0 \leq i \leq k$,
as the truth values for any larger index, for example $i=k+1$, can be reduced to evaluating the
LTL formula at the corresponding state of the model, in this case the loop state $i=j$.

By the above we know that $l_j$ is the
only loop selector variable which is true, and that the subformulas are correctly evaluated
for all indices by the induction hypothesis. If $\TSF = \Next{\LSF}$, the encoding
of $\Trans{\TSF}_k$ picks the truth value of $\LSF$ from $\Trans{\LSF}_j$
corresponding to the index $k+1$ in the $(k,j)$-loop (recall that $l_j$
is the only loop selector variable which holds), and we are done.

In the case $\TSF = \Until{\LSF}{\RSF}$ we have to do a case analysis.
First consider
case (i): $\pi^{i} \models \RSF$ for some $j \leq i \leq k$, and therefore 
$\pi^i\models\Until{\LSF}{\RSF}$. Without loss of generality, pick the smallest such $i$. 
Now clearly at index $i$ the auxiliary translation 
$\ATrans{\Until{\LSF}{\RSF}}_i$ is true. Because the auxiliary translation $\ATrans{\Until{\LSF}{\RSF}}_n$
for all indices $j \leq n \leq i$ is just the one-step identity of until,
$\ATrans{\Until{\LSF}{\RSF}}_n$ is true iff $\pi^n \models \Until{\LSF}{\RSF}$.
In particular, at the loop point $j$ we have:
$\ATrans{\Until{\LSF}{\RSF}}_j$ is true iff $\pi^j \models \Until{\LSF}{\RSF}$.
Now consider case (ii): $\pi^{i} \not \models \RSF$ for all $j \leq i \leq k$.
In this case clearly $\pi^j \not \models \Until{\LSF}{\RSF}$.
It is now easy to check
from the definition of the auxiliary encoding
that $\ATrans{\Until{\LSF}{\RSF}}_n$ is false for all
indices $j \leq n \leq k$. 
In both cases we have
$\ATrans{\Until{\LSF}{\RSF}}_j$ is true iff $\pi^j \models \Until{\LSF}{\RSF}$,
and because the encoding of $\Trans{\Until{\LSF}{\RSF}}_k$ uses 
$\ATrans{\Until{\LSF}{\RSF}}_j$ to obtain the value of $\pi^{k+1} \models \Until{\LSF}{\RSF}$,
we have $\pi^{k} \models \Until{\LSF}{\RSF}$ iff $\Trans{\Until{\LSF}{\RSF}}_k$ is true.

In the case $\TSF = \Release{\LSF}{\RSF}$ we have to do a very similar (dual) case analysis.
First consider
case (i): $\pi^{i} \not \models \RSF$ for some $j \leq i \leq k$, and therefore 
$\pi^\not\models\Release{\LSF}{\RSF}$. Without loss
of generality, pick the smallest such $i$. Now clearly at index $i$ the auxiliary translation 
$\ATrans{\Release{\LSF}{\RSF}}_i$ is false. Because the auxiliary translation $\ATrans{\Release{\LSF}{\RSF}}_n$
for all indices $j \leq n \leq i$ is just the one-step identity for release,
$\ATrans{\Release{\LSF}{\RSF}}_n$ is true iff $\pi^n \models \Release{\LSF}{\RSF}$.
In particular, at the loop point $j$ we have:
$\ATrans{\Release{\LSF}{\RSF}}_j$ is true iff $\pi^j \models \Release{\LSF}{\RSF}$.
Now consider case (ii): $\pi^{i} \models \RSF$ for all $j \leq i \leq k$. 
In this case clearly $\pi^j \models \Release{\LSF}{\RSF}$.
It is now easy to check
from the definition of the auxiliary encoding
that $\ATrans{\Release{\LSF}{\RSF}}_n$ is true for all
indices $j \leq n \leq k$. 
In both cases we have
$\ATrans{\Release{\LSF}{\RSF}}_j$ is true iff $\pi^j \models \Release{\LSF}{\RSF}$,
and because the encoding $\Trans{\Release{\LSF}{\RSF}}_k$ uses 
$\ATrans{\Release{\LSF}{\RSF}}_j$ to obtain the value of $\pi^{k+1} \models \Release{\LSF}{\RSF}$,
we have $\pi^{k} \models \Release{\LSF}{\RSF}$ iff $\Trans{\Release{\LSF}{\RSF}}_k$ is true.

Thus by forcing the top level formula $\Trans{\TLF}_0$ to be true we get that
$M$ has an initialised path $\pi$ such that
$\pi\models_k\TLF$ iff $\Trans{M,\TLF,k}$ is satisfiable, from which the full theorem
follows.
\qed

The encoding has a few desirable properties of which the most important one is that when
the encoding is seen as a Boolean circuit where the loop selector variables and 
the atomic propositions of the model are input variables, the size of the generated formula 
is $O(\vert I\vert + k\cdot\vert T\vert + k \cdot \vert\TLF\vert)$. The encoding also has 
a \emph{unique model property} in the following sense: if the $(k,l)$-loop is given 
(i.e., the computation $\pi$ together with the $l_j$ variables
are fixed), the Boolean circuit representing the LTL encoding has no free variables.
Consequently, there is no nondeterminism in evaluating the circuit that evaluates
the LTL formula, and if the encoding is satisfiable
the given $(k,l)$-loop defines a unique model of the Boolean circuit.

If the loop selector variables, atomic propositions and 
their negations are seen as inputs to the circuit, the circuit for the LTL encoding $\Trans{\TLF}_0$ 
is monotonic. This can be exploited to devise an improved encoding of the Boolean
circuit to conjunctive normal form (CNF) formulas.
A similar optimisation has been presented in the encoding of~\cite{FSW02}.


The original encoding~\cite{BCCZ99} and its improved version~\cite{CPRS02} both result 
in formulas that are at least quadratic w.r.t.\ $k$. Frisch et al.~\cite{FSW02} have presented 
an alternative encoding based on normal forms for LTL\@. This so-called fixpoint encoding is more 
efficient than previous attempts, but it produces formulas that
are non-linear w.r.t.\ $k$~\cite{LBHJ04}. An improved version of the fixpoint encoding,
which includes a generalisation to PLTL, is linear w.r.t.\ $k$
but does not provide minimal length counterexamples for PLTL formulas~\cite{CiRoSh:FMCAD04}.
\iffalse
Unfortunately, \cite{CiRoSh:FMCAD04} contains a bug in the way it handles
formulas containing past temporal operators. However, this bug can be easily fixed
(the required fix is mentioned in Sect.~\ref{sec:pltl-bmc}).
\else
Note, that \cite{CiRoSh:FMCAD04} contains an ambiguity in its description
that may lead an implementation choice that results in
wrong handling of formulas containing past temporal operators. For
details see Sect.~\ref{sec:pltl-bmc}.
\fi
The normal form used in the fixpoint encoding~\cite{FSW02} is similar to tableau methods 
for constructing a symbolic B\"uchi automaton ${\mathcal{A}}_\TLF$ representing an LTL formula $\TLF$. 
It is also possible to do BMC by applying the automata theoretic approach
and symbolically encode a product system $M\times{\mathcal{A}}_{\neg\TLF}$~\cite{MRS02,ClarkeKroeningOuaknineStrichman05}.  
BMC is performed by searching for fair loops 
in the product system. 
This approach produces a linear size encoding
if the search for fair loops is encoded with an encoding such as~\cite{CPRS02,LBHJ04} 
that can encode $\Globally{\Finally{p}}$ in linear size. 
Since this method only searches for looping counterexamples, it must sometimes go 
deeper than other methods also accepting no-loop safety counterexamples.

See~\cite{HelNie03} for earlier work on linear size bounded model checking encodings
for LTL employing logic programs with the stable model semantics instead of using SAT\@. This
work does not directly give us a linear size SAT encoding because the best known
automatic translation from logic programs with the stable model semantics into SAT
are super-linear (roughly $O(n \log_2 n)$), see~\cite{Janhunen04:ecai}.

\subsection{BMC for LTL with Eventualities}\label{sec:ltl-eventuality}
An alternative approach to encoding semantics of LTL formulas is to use
an \emph{eventuality encoding}, which in the loop case requires that 
for each until formula $\Until{\LSF}{\RSF}$ the right hand side 
formula $\RSF$ holds at some point in the loop (and dually for release).
The main idea for until formulas is to first evaluate whether the \emph{eventuality formula} $\Finally{\RSF}$ holds
in the last state $k$, and use this knowledge to evaluate the value of the main encoding.
If we know that $\Finally{\RSF}$ does not hold at $k$, then surely
$\Until{\LSF}{\RSF}$ cannot hold at $k$ either. In all other cases the one-step
LTL identities actually evaluate the bounded LTL semantics correctly.
A dual construction is applied for release
formulas.
This idea above enables one to replace the auxiliary encodings of until and release
with simpler ones, but at the same time the
encoding becomes a set of Boolean equations with cyclic dependencies between
variables instead of a Boolean circuit where no such cyclic dependencies exist.
Having cyclic dependencies allows for a slightly smaller encoding but in our
opinion makes the approach a bit harder to understand.

The eventuality encoding is quite similar to the fixpoint evaluation encoding, 
so only the LTL part of the new encoding will presented. 
%
The encoding is no longer defined as a recursive function over the LTL 
formula but as Boolean constraints over the so called \emph{formula variables}
$\Trans{\TSF}_i$, which are fresh unconstrained propositional variables.
There is a variable $\Trans{\TSF}_i$ for every subformula 
$\TSF \in \mathit{cl}(\TLF)$ and for all $0\leq i \leq k+1$. The interpretation
of $\Trans{\TSF}_i$ is still that it is true iff $\TSF$ holds at position $i$ 
in the model.
For propositional LTL formulas the encoding is as follows:
{\encsize
\[
\begin{array}{c@{\quad}|@{\quad}c@{\quad}|}
\TSF  & 0\leq i\leq k+1 \\
\hline&\\[-2ex]
p_i & \Trans{p}_i \Leftrightarrow p \in L(s_i)
\\[1ex]
\neg p_i & \Trans{\neg p}_{i} \Leftrightarrow p \not \in L(s_i)
\\[1ex]
\LSF\wedge\RSF & \Trans{\LSF\wedge\RSF}_i \Leftrightarrow \Trans{\LSF}_i\wedge\Trans{\RSF}_i  
\\[1ex]
\LSF\vee\RSF & \Trans{\LSF\vee\RSF}_i \Leftrightarrow \Trans{\LSF}_i\vee\Trans{\RSF}_i  
\end{array}
\]
}\smallskip%

\noindent
The encoding for the temporal subformulas is changed to the following
(the only thing that changes is the encoding at index $k$):
{\encsize
\[
\begin{array}{c@{\quad}|@{\quad}c@{\quad}|}
\TSF &  0\leq i \leq k \\[1ex]
\hline
&\\[-2ex]
 \Next{\SF} & \Trans{\Next{\SF}}_i\Leftrightarrow
\Trans{\SF}_{i+1} \\[1ex]
 \Until{\LSF}{\RSF} & \Trans{\Until{\LSF}{\RSF}}_i\Leftrightarrow
\Trans{\RSF}_i\vee\left(\Trans{\LSF}_i\wedge\Trans{\Until{\LSF}{\RSF}}_{i+1}\right)\\[1ex]
\Release{\LSF}{\RSF} & \Trans{\Release{\LSF}{\RSF}}_i\Leftrightarrow
\Trans{\RSF}_i\wedge\left(\Trans{\LSF}_i\vee\Trans{\Release{\LSF}{\RSF}}_{i+1}\right)\\[1ex] 
\end{array}
\]
}\smallskip%

\noindent
To compensate for the change at index $k$ we will for each subformula $\TSF \in \mathit{cl}(\TLF)$
add the following constraints $\Trans{\mathit{LastStateFormula}}_k$:
{\encsize
\[
\begin{array}{c@{\quad}|@{\quad}c@{\quad}|}
\hline& \\[-2ex]
\mathrm{Base} & \neg\mathit{LoopExists}\Rightarrow\left(\Trans{\TSF}_{k+1}\Leftrightarrow\bot\right)\\[1ex]
\hline& \\[-2ex]
1\leq i \leq k & \lvar{i}\Implies\left(\Trans{\TSF}_{k+1}\Equiv\Trans{\TSF}_i\right) \\[1ex]
\end{array}
\]
}\smallskip%

\noindent
The constraints state that if there is no loop, all formula variables at index $k+1$ should evaluate to
$\bot$. This is the same as in the fixpoint evaluation encoding and results in the 
no-loop case in the the bounded LTL semantics.
For the case when a 
loop exists, the added constraints force all formula variables at index $k+1$ to get their 
values from the loop point $j$, where $l_j$ holds. Note that this can create
a cyclic dependency between variables in the Boolean equation system as $\Trans{\TSF}_{j}$ 
can depend indirectly through the states $j+1, j+2, \ldots, k-1, k$ on the value 
of $\Trans{\TSF}_{k+1}$ 
which is constrained to be equal to $\Trans{\TSF}_{j}$ itself.

The reader might be puzzled why the
$\Trans{\mathit{LoopConstraints}}_k$
contains constraints of the form:
$\lvar{i} \Rightarrow (s_{i-1}=s_k)$ while
$\Trans{\mathit{LastStateFormula}}_k$ contains analogous
constraint with off-by-one indices:
$\lvar{i}\Implies\left(\Trans{\TSF}_{k+1}\Equiv\Trans{\TSF}_i\right)$.
This is an optimisation which allows detection of no-loop safety counterexamples
one unrolling of the system transition relation earlier. This optimisation 
(used also in~\cite{FSW02,BC03,LBHJ04,LBHJ05,HelJunLat:CAV05}) could easily be undone
changing the loop shape of $(k,l)$-loops to match that of~\cite{BCRZ99} and
requiring: $\Trans{M}_k \Leftrightarrow I(s_0)\wedge\bigwedge_{i=1}^{\modification{k+1}{\scriptstyle}} T(s_{i-1},s_i)$
and $\lvar{i} \Rightarrow (s_{\modification{i}{\scriptstyle}}=s_{\modification{k+1}{\scriptstyle}})$, thus bringing the system 
and formula indices back to synch.

There is still one final piece missing because the encoding as it stands so far has
models which do not agree with the semantics of LTL.
The constraints introduced so far allow the case where $\Trans{\Until{\LSF}{\RSF}}$ is true
at all indices of the loop even if $\Trans{\RSF}$ is true at no index of the loop
(this can happen when $\Trans{\LSF}$ is true at all indices of the loop). 
This clearly violates the semantics of until and
needs to be taken care of. In such a case the SAT solver has found a solution for the evaluation
of the cyclic dependencies between until variables mentioned above, but this solution is not the 
required least fixpoint solution (see also discussion on this topic in Sect.~\ref{sec:ltl-orig}).
For release formulas the situation is less severe. It can be
the case that $\Trans{\RSF}$ holds at all indices of the loop but $\Trans{\Release{\LSF}{\RSF}}$
holds in no index. This is not fatal in the sense that in this case
the semantics of release have been under-approximated (as is also done by the no-loop safety case).
In addition, the encoding has a satisfying truth assignment where the semantics of release is, in fact,
evaluated correctly.

To disallow assignments as described above, where the eventualities of until and
release are not fulfilled, we use a set of auxiliary constraints for until 
and release subformulas. 
The constraints perform a similar function to the auxiliary encoding
of until and release in the fixpoint encoding. In the table below 
$\ATrans{\TSF}_i$ are new auxiliary formula variables used by the constraints. 
{\encsize
\[
\begin{array}{c@{\quad}|@{\quad}c@{\quad}|@{\quad}c|}
 & \TSF  & \\
\hline & &\\[-2ex]
\mathrm{Base} & \Until{\LSF}{\RSF} &\mathit{LoopExists}\Rightarrow\left(
\Trans{\Until{\LSF}{\RSF}}_k\Rightarrow\ATrans{\Finally{\RSF}}_k\right)\\[1ex] 
& \Release{\LSF}{\RSF} & \mathit{LoopExists}\Rightarrow\left(
\Trans{\Release{\LSF}{\RSF}}_k\Leftarrow\ATrans{\Globally{\RSF}}_k\right)\\[1ex]
&  \Until{\LSF}{\RSF} & \ATrans{\Finally{\RSF}}_0 \Leftrightarrow \bot\\[1ex] 
&   \Release{\LSF}{\RSF} & \ATrans{\Globally{\RSF}}_0 \Leftrightarrow \top \\[1ex]
\hline & &\\[-2ex]
 1\leq i\leq k
&  \Until{\LSF}{\RSF} & \ATrans{\Finally{\RSF}}_i \Leftrightarrow 
\ATrans{\Finally{\RSF}}_{i-1} \vee \left(\InLoop{i}\wedge\Trans{\RSF}_i\right)\\[1ex]  
&  \Release{\LSF}{\RSF} &  \ATrans{\Globally{\RSF}}_i \Leftrightarrow 
\ATrans{\Globally{\RSF}}_{i-1} \wedge \left(\neg\InLoop{i}\vee\Trans{\RSF}_i\right)\\[1ex]
\end{array}
\]
}\smallskip%

\noindent
We use the names $\ATrans{\Finally{\RSF}}_{i}$ and $\ATrans{\Globally{\RSF}}_{i}$ for the
auxiliary variables because it describes the function of the constraints well.
The constraint 
$\mathit{LoopExists}\Rightarrow\left(\Trans{\Until{\LSF}{\RSF}}_k\Rightarrow\ATrans{\Finally{\RSF}}_k\right)$
intuitively ensures that in the loop case if $\Until{\LSF}{\RSF}$ holds at $k$, then
there is some index in the loop where $\RSF$ holds. This is quite similar to,
but not technically identical to, the use of B{\"u}chi acceptance sets for ensuring the correct semantics
for until, as will be shown later. The encoding for release is only required to get the exact LTL 
semantics for release formulas. The constraint
$\mathit{LoopExists}\Rightarrow\left(\Trans{\Release{\LSF}{\RSF}}_k\Leftarrow\ATrans{\Globally{\RSF}}_k\right)$
could be safely dropped if we allow the satisfying models of the encoding to safely
under-approximate the bounded semantics instead of exactly capturing
it.\footnote{This is similar to the fact that most LTL to B{\"u}chi automata translations
do not employ acceptance sets for release.}
Dropping the auxiliary constraints could also be done for the fixpoint encoding of
Sect.~\ref{sec:ltl-fixe} by adding $\Trans{\mathit{LastStateFormula}}_k$ constraints
for release subformulas.
The intuitive idea of the auxiliary encoding is that if a loop exists,
$\ATrans{\Finally{\RSF}}_k$ ($\ATrans{\Globally{\RSF}}_k$) is the
evaluation of the formula $\Finally{\RSF}$ ($\Globally{\RSF}$) at $\pi^k$.

We denote the constraints on the formula variables and the auxiliary variables above with
$\Trans{\mathit{EventuallyLTL}}_k$.
The conjunction of these four sets of constraints and requiring that the formula
holds in the initial state forms the full \emph{eventuality encoding} of the
boun\-ded model checking problem into SAT:
{\encsize
\[
\Trans{M,\TLF,k} \Leftrightarrow \Trans{M}_k\wedge\Trans{\mathit{LoopConstraints}}_k\wedge\Trans{\mathit{LastStateFormula}}_k\wedge\Trans{\mathit{EventuallyLTL}}_k\wedge\Trans{\TSF}_0.
\]
}\smallskip%

\begin{theorem}\label{thm:ltl-eventuality}
Given a Kripke structure $M$ and an LTL formula $\TLF$, 
$M$ has an initialised path $\pi$ such that $\pi\models\TLF$ iff there exists a $k\in\mathbb{N}$ such
that the eventuality encoding $\Trans{M,\TLF,k}$ is satisfiable. In particular, if 
$\pi\models_k\TLF$ then the eventuality encoding $\Trans{M,\TLF,k}$ is satisfiable.
\end{theorem}
\proof
We proceed similarly to the proof of Thm.~\ref{thm:ltl-fixe}, and will only give the
changes to the proof needed to reflect changes in the encoding.
The only changes are the encoding of temporal subformulas at index $k$,
the use of proxy variables $\Trans{\TSF}_{k+1}$,
the new auxiliary encoding $\Trans{\mathit{EventuallyLTL}}_k$,
and the new $\Trans{\mathit{LastStateFormula}}_k$ constraints.

We will now prove by induction on the structure
of the LTL formula $\TLF$ that the eventuality encoding is satisfiable and
for all  $\TSF \in \mathit{cl}(\TLF), 0 \leq i \leq k$:
$\pi^i\models_k \TSF$ iff in the unique satisfying truth assignment of the
eventuality encoding $\Trans{\TSF}_i$ is true.

First consider the no-loop case (b): In this case, 
because $\LoopExists$ is false,
it is easy to see that
the new $\Trans{\mathit{LastStateFormula}}_k$
constraints will force the proxy variables $\Trans{\TSF}_{k+1}$ to $\bot$, and
the encoding becomes exactly the same as in the
fixpoint encoding case and thus has a unique satisfying truth assignment.
Also the new auxiliary encoding constraints will lead to a unique satisfying truth
assignment as as $\mathit{LoopExists}$ is false.

Now consider the $(k,j)$-loop case (a): 
Recall from the proof of Thm.~\ref{thm:ltl-fixe} that
in a $(k,j)$-loop we need to
only consider the truth value of LTL formulas at indexes $0 \leq i \leq k$,
as the truth values for any larger index, for example $i=k+1$, can be reduced to evaluating the
LTL formula at the corresponding state of the model, in this case the loop state $i=j$.

By earlier analysis we know that $l_j$ is the
only loop selector variable which is true, and that the the encoding for all
subformulas are correctly evaluated
for all indices by the induction hypothesis. In this case the $\Trans{\mathit{LastStateFormula}}_k$
constraints are satisfiable and uniquely
set the value of the proxy variable $\Trans{\TSF}_{k+1}$
for every subformula $\TSF \in \mathit{cl}(\TLF)$ to be equivalent to the
value of the subformula at the loop point $j$, namely $\Trans{\TSF}_j$.
Therefore we do not need to consider the index $i=k+1$ in our proofs provided
that the index $i = j$ is evaluated correctly. 

If $\TSF = \Next{\LSF}$, the encoding differs from the fixpoint evaluation encoding
only at the index $k$. The encoding
of $\Trans{\TSF}_k$ together with $\Trans{\mathit{LastStateFormula}}_k$
picks the truth value of $\LSF$ from $\Trans{\LSF}_j$
corresponding to the index $k+1$ in the $(k,j)$-loop (recall that $l_j$
is the only loop selector variable which holds), 
all constraints are satisfiable in a unique way,
and we are done.

For until and release formulas our proof strategy is the following.
We first prove that if there is a satisfying truth assignment then it must for
the end point $n=k$ have the property that $\pi^n \models_k \TSF$ iff $\Trans{\TSF}_n$ is true.
After this we observe that for both until and release formulas the following holds:
the truth assignment that matches the bounded semantics of LTL for all indexes is satisfiable.
We will simultaneously prove uniqueness by noting that any truth assignment which matches the
bounded semantics of LTL at the index $n = k$ will force all other variables of the
truth assignment to a unique value that matches the semantics of LTL for all indexes,
and also satisfies the auxiliary encoding in a unique way.
This is the case because when the truth value of $\Trans{\TSF}_k$ is fixed, for all
other $0 \leq n < k$ the formula $\Trans{\TSF}_n$ will obtain a truth value in a functional way based on
the value of $\Trans{\TSF}_{n+1}$ matching the bounded semantics of LTL.
Also the encoding of $\Trans{\TSF}_k$ is satisfiable,
as it matches the bounded semantics of LTL, and $\Trans{\TSF}_{k+1}$ matches
the value of $\Trans{\TSF}_{j}$ at the loop point $j$.
The auxiliary encoding also has a unique satisfying truth assignment
as the auxiliary encoding contains no cyclic dependencies.

In the case $\TSF = \Until{\LSF}{\RSF}$ we have to do a case analysis.
First consider
case (i): $\pi^{i} \models \RSF$ for some $j \leq i \leq k$. Without loss
of generality, pick the smallest such $i$
(intuition: the cyclic dependency over until subformulas is broken
at index $i$).
Clearly at index $i$ the auxiliary translation 
$\ATrans{\Finally{\RSF}}_i$ is true. Because of this,
the auxiliary translation $\ATrans{\Finally{\RSF}}_k$ is true, and the
corresponding auxiliary translation Base constraint is satisfied.
Therefore, $\pi^{i} \models \Until{\LSF}{\RSF}$ and 
$\Trans{\Until{\LSF}{\RSF}}_i$ is also true.
Because the encoding follows the one-step identity of until we also get for all
$j \leq n \leq i$: $\Trans{\Until{\LSF}{\RSF}}_n$ iff $\pi^{n} \models \Until{\LSF}{\RSF}$,
and from encoding at $k$ together with  $\Trans{\mathit{LastStateFormula}}_k$
that $\Trans{\Until{\LSF}{\RSF}}_k$ iff $\pi^{k} \models \Until{\LSF}{\RSF}$.
Thus we have established that for all indexes $j \leq n \leq i$ and $n = k$ the encoding matches
the semantics of LTL, and because of this and our proof strategy, 
the encoding has a unique satisfying truth assignment 
that matches the semantics of LTL for all $0 \leq n \leq k$.
Now consider case (ii): $\pi^{i} \not \models \RSF$ for all $j \leq i \leq k$.
In this case the auxiliary translation $\ATrans{\Finally{\RSF}}_k$ is false.
We have that $\pi^k \not \models \Until{\LSF}{\RSF}$, and if we set
$\Trans{\Until{\LSF}{\RSF}}_k$ to be true then
the auxiliary translation Base constraint is not satisfied. Therefore
we must set $\Trans{\Until{\LSF}{\RSF}}_k$ to false (intuition:
the cyclic dependency over until subformulas is broken
at index $k$)
which matches the bounded LTL semantics
of until at $n=k$ and also satisfies the auxiliary constraints. By our proof strategy
all other indices $0 \leq i < k$ have a unique satisfying truth assignment
obtained from the one-step identity of until based on $\Trans{\Until{\LSF}{\RSF}}_k$
matching the semantics of LTL, and also
leading to the satisfaction of the constraints $\Trans{\mathit{LastStateFormula}}_k$.
In both cases (i) and (ii) we
have for all $0 \leq n \leq k$ that
$\pi^{n} \models \Until{\LSF}{\RSF}$ iff in the unique satisfying truth assignment
$\Trans{\Until{\LSF}{\RSF}}_n$ is true.

In the case $\TSF = \Release{\LSF}{\RSF}$ we have to do a very similar (dual) case analysis.
First consider
case (i): $\pi^{i} \not \models \RSF$ for some $j \leq i \leq k$. Without loss
of generality, pick the smallest such $i$. Now clearly at index $i$ the auxiliary translation 
$\ATrans{\Globally{\RSF}}_i$ is false. Because of this,
the auxiliary translation $\ATrans{\Globally{\RSF}}_k$ is false, and the
corresponding auxiliary translation Base constraint is satisfied.
Hence $\pi^{i} \not \models \Release{\LSF}{\RSF}$, and 
$\Trans{\Release{\LSF}{\RSF}}_i$ is also false.
Because the encoding follows the bounded LTL semantics of release we also get for all
$j \leq n \leq i$: $\Trans{\Release{\LSF}{\RSF}}_n$ iff $\pi^{n} \models \Release{\LSF}{\RSF}$,
and from the encoding at $k$ together with $\Trans{\mathit{LastStateFormula}}_k$
that $\Trans{\Release{\LSF}{\RSF}}_k$ iff $\pi^{k} \models \Release{\LSF}{\RSF}$,
and we can proceed similarly to the until case.
Now consider case (ii): $\pi^{i} \models \RSF$ for all $j \leq i \leq k$.
In this case the auxiliary translation $\ATrans{\Globally{\RSF}}_k$ is true.
We have that $\pi^k \models \Release{\LSF}{\RSF}$, and if we set
$\Trans{\Release{\LSF}{\RSF}}_k$ to be false then
the auxiliary translation Base constraint is not satisfied. Therefore
we must set $\Trans{\Release{\LSF}{\RSF}}_k$ to true, satisfying the auxiliary constraints,
and we can proceed similarly to the until case.
In both cases (i) and (ii) we
have for all $0 \leq n \leq k$ that
$\pi^{n} \models \Release{\LSF}{\RSF}$ iff in the unique satisfying truth assignment
$\Trans{\Release{\LSF}{\RSF}}_n$ is true.

Now proceed similarly to the proof of Thm.~\ref{thm:ltl-fixe} to complete the proof.
\qed

The eventuality encoding has the unique model property in a very similar sense
as the fixpoint evaluation encoding: after fixing the loop point and the valuation of atomic
propositions at all time points there can only be a unique valuation
of the variables of the encoding that satisfies all the constraints. 
However, this cannot be explained by the fact that the encoding is a Boolean circuit 
(it is not, as it contains cyclic dependencies
between variables); it follows from Thm.~\ref{thm:ltl-eventuality}
that all the formula variables of the encoding are uniquely determined by the 
bounded semantics of LTL\@.

There is some similarity with the separated normal form (SNF) encodings of~\cite{FSW02,CiRoSh:FMCAD04}
for BMC and the eventuality encoding presented here in the sense that the SNF encodings
first split a (strong) until to a conjunction of a weak until and an eventuality formula, and use this to
devise the BMC encoding for all time steps. We instead use the eventuality formula to evaluate
the correct value for the (strong) until formula at the last state $k$ only.

\subsection{BMC for LTL with B{\"u}chi Automata}\label{sec:ltl-buchi}
The knowledgeable reader has certainly noticed the close
correspondence between our eventuality encoding and the use of
B{\"u}chi automata symbolically implementing the tableau construction
\cite{LP85} for LTL model checking,
such as \cite{BurchClarkeMcMillanDillHwang92,CGH97,KPR98,Schneider01}.
Wolper, Vardi and Sistla were the first to show how to compile LTL
directly into B{\"u}chi automata~\cite{WolperVardiSistla83,VW94}.
\iffalse
Improved versions are used today mainly in explicit-state (e.g., SPIN~\cite{Holzmann03})
\else
Gerth et al.~\cite{GPVW95} suggested an algorithm
that produces smaller automata. It has subsequently been
improved by a number of authors~\cite{Cou99,DGV99,SB00,EH00,GO01,GO03,SebastianiTonetta03}. These improved
versions are used today mainly in explicit-state (e.g., SPIN~\cite{Holzmann03})
\fi
but also in some symbolic model checkers (e.g., VIS~\cite{VIS}).

In symbolic treatment of LTL, a compact symbolic representation of
the automaton has mostly been preferred to a small number of
states. B{\"u}chi automata for that purpose are usually symbolic
implementations of the tableau construction in~\cite{LP85}. A first
application of the tableau in symbolic context is given by Burch et
al.~\cite{BurchClarkeMcMillanDillHwang92}; for proofs and an
experimental evaluation see~\cite{CGH97}. A self-contained
presentation of symbolic model checking of LTL with past can be found
in~\cite{KPR98}. Schneider exploits the temporal hierarchy
for further optimisations~\cite{Schneider01}.

Another consideration is the depth at which the verification procedure
stops. A tight B{\"u}chi automaton is required to accept shortest
witnesses~\cite{SB05,Schuppan06,KV01}. B{\"u}chi automata constructed with
an algorithm based on~\cite{GPVW95} typically fail this criterion;
methods based on~\cite{LP85} such as~\cite{BurchClarkeMcMillanDillHwang92,CGH97} fulfil it for the future
fragment only~\cite{SB05,Schuppan06}. In Sect.~\ref{sec:pltl-buchi}
we apply the idea of virtual unrolling (see
Sect.~\ref{sec:pltl-eventuality}) to B{\"u}chi automata to obtain a
B{\"u}chi automaton with a small symbolic representation that is tight
for PLTL\@. On the other hand, Awedh and Somenzi~\cite{AS06} showed experimentally
that bounded model checking with constructions based on~\cite{LP85}
often lead to larger termination depths than with those based on~\cite{GPVW95}
if the property holds~\cite{AS06}.

Following the automata-theoretic approach~\cite{VW86b}, B{\"u}chi
automata are employed for bounded model checking of infinite state systems
in de Moura et al.~\cite{MRS02}, instead of using a dedicated encoding.
Clarke et al.~employ B{\"u}chi automata to obtain
completeness bounds for arbitrary $\omega$-regular properties~\cite{ClarkeKroeningOuaknineStrichman05}.
Awedh and Somenzi present a
complete bounded model checking procedure based on such an encoding~\cite{AS04, AS06}.
In Sect.~\ref{sec:experiments} we report on
experiments comparing the performance of a dedicated encoding with the
automata-theoretic approach in bounded model checking.
Below we first slightly modify the eventuality encoding to obtain an
encoding along the lines of
\cite{BurchClarkeMcMillanDillHwang92,CGH97}. This approach is then
generalised to show how to encode emptiness checking of the product of
a model with an arbitrary B{\"u}chi automaton.

\subsubsection{Modifying the Eventuality Encoding}

Only minor changes are needed to obtain a B{\"u}chi automata-based LTL
encoding
from the eventuality
encoding. For every until and release subformula we introduce 
new auxiliary variables $\ATrans{Acc(\cdot)}_i$. 
The auxiliary eventuality encoding needs to be replaced by
the auxiliary B{\"u}chi encoding defined as follows:
{\tightencsize
\[
\begin{array}{c@{\quad}|@{\quad}c@{\quad}|@{\quad}c|}
 & \TSF  &  1\leq i\leq k \\
\hline & &\\[-2ex]
\mathrm{Base} & \Until{\LSF}{\RSF} &\mathit{LoopExists}\Rightarrow
\ATrans{Acc(\Until{\LSF}{\RSF})}_k, \ATrans{Acc(\Until{\LSF}{\RSF})}_0 \Leftrightarrow \bot \\[1ex]
& \Release{\LSF}{\RSF} &\mathit{LoopExists}\Rightarrow
\ATrans{Acc(\Release{\LSF}{\RSF})}_k,  \ATrans{Acc(\Release{\LSF}{\RSF})}_0 \Leftrightarrow \bot \\[1ex]
\hline & &\\[-2ex]
& \Until{\LSF}{\RSF} & \ATrans{Acc(\Until{\LSF}{\RSF})}_i \Leftrightarrow 
\ATrans{Acc(\Until{\LSF}{\RSF})}_{i-1} \vee \left(\InLoop{i}\wedge \left(\Trans{\RSF}_i \vee \neg \Trans{\Until{\LSF}{\RSF}}_i \right)\right)\\[1ex]  
& \Release{\LSF}{\RSF} & \ATrans{Acc(\Release{\LSF}{\RSF})}_i \Leftrightarrow 
\ATrans{Acc(\Release{\LSF}{\RSF})}_{i-1} \vee \left(\InLoop{i}\wedge \left(\neg \Trans{\RSF}_i \vee \Trans{\Release{\LSF}{\RSF}}_i \right)\right)\\[1ex]  
\end{array}
\]
}\smallskip

\noindent
We denote the full set of modified LTL constraints with $\Trans{\mathit{B\ddot{u}chiLTL}}_k$
The conjunction of the five sets of constraints forms the full \emph{B{\"u}chi encoding} of the
boun\-ded model checking problem into SAT:
{\encsize
\[
\Trans{M,\TLF,k} \Leftrightarrow \Trans{M}_k\wedge\Trans{\mathit{LoopConstraints}}_k\wedge\Trans{\mathit{LastStateFormula}}_k\wedge\Trans{\mathit{B\ddot{u}chiLTL}}_k\wedge\Trans{\TSF}_0.
\]
}\smallskip%

By comparing to the general B{\"u}chi encoding on $(k,l)$-loops below, it is easy to see that our B{\"u}chi encoding is nothing else than an emptiness checker
for a symbolic B{\"u}chi automaton following~\cite{BurchClarkeMcMillanDillHwang92,CGH97}. The initial state predicate $\Trans{\TLF}_0$ requires
the top level formula to hold at the initial state,
the symbolic transition relation is given by the encoding rules
for propositional and temporal operators, and
acceptance sets are defined by the auxiliary translation as follows.
For each until formula $\Until{\LSF}{\RSF}$ we add an acceptance set
$F_{\Until{\LSF}{\RSF}}$ into which the states satisfying
$\Trans{\RSF}_i \vee \neg \Trans{\Until{\LSF}{\RSF}}_i$ belong to, and 
for each release formula 
$\Release{\LSF}{\RSF}$ we add an acceptance set
$F_{\Release{\LSF}{\RSF}}$ into which the states satisfying 
$\neg \Trans{\RSF}_i \vee \Trans{\Release{\LSF}{\RSF}}_i$ belong to.

\begin{theorem}\label{thm:ltl-buchi}
Given a Kripke structure $M$ and an LTL formula $\TLF$, 
$M$ has an initialised path $\pi$ such that $\pi\models\TLF$ iff there exists a $k\in\mathbb{N}$ such
that the B{\"u}chi encoding $\Trans{M,\TLF,k}$ is satisfiable. In particular, if
$\pi\models_k\TLF$ then the B{\"u}chi encoding $\Trans{M,\TLF,k}$ is satisfiable.
\end{theorem}
\proof
We prove that the auxiliary eventuality encoding is satisfiable iff
the auxiliary B{\"u}chi encoding is. The claim then follows from Thm.~\ref{thm:ltl-eventuality}.

We only show that $\LoopExists \Rightarrow
(\Trans{\Until{\LSF}{\RSF}}_k \Rightarrow \ATrans{\Finally{\RSF}}_k)$
is satisfiable if and only if $\LoopExists \Rightarrow
\ATrans{Acc(\Until{\LSF}{\RSF})}_k$ is satisfiable. The proof for
$\Release{}{}$ is similar.

The case $\neg \LoopExists$ is clear. Hence, assume $\LoopExists$ is
true. We start with the direction from left to right. First, let
$\ATrans{\Finally{\RSF}}_k$ be true. There must be $0 \le i \le k$
such that ${\InLoop{}}_i \wedge \Trans{\RSF}_i$ is true. This
immediately gives that $\ATrans{Acc(\Until{\LSF}{\RSF})}_i$ and also
$\ATrans{Acc(\Until{\LSF}{\RSF})}_k$ is true. Now, let $\neg
\Trans{\Until{\LSF}{\RSF}}_k$ be true. With $\LoopExists
\Leftrightarrow {\InLoop{}}_k$ we have
$\ATrans{Acc(\Until{\LSF}{\RSF})}_k$.

For the other direction assume $\ATrans{Acc(\Until{\LSF}{\RSF})}_k$ is
true. Hence, there exists $0 \le i \le k$ such that ${\InLoop{}}_i
\wedge (\Trans{\RSF}_i \vee \neg \Trans{\Until{\LSF}{\RSF}}_i)$ is
true. If ${\InLoop{}}_{i'} \wedge \Trans{\RSF}_{i'}$ for some $i'$ we
have $\ATrans{\Finally{\RSF}}_{i'}$ and, therefore,
$\ATrans{\Finally{\RSF}}_k$. Otherwise, there is $0 \le i' \le k$ such
that ${\InLoop{}}_{i'} \wedge \neg
\Trans{\Until{\LSF}{\RSF}}_{i'}$. By definition of the encoding for
$\Until{}{}$ we obtain ${\InLoop{}}_j
\Rightarrow \neg \Trans{\Until{\LSF}{\RSF}}_j$ for all $0 \le j < i'$ and, via $\neg
\Trans{\Until{\LSF}{\RSF}}_{k+1}$, $\neg
\Trans{\Until{\LSF}{\RSF}}_k$.
\qed

Notice that the B{\"u}chi encoding above also generates no-loop safety
counterexamples. It also  has the unique model property unlike in
most other B{\"u}chi automata constructions which do not employ acceptance sets
for release formulas. The unique model property allows us to read the exact bounded
semantics for all LTL subformulas and all time indexes considered
directly from the truth assignment given
by the SAT engine. As in the other encodings, if the unique model property
is not of interest to us, we can do what most other B{\"u}chi automata constructions
do and drop the constraint $\mathit{LoopExists}\Rightarrow
\ATrans{Acc(\Release{\LSF}{\RSF})}_k$ and the auxiliary translation for release
to obtain a slightly smaller encoding. 

\subsubsection{General Approach}\label{sec:buchi-general}

The above approach can easily be generalised to obtain an encoding to
check existence of an initialised fair path in a fair Kripke structure. If
$M=(S,T,I,L,F=\{F_0,\ldots,F_f\})$ is a fair Kripke structure, it is
sufficient to extend the loop constraints with the following B{\"u}chi
loop constraints:
{\encsize
\[
\begin{array}{c@{\quad}|@{\quad}c|}
& 0 \leq m \leq f\\[1ex]
\hline &\\[-2ex]
\mathrm{Base} & \LoopExists \Leftrightarrow \top\\[1ex]
& \LoopExists \Rightarrow \ATrans{Acc_m}_k\\[1ex]
& \ATrans{Acc_m}_0 \Leftrightarrow \bot\\[1ex]
\hline &\\[-2ex]
 1\leq i\leq k & \ATrans{Acc_m}_i \Leftrightarrow\ATrans{Acc_m}_{i-1} \vee \left(\InLoop{i}\wedge s_i \in F_m \right)\\[1ex]
\end{array}
\]
}\smallskip%

\noindent
For each acceptance set $F_m$ an additional
constraint $\ATrans{Acc_m}$ is introduced to check satisfaction of
$F_m$ in the loop. Hence, the following conjunction forms the
\emph{general B{\"u}chi encoding} of the bounded model checking
problem into SAT:
{\encsize
\[
\Trans{M,k} \Leftrightarrow \Trans{M}_k\wedge\Trans{\mathit{LoopConstraints}}_k\wedge\Trans{\mathit{B\ddot{u}chiLoopConstraints}}_k.
\]
}\smallskip%

\begin{theorem}\label{thm:ltl-gen-buchi}
Given a Kripke structure $M$, $M$ has a fair $(k,l)$-loop $\pi$ for
some $0 < l \le k$ iff there exists a $k\in\mathbb{N}$ such that the
general B{\"u}chi encoding $\Trans{M,k}$ is satisfiable.
\end{theorem}

\begin{proof}
First we show that if $\Trans{M,k}$ is satisfiable then
$M$ has a fair loop. Assume $\Trans{M,k}$ is satisfiable for some
$k$. Fix an arbitrary satisfying assignment. As $\LoopExists$ is true,
there is a unique $0 < l \le k$ such that $l_l$ is true. It follows
that $s_{l-1} = s_k$. Hence, $s_0 \ldots s_{l-1} (s_l \ldots
s_k)^\omega$ is an initialised $(k,l)$-loop in $M$. Further, the loop
is fair, as for each acceptance set $F_m$, $0 \le m \le f$ there is
some $0 \le j \le k$ such that $InLoop_{j}$ is true and $s_{j}
\in F_m$.

In the second case let
$M$ have a fair loop.  We need to prove that
$\Trans{M,k}$ is satisfiable for some $k \in \Nat$. Assume $\pi = s_0
\ldots s_{l-1} (s_{l} \ldots s_k)^\omega$ with $s_{l-1} = s_k$ is a
fair loop in $M$. For each $0 \le m \le f$ there is $l \le i_m \le k$
such that $s_{i_m} \in F_m$. With $s_0 \ldots s_k$, $LoopExists
\Leftrightarrow \top$, $l_i \Leftrightarrow i=l$, $InLoop_i
\Leftrightarrow i \ge l$, and 
$\ATrans{Acc_m}_i \Leftrightarrow i \ge i_m$
for all $0 \le m \le f$
we obtain a satisfying
assignment for $\Trans{M,k}$.
\end{proof}

Note that the above variant only considers looping witnesses, as is
often done in the automata-theoretic approach to model checking of LTL\@.
Finite (no-loop) witnesses to safety properties help, as they do not
need to close a loop, focus attention on that part of an infinite path
that is most relevant for violation of a safety property. In addition,
minimising a no-loop witness to a safety property minimises the
distance between an initial state and the actual point of violation. In
contrast, minimising a looping witness just minimises the total length
of the looping path, regardless of where the property fails.
To also obtain finite witnesses, $M$ can be given as the product of the
model, a B{\"u}chi automaton accepting looping witnesses, and an
automaton on finite words accepting finite witnesses to the
property.

\section{Liveness Checking as Safety Checking}\label{sec:l2s}
While verification of safety properties can be handled using (simple)
reachability checking, verification of liveness or, more generally,
$\omega$-regular properties requires detection of fair
loops. Traditionally, loop detection is an integral part of the search
algorithm~\cite{LP85,VW86b,EL87}. Bounded model checking has to pull
the algorithm out of the search procedure, i.e., the SAT solver, by
making it part of the propositional formula submitted to the SAT
solver~\cite{BCCZ99}. Building on that, we below present an approach
that fully integrates loop detection into the model.

The \emph{liveness-to-safety transformation} takes a fair Kripke
structure $M$ and transforms it into another Kripke structure
$\ltos{M}$ such that there is an initialised fair path in $M$ iff a
certain set of states is reachable in $\ltos{M}$. This method makes
techniques available for arbitrary $\omega$-regular properties that
have only been applicable to safety properties so far. It has already
proven to be useful as a method to find shortest looping
counterexamples with a BDD-based model checker~\cite{SB05}, and to
extend SAT-based interpolation \cite{McM03} and large-scale directed
model checking \cite{EdelkampJabbar06} to $\omega$-regular
properties. On selected examples, an exponential speedup can be
observed compared to traditional BDD-based model
checking~\cite{SB04}. Still, because of its impact on the size of the
state space (see below), this approach may in many cases not be able
to replace dedicated methods for verifying $\omega$-regular
properties. In Sect.~\ref{sec:experiments} we evaluate experimentally
how invariant checking of a transformed model performs in comparison
to dedicated encodings for PLTL properties.
The liveness-to-safety transformation was originally proposed in
\cite{BiereArthoSchuppan02} and has been further developed in
\cite{SB04,SchuppanBiere05INFINITY,Schuppan06}. Bouajjani et
al.~independently applied the same technique in the context of regular
model checking~\cite{BouajjaniHabermehlVojnar04}. The presentation
below contains no new results, but deviates from previous work to
emphasise similarities with the bounded model checking approach at the
core of this paper.

\subsection{Transformation}

A typical modelling language of a model checker allows only access to
the current and next states of a path. It is not directly possible to
ask whether the current state has been seen before, thus preventing a
loop check in the model. On the other hand, a bounded model checker
has all states of the current path available on the propositional formula level. Hence,
in the latter situation the loop check is easy. The key idea of the
transformation is now to augment the model $M$ with a second instance
of the state variables to hold a copy of one previously seen state of
$M$. This 
avoids storing every state of a (then necessarily bounded)
path. Triggered by an oracle, the augmented model $\ltos{M}$ then at
some point of a forward exploration guesses the loop start and records
that guess in the second instance of the state variables of $M$. Once
the guess has been made, the forward search proceeds as if moving
forward from time point $l$ to $k$ in a witness for a bounded model
checker: record which acceptance sets have been visited ($\ltos{M}$
contains a corresponding set of flags), and, once all of them have been visited,
try to close the loop by comparing the current state with the recorded
guess.

Formally, the transformation is defined in Fig.~\ref{fig:defl2s}. Let
$M=(S,T,I,L,F=\{F_0,\ldots,F_f\})$ be a fair Kripke
structure. Assume, its state space $S$ is made up of a single state
variable $v$ with range $S$. We construct
$\ltos{M}=(\ltos{S},\ltos{T},\ltos{I},\ltos{L})$ as follows. The set
of state variables in $\ltos{M}$ consists of $v$, $\ltoscopy{v}$, $\lstart$,
$\InLoop{}$, $\LoopClosed$, and, for each acceptance set $F_m$,
$\ATrans{Acc_m}$. $v$ and $\ltoscopy{v}$ have range $S$, all other
variables are Booleans. $\ltos{S}$, $\ltos{T}$ and $\ltos{I}$ are the
maximal subsets of $S \times S \times \Bool^3 \times \Bool^{f+1}$,
$\ltos{S} \times \ltos{S}$, and $\ltos{S}$, respectively, which
fulfil the constraints in the following table. $\ltos{L}$ is $L$
extended with $\LoopClosed$: $\ltos{L}(\ltos{s}) = L(s)$ if $\neg
\LoopClosed(\ltos{s})$, $L(s) \cup \{\LoopClosed\}$ otherwise.

\begin{figure}
{\encsize
\[
\begin{array}{cc|c}
\mbox{line no} &\mbox{constraint} & \mbox{applies to}\\[1ex]
\hline&&\\[-2ex]
1 & S & \ltos{S}\\[1ex]
2 & T & \ltos{T}\\[1ex]
3 & I & \ltos{I}\\[1ex]
\hline&&\\[-2ex]
4 & \lstart \Leftrightarrow \bot & \ltos{I}\\[1ex]
5 & \InLoop{} \Leftrightarrow \bot & \ltos{I}\\[1ex]
6 & \Succs{\InLoop{}} \Leftrightarrow \InLoop{} \vee \Succs{\lstart} & \ltos{T}\\[1ex]
7 & \InLoop{} \Rightarrow \neg \Succs{\lstart} & \ltos{T}\\[1ex]
8 & (\Succs{\lstart} \Rightarrow \Succs{\ltoscopy{v}} = v) \wedge (\neg \Succs{\lstart} \Rightarrow \Succs{\ltoscopy{v}} = \ltoscopy{v}) & \ltos{T}\\[1ex]
\hline&&\\[-2ex]
9 & \forall 0 \le m \le f : (\ATrans{Acc_m} \Leftrightarrow \bot) & \ltos{I}\\[1ex]
10 & \forall 0 \le m \le f : (\Succs{\ATrans{Acc_m}} \Leftrightarrow \ATrans{Acc_m} \vee (\Succs{\InLoop{}} \wedge \Succs{v} \in F_m)) & \ltos{T}\\[1ex]
\hline&&\\[-2ex]
11 & \LoopClosed \Rightarrow \InLoop{} & \ltos{S}\\[1ex]
12 & \LoopClosed \Rightarrow v = \ltoscopy{v} & \ltos{S}\\[1ex]
13 & \forall 0 \le m \le f : (\LoopClosed \Rightarrow \ATrans{Acc_m}) & \ltos{S}\\[1ex]
\end{array}
\]
}\smallskip%
\caption{\label{fig:defl2s}Formal definition of the liveness-to-safety transformation}
\end{figure}

The original instance of the state variables, $v$, is subject to the
same constraints in $\ltos{M}$ as in $M$ (lines 1--3). For example, if
$\ltos{s} \in \ltos{S}$, then it must also be the case that
$v(\ltos{s}) \in S$. Similarly, $(\ltos{s},\Succs{\ltos{s}}) \in
\ltos{T}$ only if $(v(\ltos{s}),v(\Succs{\ltos{s}})) \in
T$. $\ltoscopy{v}$ is the second instance of the state variables. When
the oracle $\lstart$ becomes true, the loop start is guessed by
recording the previous value of $v$ in $\ltoscopy{v}$ (line
8). $\InLoop{}$ then becomes and remains true to signal the fact that
the loop has been started (line 6). It prevents $\lstart$ from
becoming true for a second time (line 7), which, in turn, ensures that
the recorded value in $\ltoscopy{v}$ will not be overwritten (line
8). When $\InLoop{}$ is true, visiting an accepting set $F_m$ is
recorded in $\ATrans{Acc_m}$ (line 10). $\LoopClosed$ can finally
become true to signal that a fair loop has been found when $\ltos{M}$
is in the loop, all acceptance sets have been seen, and the valuation
of the original instance of the state variables, $v$, is equal to the
guess kept in $\ltoscopy{v}$ (lines 11--13).

Note the similarity with the encodings for bounded model checking
presented in Sect.~\ref{sec:ltl-bmc}. Lines 4--7 and line 11
correspond to the loop constraints. Lines 9, 10, and 13 are equivalent
to the part of the general B{\"u}chi encoding that handles acceptance
sets. $\LoopExists{}$ has been renamed to $\LoopClosed$ to emphasise
that there is no implication from $\InLoop{}$ to $\LoopClosed$ and
$\LoopClosed$ is present in each state while there is only a single
instance of $\LoopExists{}$ in BMC\@. $l_i$ has been turned into 
oracle $\lstart$, i.e., rather than indicating that a loop exists between states
with index $l-1$ and $k$, it triggers saving the previous value of $v$
in $\ltoscopy{v}$. The corresponding check for equality of $v$ and
$\ltoscopy{v}$ has been shifted to $\LoopClosed$.\footnote{We state
without proof that for a fair $(k,l)$-loop $\pi$ there is an initialised path in
the transformed model and a satisfying assignment of the general B{\"u}chi
encoding such that the valuations of $l_s$, $\InLoop{}$, and
$\ATrans{Acc_m}$ coincide on corresponding indices of the path.}

Theorem~\ref{thm:l2s-correctness} states correctness of the
construction.
\begin{theorem}\label{thm:l2s-correctness}
Let $M=(S,T,I,L,F=\{F_0,\ldots,F_f\})$ be a fair Kripke
structure, let $\ltos{M}$ be defined as above. $M$ has an initialised fair path
$\pi$ iff some state $\ltos{s}$ is reachable in $\ltos{M}$ such that
$\LoopClosed(\ltos{s})$ is true.
\end{theorem}

\begin{sloppypar}
\proof
For simplicity, we restrict the proof to a single acceptance set
$F_0$. Generalisation to multiple acceptance sets is
straightforward. States in $\ltos{M}$ are written as tuples
$(v,\ltoscopy{v},\lstart,\InLoop{},\LoopClosed,\ATrans{Acc_0})$. Further, it
is sufficient to prove the following bi-implication~\cite{VW94}:
\[
\begin{array}{c}
\exists \pi = (s_0 \ldots s_{l-1}) (s_l \ldots s_m \ldots s_k)^\omega
\mbox{ initialised fair path in } M\\
\mbox{ with } k \ge m \ge l > 0 \wedge s_{l-1} = s_k \wedge s_l,\ldots,s_{m-1} \not \in F_0 \wedge s_m \in F_0\\
\Leftrightarrow\\
\exists \ltos{s} \mbox{ reachable in } \ltos{M} \mbox{ such that } \LoopClosed(\ltos{s}) \Leftrightarrow \top\\
\end{array}
\]

\begin{itemize}
\item[``$\Rightarrow$''] Let $\pi = (s_0 \ldots s_{l-1})(s_l \ldots
s_m \ldots s_k)^\omega$ be an initialised fair path in $M$ with $k \ge m \ge l >
0$, $s_{l-1} = s_k$, $s_l,\ldots,s_{m-1} \not \in F_0$, and $s_m \in
F_0$. Clearly, for arbitrary $\ltoscopy{s}_0 \in S$,
$(s_0,\ltoscopy{s}_0,\bot,\bot,\bot,\bot) \ldots
(s_{l-1},\ltoscopy{s}_0,\bot,\bot,\bot,\bot)$ is an initialised finite path in
$\ltos{M}$. We extend that prefix to reach a state $\ltos{s}_k$ with
$\LoopClosed(\ltos{s}_k) \Leftrightarrow \top$ by distinguishing four
cases:
\begin{enumerate}
\item $k = m = l$: Set $\ltos{s}_k = \ltos{s}_m = \ltos{s}_l$ to
$(s_k,s_{l-1},\top,\top,\top,\top)$.
\item $k = m > l$: Proceed from $\ltos{s}_l =
(s_l,s_{l-1},\top,\top,\bot,\bot)$ via
$(s_{l+1},s_{l-1},\bot,\top,\bot,\bot) \ldots
(s_{k-1},s_{l-1},\bot,\top,\bot,\bot)$ to $\ltos{s}_m = \ltos{s}_k =
(s_k,s_{l-1},\bot,\top,\top,\top)$.
\item $k > m = l$: Continue from $\ltos{s}_m = \ltos{s}_l =
(s_l,s_{l-1},\top,\top,\bot,\top)$ via
$(s_{l+1},s_{l-1},\bot,\top,\bot,\top) \ldots
(s_{k-1},s_{l-1},\bot,\top,\bot,\top)$ to $\ltos{s}_k =
(s_k,s_{l-1},\bot,\top,\top,\top)$.
\item $k > m > l$: Combine cases (2) and (3) to obtain
\[
\begin{array}{c}
(s_l,s_{l-1},\top,\top,\bot,\bot) (s_{l+1},s_{l-1},\bot,\top,\bot,\bot) \ldots (s_{m-1},s_{l-1},\bot,\top,\bot,\bot) \circ{}\\
\hspace{2em}{}\circ (s_m,s_{l-1},\bot,\top,\bot,\top) \ldots (s_{k-1},s_{l-1},\bot,\top,\bot,\top) (s_k,s_{l-1},\bot,\top,\top,\top)
\end{array}
\]
\end{enumerate}
\item[``$\Leftarrow$''] Let $\widetilde{\ltos{s}}$ be a reachable
state in $\ltos{M}$ with $\LoopClosed(\widetilde{\ltos{s}})
\Leftrightarrow \top$. Hence, there is an initialised finite path
$\widetilde{\ltos{\pi}}$ that ends in $\widetilde{\ltos{s}}$. Let
$\ltos{\pi} = \ltos{s}_0 \ldots \ltos{s}_k$ be the prefix of
$\widetilde{\ltos{\pi}}$ such that $\ltos{s}_k$ is the first (and
only) state in $\ltos{\pi}$ with $\LoopClosed(\ltos{s}_k)
\Leftrightarrow \top$. By definition of $\ltos{M}$,
$\InLoop{}(\ltos{s}_k) \Leftrightarrow \top$,
$\ltoscopy{v}(\ltos{s}_k) = v(\ltos{s}_k)$, and
$\ATrans{Acc_0}(\ltos{s}_k) \Leftrightarrow \top$. Further,
$\InLoop{}$ starts off false at $\ltos{s}_0$, switches to true when
$\lstart$ becomes true at some index $l > 0$, and remains true up to
$\ltos{s}_k$. Note, that $\lstart$ is true only at index $l > 0$. This
ensures, that $\ltoscopy{v}$ contains an arbitrary
$\ltoscopy{v}(\ltos{s}_0)$ from index 0 to $l-1$ and
$v(\ltos{s}_{l-1})$ from index $l$ onward. Thus, we have
$v(\ltos{s}_k) = \ltoscopy{v}(\ltos{s}_k) = \ltoscopy{v}(\ltos{s}_l) =
v(\ltos{s}_{l-1})$. $\ATrans{Acc_0}$ also is false initially and
changes at some index $l \le m \le k$ to true to remain there up to
index $k$. From the definition of $\ATrans{Acc_0}$ we have
$v(\ltos{s}_m) \in F_0$ and $\forall l \le i < m : v(\ltos{s}_i) \not
\in F_0$. It follows that, depending on the values of $k$, $m$, and
$l$, $\ltos{\pi}$ corresponds to one of the shapes (1) -- (4) outlined
in the first part of the proof. By the construction of $\ltos{M}$, in
all cases $\pi' = s_0 \ldots s_l \ldots s_m \ldots s_k = v(\ltos{s}_0)
\ldots v(\ltos{s}_l) \ldots v(\ltos{s}_m) \ldots v(\ltos{s}_k)$ is an
initialised finite path in $M$ with $s_{l-1} = s_k$, $s_l,\ldots,s_{m-1} \not \in
F_0$, and $s_m \in F_0$. Hence, $\pi = (s_0 \ldots s_{l-1})(s_l \ldots
s_m \ldots s_k)^\omega$ is an initialised fair path in $K$ as desired.\qed
\end{itemize}
\end{sloppypar}

The following immediate corollary enables using methods such as
\cite{SSS00,ES03,McM03,DBLP:journals/entcs/ArmoniFFHPV05} to obtain a
complete bounded model checking procedure for PLTL:
\begin{corollary}
Given a fair Kripke structure $M$, $M$ has an initialised fair path $\pi$
iff there exists a $k\in\mathbb{N}$ such that $\Trans{\ltos{M}}_k
\wedge \LoopClosed(\ltos{s}_k)$ is satisfiable.
\end{corollary}

The liveness-to-safety transformation roughly doubles the number of
state variables in the model. It can be shown that, with a small
modification of the way acceptance sets are handled, radius and
diameter of $\ltos{M}$ increase only by a small, constant factor
\cite{Schuppan06}.
If forward breadth-first search is used for reachability analysis of
$\ltos{M}$, the proof of Thm.~\ref{thm:l2s-correctness} implies
that a shortest fair looping path in $M$ is found. If $M$ is the
product of a model $\tilde{M}$ and a tight B{\"u}chi automaton $B$
for some property $\TLF$, that implies that the path is a
shortest witness with respect to $\TLF$ in $\tilde{M}$.

\subsection{Optimising the Transformation}

\subsubsection*{BDD Variable Order}\label{sec:l2sbddordopt}

If a BDD-based model checker is used to determine reachability in a
transformed model it is important to use a variable order that
interleaves the Boolean variables making up $s$ and
$\ltos{s}$. Otherwise the sizes of the BDDs representing $\ltos{M}$
may explode~\cite{SB04}.

\subsubsection*{Variable Optimisation}\label{sec:l2svaropt}

The overhead induced by the transformation of $M$ into $\ltos{M}$
mostly stems from the additional instance of the state variables of
$M$ present in $\ltos{M}$. Hence, leaving some of $M$'s state
variables out of loop detection might reduce that overhead. Kroening
and Strichman proved in the context of bounded model checking that
input variables can be ignored when computing the recurrence diameter
for simple liveness properties of the form $\Finally{p}$
\cite{KS03}. E{\'en} and S{\"o}rensson \cite{ES03} use the same idea in temporal
induction for safety properties in incremental BMC\@. We show below that
this idea can be extended to the liveness-to-safety transformation.

We call a state variable $v_i$ a \emph{transition input variable} iff
its value in the next state, $x_i'$, is not constrained by its value
in the current state, $x_i$, and the values of other variables in the
current and next state: if $((x_0,x_1,\ldots,
x_i,\ldots),(x_0',x_1',\ldots,x_i',\ldots))$ is a transition in $T$,
then, for all $\widetilde{x_i'}$ in the range of $v_i$,
$((x_0,x_1,\ldots,x_i,\ldots)(x_0',x_1',\ldots,\widetilde{x_i'},\ldots))$
is also in $T$.

A state variable $v_i$ is \emph{irrelevant for fairness} iff its value
$x_i$ does not influence whether a state is in an acceptance set or
not: for all acceptance sets $F_m$, for all $v_i$ in $V_i$,
we have that $(x_0,x_1,\ldots,x_i,\ldots)$ is in $F_m$ iff for all
$\widetilde{x_i}$ in the range of $v_i$,
$(x_0,x_1,\ldots,\widetilde{x_i},\ldots)$ is also in $F_m$.

Let $V_i$ be the set of transition input variables that are irrelevant
for fairness. Elements of $V_i$ can be left out of loop detection:
\begin{proposition}
Let $M$ be a fair Kripke structure with set of state variables
$V$ and set of transition input variables that are irrelevant for
fairness $V_i \subseteq V$. Let $\ltos{M}$ be defined as above, let
$\widetilde{\ltos{M}}$ be the variant of $\ltos{M}$ that restricts
loop detection (i.e., lines 8 and 12 in the definition of $\ltos{M}$)
to the variables in $V \setminus V_i$. There is a reachable state
$\ltos{s}$ such that $\LoopClosed(\ltos{s})$ is true in $\ltos{M}$ iff
there is one in $\widetilde{\ltos{M}}$.
\end{proposition}

\begin{proof}
The ``$\Rightarrow$''-direction is trivial. For ``$\Leftarrow$'' it is
sufficient to prove the following implication: if $\widetilde{\pi}=
s_0 \ldots s_{l-1} \ldots s_m \ldots \widetilde{s_k}$ is an initialised finite path
in $M$ with $k \ge m \ge l > 0$, $v(\widetilde{s_k}) = v(s_{l-1})$ for
all variables $v \in V \setminus V_i$, and $s_m \in F_0$, then
$\widetilde{\pi}$ with its last state replaced by $s_{l-1}$ is an initialised
finite path in $M$ with $k \ge m \ge l > 0$, $s_k = s_{l-1}$, and $s_m
\in F_0$.
\begin{enumerate}
\item By assumption, $(s_{k-1},\widetilde{s_k}) \in T$. Construct a
sequence of states $\widetilde{s_k}=t_0,t_1,\ldots,t_{|V_i|}=s_{l-1}$
such that all $t_j,t_{j+1}$ differ at most by the value of one
variable in $V_i$. By definition, for each $t_j,t_{j+1}$,
$(s_{k-1},t_j) \in T$ iff $(s_{k-1},t_{j+1}) \in T$. Hence,
$(s_{k-1},s_{l-1}) \in T$.
\item If $k > m$, $s_m \in F_0$. Otherwise, use the same sequence of
states $\widetilde{s_k}=t_0,t_1,\ldots,t_{|V_i|}=s_{l-1}$ to show that
$s_m = \widetilde{s_k} \in F_0$ iff $s_{l-1} \in F_0$.
\end{enumerate}
\end{proof}

Note that the restriction w.r.t.~acceptance sets can be dropped if
visiting an acceptance set is detected from index $l-1$ to $k-1$
rather than from $l$ to $k$.

We remark that if the Kripke structure being transformed is the
product of a model and a B{\"u}chi automaton generated from a PLTL
formula, the set of input variables must be determined with respect to
both. Hence, input variables of the model that appear in the PLTL
property to be verified may need to be included in the loop detection.
Clearly, variables that remain constant after initialisation need not
be considered for loop detection either. Leaving constant and input
variables out of loop detection as described above is referred to as
\emph{variable optimisation}.\footnote{Note that variable optimisation
could also be applied in specialised algorithms for bounded model
checking such as the one presented in Sect.~\ref{sec:incbmc} but this
is not currently implemented.} For more aggressive optimisations,
which, however, may not preserve length of counterexamples or even
lead to false positives, see~\cite{Schuppan06}.

Kroening and Strichman assume that input variables are a separate
syntactic entity. While a corresponding {\tt IVAR} declaration is
available in the NuSMV input language \cite{NuSMVManual}, many
benchmarks were written before NuSMV was available or don't make use
of this feature to retain compatibility to the original version of SMV
\cite{McMillan93,CMUSMVURL}. Therefore, Kroening and Strichman also
use an approach based on the transition relation of the
system. E{\'e}n and S{\"o}rensson \cite{ES03} additionally remove
output variables. As ignoring these may lead to shorter
counterexamples on the reduced set of variables in our approach
(though only by one state), they are handled by the more aggressive
optimisations in \cite{Schuppan06}.

\section{BMC for PLTL}\label{sec:pltl-bmc}
PLTL has features which impact the way model checking can be done. We illustrate
these features through a running example, taken from~\cite{BC03} and adapted to better suit our setting.
In this example the system to be model checked is a counter which uses a variable \(x\)
to store the counter value. The counter is initialised to \(0\), and the system adds one to the 
counter variable \(x\) at each time step until the highest value \(5\) is reached. 
After this the counter is reset to the value \(2\) in the
next time step and the system starts looping as illustrated in Fig.~\ref{fig:counter1}.
Thus the system is deterministic and the counter values can be seen as an infinite 
sequence \((012)(3452)^\omega\) corresponding to a \((6,3)\)-loop of the system. 
\begin{figure}
\centerline{
\scalebox{0.6}{
\input{figs/counter1.\texpic}
}}
\caption{Execution of the counter system}\label{fig:counter1}
\end{figure}
Consider the $(6,3)$-loop of the counter system. The formula
{\encsize
\begin{equation*}
((x = 3) \wedge \Prev \Prev \Prev (x = 0))
\end{equation*}
}\smallskip%

\noindent
holds only at time point \(3\) but not at any later time point.
This demonstrates the (quite obvious) fact that
unlike pure future LTL formulas, the
PLTL past formulas can distinguish
states which belong to different unrollings of the loop.
We introduce the notion of a time point belonging to a \(d\)-unrolling of the loop
to distinguish between different copies of each state in the unrolling of the loop part.
\begin{definition}
For a \((k,l)\)-loop \(\pi\) we say that
the \emph{period \(p(\pi)\)} of \(\pi\) is \((k-l)+1\),
i.e., the number of states the loop consists of. We
define that a time point \(i\geq 0\) in \(\pi\)
belongs to the \emph{\(d\)-unrolling of the loop}
iff $d\geq 0$ is the smallest integer such that \(i < l + ((d+1) \cdot p(\pi))\).
\end{definition}
The formula \(\Prev \Prev \Prev (x = 0)\) holds
at time point \(3\), which belongs to the \(0\)-unrolling of the loop.
However, at time point \(7\) belonging to the \(1\)-unrolling of the loop
the formula \(\Prev \Prev \Prev (x = 0)\) does not hold even though they both
correspond to the first state in the unrolling of the loop.

Benedetti and Cimatti~\cite{BC03} observed that encoding the BMC problem for
PLTL when the bounded path has no loop was fairly straightforward. 
It is simple to generalise the no-loop case of Biere et al.~\cite{BCCZ99} to 
include past operators, as they have simple semantics.
In the no loop case our encoding reduces to essentially the same as~\cite{BC03}. 
When loops are allowed the matter is more complicated, and therefore we will focus
on this part in the rest of this section.
The fact which enables us to do bounded model checking of PLTL formulas
(containing past operators in the loop case) is the following property 
first observed by~\cite{LMS02} and later independently by~\cite{BC03}:
for \((k,l)\)-loops the ability to distinguish between time points
in different \(d\)-unrollings in the past is limited by the past operator depth
\(\delta(\TSF)\) of a formula \(\TSF\).
\begin{proposition}\label{prop-depth}
Let \(\TSF\) be a PLTL formula and \(\pi\) be a \((k,l)\)-loop.
For all \(i \geq l\) it holds that if
the time point \(i\) belongs to a $d$-unrolling of the loop with
\(d \geq \delta(\TSF)\)
then: $\pi^{i} \models \TSF$ iff $\pi^{j} \models \TSF$,
where $j = i- ((d-\delta(\TSF)) \cdot p(\pi))$.
\end{proposition}
\begin{proof}
The proposition directly follows from Thm.~1 and Lemma~2 of~\cite{BC03}.
\end{proof}
The proposition above can be interpreted saying that after unrolling the
loop \(\delta(\TSF)\) times the formula cannot distinguish different
unrollings of the loop from each other. Therefore if we want
to evaluate a formula at an index \(i\) belonging to
a \(d\)-unrolling with \(d > \delta(\TSF)\), it is equivalent to 
evaluate the formula at the corresponding state of the
\(\delta(\TSF)\)-unrolling.

Consider again the running example where
we next want to evaluate whether the formula
{\encsize
\begin{equation}\label{formula:main-rex}
\Finally \left(\left(x = 3\right) \wedge \Once \left(\left(x = 4\right) \wedge \Once \left(x = 5 \right)\right)\right)
\end{equation}
}\smallskip%

\noindent
holds in the counter system. The formula expresses that it is possible to reach a point
at which the counter has had the values \(3,4,5\) in decreasing order in the past.
By using the semantics of PLTL it is easy to check that this
indeed is the case. The earliest time where
the subformula
\(((x = 3) \wedge \Once ((x = 4) \wedge \Once (x = 5 )))\)
holds is time \(11\) and thus the top-level formula holds at time \(0\).
In fact the mentioned subformula holds for all time points of the
form \(11 + i \cdot 4\), where \(i \geq 0\) and \(4 = p(\pi)\) is the period of
the loop \(3452\). The time point \(11\) corresponds to
a time step which is in the \(2\)-unrolling of
the loop \(3452\).
This stabilisation at the second unrolling
is guaranteed by the past operator depth of the formula in question, which is two.
The subformula 
\(((x = 4) \wedge \Once (x = 5))\)
has past operator depth \(\delta(\TSF) = 1\) and it
holds for the first time at time step \(8\) which is in the
\(1\)-unrolling of the loop. Again the stabilisation of the formula value
is guaranteed by the past operator depth of
one of the formula in question.
It will also hold for all time steps of the form
\(8 + i \cdot 4\), where \(i \geq 0\).
Thus, if we need to evaluate any subformula at a time step
which belongs to a deeper unrolling than its past operator depth,
e.g., if we want to evaluate
\(((x = 4) \wedge \Once (x = 5))\)
at time step \(16\) in \(3\)-unrolling, we can just take a look at the truth value of that
formula at the time step corresponding to the unrolling of the
formula to its past operator depth, in this case at time step \(8 = 16 - (3-1) \cdot 4\).

The previous discussion suggests the following extension of the
encodings presented in Sect.~\ref{sec:ltl-bmc}.
Intuitively, past temporal
operators can be encoded in a similar way as the future operators by using their
characterisation in terms of previous and current state
values. The issue of stabilisation needs to be dealt with though. Otherwise
a subformula can have different truth values at equivalent positions in
the path, which can lead to other subformulas being incorrectly evaluated.
One way to ensure stabilisation is to extend the loop check $l_i \Rightarrow (s_{i-1} = s_k)$ 
to also include the truth values of all formula variables (see~\cite{KPR98}).
While being intuitive and straightforward to implement, the
approach just sketched requires that the model is unrolled deep enough 
so that loop in the model is unrolled to guarantee the stabilisation of 
all temporal formulas.

Benedetti and Cimatti~\cite{BC03} suggested an alternative. The
transition relation of the model is only unrolled virtually. Rather
than having one variable representing the truth of a subformula at a
given index in the loop, several variables $\Trans{\TSF}_i^d$ are used
per subformula $\TSF$, which represent the truth of $\TSF$ at the same
relative position $i$ to the underlying finite path but
at different unrollings $d$, see Fig.~\ref{fig:counter3}.
The number of such variables required for each subformula
can be limited by Proposition~\ref{prop-depth}.

The bound $k$ at which a particular witness is reported may be
different for both variants. The first variant cannot guarantee that
the minimal length witnesses are found. However, if the bound required by the
first variant is not much larger than that of the alternative, 
even with a higher bound the first encoding may be more compact as only
one variable per subformula and index is introduced. On the other
hand, if several unrollings of the loop are required for
stabilisation, the second variant may be more compact: in that case,
savings due to having fewer instances of the transition relation of
the model will more than compensate for the overhead introduced by
the virtual unrolling of the formula variables.

Below we develop a propositional encoding of the BMC problem for PLTL
that integrates both variants. We first use the idea of Benedetti and
Cimatti~\cite{BC03} to extend the eventuality encoding for LTL with
past formulas as it is our encoding of choice for an incremental SAT
encoding to be presented in Sect.~\ref{sec:incbmc}. Based on that we
briefly discuss how the other LTL encodings can also be extended to
PLTL along similar lines. In fact, the encoding presented in this
section is essentially a non-incremental version of the incremental PLTL encoding
presented in~\cite{HelJunLat:CAV05}. We then show that by adding a
check for stabilisation of all temporal subformulas, the level of
virtual unrolling can be chosen freely between full and no
unrolling. Finally, we extend the idea of virtual unrolling to
B{\"u}chi automata.

\subsection{BMC for PLTL with Eventualities}\label{sec:pltl-eventuality}

The basic idea of the encoding is to virtually unroll the path by making several copies 
of the original finite path. A copy of the original path corresponds to a certain $d$-unrolling. 
If all loop selector 
variables $l_i$ are false the encoding collapses to the original path without a loop. 
The number of copies of the path for a 
PLTL subformula $\TSF$ is dictated by its past operator depth $\delta(\TSF)$. Since different subformulas
have different past depths, the encoding is such that subformulas with different past depths see 
different Kripke structures. Figure~\ref{fig:counter3} shows the running example unrolled to depth $d=2$,
for evaluating the formula~(\ref{formula:main-rex}).


\begin{figure}
\centerline{
\scalebox{0.6}{
\input{figs/counter3.\texpic}
}}
\caption{Black arcs show the Kripke structure induced by virtual unrolling of the loop
for $k=6$ up to depth \(2\) (i.e., \(\delta(\TSF) = 2\)) when \(l_3\) holds}\label{fig:counter3}
\end{figure}

First of all the \emph{PLTL eventuality encoding} contains the
model constraints $\Trans{M}_k$
and the loop constraints $\Trans{\mathit{LoopConstraints}}_k$ which
are both encoded exactly as in the LTL case.

To represent the original path and its copies, 
the PLTL formula variables $\Trans{\TSF}^d_i$ have two parameters: $d$ is the current 
$d$-unrolling and $i$ is the index in the current $d$-unrolling. The case where $d=0$ 
corresponds to the original $k$-step path. 
Subformulas at virtual unrolling depth beyond their past operator depth can 
by Proposition~\ref{prop-depth} be mapped to the depth corresponding 
to the past operator depth. From this we get our first rule
for each subformula $\TSF \in \mathit{cl}(\TLF)$:
{\encsize
\begin{equation*}
\Trans{\TSF}_i^d = \Trans{\TSF}_i^{\delta(\TSF)}, \ \mathrm{when} \ d> \delta(\TSF).
\end{equation*}
}\smallskip%

The rest of the encoding is split into cases based on the values of $i$ and $d$.
The encoding for propositional formulas is the same as in the LTL case except that
each subformula has constraints for several different $d$-unrollings.
Constraints for atomic propositions and their negation are straightforward. We simply project 
the atomic propositions onto the original path. The Boolean operators $\vee$ 
and $\wedge$ are encoded to stay in the current $d$-unrolling.
{\encsize
\[
\begin{array}{c@{\quad}|@{\quad}c@{\quad}|}
\TSF & 0\leq i\leq k , 0\leq d\leq \delta(\TSF)\\
\hline&\\[-2ex]
p & \Trans{p}^d_i \Leftrightarrow
  p \in L(s_i)
\\[1ex]
\neg p & \Trans{\neg p}^d_{i} \Leftrightarrow
  p \not \in L(s_i)
\\[1ex]
\LSF\wedge\RSF & \Trans{\LSF\wedge\RSF}^d_i \Leftrightarrow
\Trans{\LSF}^d_i\wedge\Trans{\RSF}^d_i  
\\[1ex]
\LSF\vee\RSF & \Trans{\LSF\vee\RSF}^d_i \Leftrightarrow
\Trans{\LSF}^d_i\vee\Trans{\RSF}^d_i  
\end{array}
\]
}\smallskip%

The translation of the future operators is also a very straightforward generalisation of
the pure future LTL encoding of Sect.~\ref{sec:ltl-eventuality};
we just have introduce constraints for all $d$-unrollings.
{\encsize
\[
\begin{array}{@{\quad}c@{\quad}|@{\quad}c|}
\TSF & 0\leq i \leq k, 0\leq d \leq \delta(\TSF) \\[1ex]
\hline & \\[-2ex]
\Next{\SF} & \Trans{\Next{\SF}}^d_i\Leftrightarrow\Trans{\SF}_{i+1}^d \\[1ex]
\Until{\LSF}{\RSF} & \Trans{\Until{\LSF}{\RSF}}^d_i\Leftrightarrow\Trans{\RSF}_i^d\vee\left(
\Trans{\LSF}^d_i\wedge\Trans{\Until{\LSF}{\RSF}}^d_{i+1}\right)\\[1ex]
\Release{\LSF}{\RSF} & \Trans{\Release{\LSF}{\RSF}}_i^d\Leftrightarrow\Trans{\RSF}_i^d\wedge\left(
\Trans{\LSF}_i^d\vee\Trans{\Release{\LSF}{\RSF}}^d_{i+1}\right)\\[1ex] 
\end{array}
\]
}\smallskip%

The $\Trans{\mathit{LastStateFormula}}_k$ constraints of the LTL case
have to be changed in the PLTL case to take care of binding
the different unrollings of the encoding together in the way
shown by following the black arcs of Fig.~\ref{fig:counter3}
in the forward direction.
The truth values of $\Trans{\TSF}_{k+1}^d$ are picked
from the loop point $i$ of the next unrolling level
$\Trans{\TSF}_i^{d+1}$, or if we are at the last
level $d=\delta(\TSF)$ then from the loop point at the last level
$\Trans{\TSF}_i^{\delta(\TSF)}$. This is achieved by the
expression $\Trans{\TSF}_i^{\mathit{min}(d+1,\delta(\TSF))}$.
For all $\TSF\in\mathit{cl}(\TLF)$ the following constraints are created:
{\encsize
\[
\begin{array}{c@{\quad}|@{\quad}c@{\quad}|}
  & 0\leq d\leq \delta(\TSF) \\
\hline& \\[-2ex]
\mathrm{Base} & \neg\mathit{LoopExists}\Rightarrow\left(\Trans{\TSF}^d_{k+1}\Leftrightarrow\bot\right)\\[1ex]
\hline& \\[-2ex]
1\leq i \leq k & \lvar{i}\Implies\left(\Trans{\TSF}^d_{k+1}\Equiv\Trans{\TSF}^{\mathit{min}(d+1,\delta(\TSF))}_i\right) \\[1ex]
\end{array}
\]
}\smallskip%

When $d=\delta(\TSF)$ we have reached the $d$-unrolling where the Kripke structure
loops back. At this depth we can guarantee that the satisfaction of all subformulas has stabilised 
(see Proposition~\ref{prop-depth}). Therefore at the maximum unrolling depth we add the
auxiliary translation constraints which, similarly to the LTL case,
are needed to correctly evaluate the until and release formulas along the loop.
{\encsize
\[
\begin{array}{c@{\quad}|@{\quad}c@{\quad}|@{\quad}c|}
 & \TSF  & \\
\hline & &\\[-2ex]
\mathrm{Base} & \Until{\LSF}{\RSF} &\mathit{LoopExists}\Rightarrow\left(
\Trans{\Until{\LSF}{\RSF}}^{\delta(\TSF)}_k\Rightarrow\ATrans{\Finally{\RSF}}^{\delta(\RSF)}_k\right)\\[1ex] 
& \Release{\LSF}{\RSF} & \mathit{LoopExists}\Rightarrow\left(
\Trans{\Release{\LSF}{\RSF}}^{\delta(\TSF)}_k\Leftarrow\ATrans{\Globally{\RSF}}^{\delta(\RSF)}_k\right)\\[1ex]
& \Until{\LSF}{\RSF} & \ATrans{\Finally{\RSF}}^{\delta(\RSF)}_0 \Leftrightarrow \bot\\[1ex] 
& \Release{\LSF}{\RSF} & \ATrans{\Globally{\RSF}}^{\delta(\RSF)}_0 \Leftrightarrow \top \\[1ex]
\hline & &\\[-2ex]
 1\leq i\leq k
& \Until{\LSF}{\RSF} & \ATrans{\Finally{\RSF}}^{\delta(\RSF)}_i \Leftrightarrow 
\ATrans{\Finally{\RSF}}^{\delta(\RSF)}_{i-1} \vee \left(\InLoop{i}\wedge\Trans{\RSF}^{\delta(\RSF)}_i\right)\\[1ex]  
& \Release{\LSF}{\RSF} &  \ATrans{\Globally{\RSF}}^{\delta(\RSF)}_i \Leftrightarrow 
\ATrans{\Globally{\RSF}}^{\delta(\RSF)}_{i-1} \wedge \left(\neg\InLoop{i}\vee\Trans{\RSF}^{\delta(\RSF)}_i\right)\\[1ex]
\end{array}
\]
}\smallskip%

\noindent
The starting point for the encoding for the past operators is using their characterisation
in terms of the current and the previous state.
This enables the encoding of the past operators to 
fit in nicely with the future encoding. Since past operators look backwards, we must 
encode the move from one copy of the path to the previous copy efficiently.
%

The simplest case of the encoding for past operators occurs at $d=0$. At this depth,
the past is unique in the sense that the path cannot jump to a lower depth. We do not need to 
take into account the loop edge, so the encoding follows from the
characterisation $\Since{\LSF}{\RSF}$ and $\Trigger{\LSF}{\RSF}$
in terms of the current and the previous state.
Encoding $\Prev{\SF}$ and $\PrevZ{\SF}$ is trivial.\footnote{The column $i=0$ has been included to make
all unrollings evaluate exactly the same truth values in the no-loop case, which has a slight advantage
if the encoding is used in a complete model checking procedure as described in Section~\ref{sec:completeness}.}
{\encsize
\[
\begin{array}{c@{\quad}|@{\quad}c@{\quad}|@{\quad}c}
\TSF & i=0,0\leq d\leq \delta(\TSF) & 1 \leq i\leq k,d=0 \\
\hline&&\\[-2ex]
\Since{\LSF}{\RSF} &
\Trans{\Since{\LSF}{\RSF}}^d_{i}\Equiv\Trans{\RSF}^d_{i} &
\Trans{\Since{\LSF}{\RSF}}^d_{i}\Equiv\Trans{\RSF}^d_{i}\vee
\left(\Trans{\LSF}^d_i\wedge\Trans{\Since{\LSF}{\RSF}}^d_{i-1}\right) 
\\[1ex]
\Trigger{\LSF}{\RSF} &
\Trans{\Trigger{\LSF}{\RSF}}^d_{i}\Equiv \Trans{\RSF}^d_{i}  &
\Trans{\Trigger{\LSF}{\RSF}}^d_{i}\Equiv\Trans{\RSF}^{d}_{i}\wedge\left(\Trans{\LSF}^d_i\vee\Trans{\Trigger{\LSF}{\RSF}}^d_{i-1}\right) 
\\[1ex]
\Prev{\LSF} &
\Trans{\Prev{\LSF}}^d_{i}\Equiv \bot &
\Trans{\Prev{\LSF}}^d_{i}\Equiv\Trans {\LSF}^d_{i-1} 
\\[1ex]
\PrevZ{\LSF} &
\Trans{\PrevZ{\LSF}}^d_{i}\Equiv\top &
\Trans{\PrevZ{\LSF}}^d_{i}\Equiv\Trans {\LSF}^d_{i-1} 
\end{array}
\]
}\smallskip%

When $d>0$ the key challenge of the encoding is to decide whether the past operator should 
consider the path to continue in the current unrolling of the path
or in the last state of the previous unrolling. The decision is taken based on the loop selector variables,
which indicate whether we are in the loop state. In terms of our running example, we need
to traverse the straight black arrows of Fig.~\ref{fig:counter3} in the reverse direction. We implement the choice with
an if-then-else construct $\left(l_i\wedge\varphi_1\right)\vee\left(\neg l_i\wedge\varphi_2\right)$. 
The expression encodes the choice if $l_i$ is true then the truth value of the expression is decided by $\varphi_1$, 
otherwise $\varphi_2$ decides the truth value of the expression.
{\tightencsize
\[
\begin{array}{c@{\quad}|@{\quad}c@{\quad}}
\TSF & 1 \leq i \leq k, 1\leq d \leq \delta(\TSF)\\
\hline&\\[-2ex]
\Since{\LSF}{\RSF}&
\Trans{\Since{\LSF}{\RSF}}^d_{i}\Equiv\Trans{\RSF}^d_{i}\vee\left(\Trans{\LSF}^d_i\wedge
\left(\left(l_i\wedge\Trans{\TSF}_k^{d-1}\right)\vee\left(\neg l_i\wedge\Trans{\TSF}^{d}_{i-1}\right)\right)\right) 
\\[1ex]
\Trigger{\LSF}{\RSF}&
\Trans{\Trigger{\LSF}{\RSF}}^d_{i}\Equiv\Trans{\RSF}^d_{i}\wedge\left(\Trans{\LSF}^d_i\vee
\left(\left(l_i\wedge\Trans{\TSF}_k^{d-1}\right)\vee\left(\neg l_i\wedge\Trans{\TSF}^{d}_{i-1}\right)\right)\right)
\\[1ex]
\Prev{\LSF}&
\Trans{\Prev{\LSF}}^d_{i}\Equiv\left(\mathit{l}_i\wedge\Trans{\LSF}^{d-1}_{k}\right)\vee
\left(\neg\mathit{l}_i\wedge\Trans{\LSF}^{d}_{i-1}\right)\\[1ex]
\PrevZ{\LSF}&
\Trans{\PrevZ{\LSF}}^d_{i}\Equiv\left(\mathit{l}_i\wedge\Trans{\LSF}^{d-1}_{k}\right)\vee
\left(\neg\mathit{l}_i\wedge\Trans{\LSF}^{d}_{i-1}\right)\\[1ex]
\end{array}
\]
}\smallskip%

Combining the tables above we get the full PLTL encoding $\Trans{\mathit{EventualityPLTL}}_k$ for $\TLF$.
Given a Kripke structure $M$, a PLTL formula $\TLF$, and a bound $k$, the
\emph{PLTL eventuality encoding} as a propositional formula is given by: 
{\encsize
\[
\Trans{M,\TLF,k}=\Trans{M}_k\wedge\Trans{\mathit{LoopConstraints}}_k\wedge\Trans{\mathit{LastStateFormula}}_k\wedge\Trans{\mathit{EventualityPLTL}}_k\wedge\Trans{\TLF}^0_0.
\]
}\smallskip%

The correctness of our encoding is established by the following theorem. 
\begin{theorem}\label{thm:pltl-eventuality}
Given a Kripke structure $M$ and a PLTL formula $\TLF$, 
$M$ has an initialised path $\pi$ such that $\pi\models\TLF$ iff there exists a $k\in\mathbb{N}$ such
that the PLTL eventuality encoding $\Trans{M,\TLF,k}$ is satisfiable. In particular, if
$\pi\models_k\TLF$ then the PLTL eventuality encoding $\Trans{M,\TLF,k}$ is satisfiable.
\footnote{As immediate corollary minimal length $(k,l)$-loop counterexamples for PLTL can be detected.
The encoding also detects minimal length informative safety counterexamples for PLTL\@.}
\end{theorem}
\proof
We proceed similarly to the proof of
Thm.~\ref{thm:ltl-eventuality}, only changes are given below.
The main change to the future only LTL encoding is that all the subformulas
$\TSF \in \mathit{cl}(\TLF)$ are virtually unrolled to their past operator depth
$\delta(\TSF)$. In addition the new past formula encodings have been introduced.

First consider the $(k,j)$-loop case (a):
We have the same induction scheme as in the proof of Thm.~\ref{thm:ltl-eventuality}.
The main change is that
we have to take the virtual unrolling into account.
We will prove by induction on the structure
of the PLTL formula $\TLF$ that the PLTL eventuality encoding is satisfiable
with a unique satisfying truth assignment.
Moreover, for all pairs of indices $i,d$ in
$0 \leq i \leq k, 0 \leq d \leq \delta(\TSF)$
such that $d=0$ or $i \geq j$ (we are in the black nodes of Fig.~\ref{fig:counter3})
it holds that
$\pi^{i + (d \cdot p(\pi))} \models_k \TSF$ iff in the unique satisfying truth assignment of the
PLTL eventuality encoding $\Trans{\TSF}^d_i$ is true.

For a future subformula $\TSF \in \mathit{cl}(\TLF)$ (the top-level subformula of $\TSF$ is a future time formula)
we do this by first proving that if the encoding is satisfiable, the
variable $\Trans{\TSF}^{\delta(\TSF)}_k$ for the last state of the top unrolling of Fig.~\ref{fig:counter3}
is true iff $\pi^{k + (\delta(\TSF) \cdot p(\pi))} \models_k \TSF$.
This is done similarly to the proof of Thm.~\ref{thm:ltl-eventuality};
only small indexing changes are needed in order to always refer to states
in the unrolling $\delta(\TSF)$ both for the encoding and for the PLTL semantics.
All formulas referred to in the proof have in the unrolling $\delta(\TSF)$
stabilised by Proposition~\ref{prop-depth} and thus we get that if the encoding is satisfiable,
$\Trans{\TSF}^{\delta(\TSF)}_k$ is true iff $\pi^{k + (\delta(\TSF) \cdot p(\pi))} \models_k \TSF$.
Now it is also easy to check that the encoding for all other pairs of indices $i,d$
in $0 \leq i \leq k, 0 \leq d \leq \delta(\TSF)$
such that $d=0$ or $i \geq j$ follows the one-step identities of the bounded PLTL semantics for $\TSF$
in a functional manner (proof by induction following the straight black arcs of
Fig.~\ref{fig:counter3} in the reverse direction
jumping from one unrolling to the previous as shown by the arcs) and thus the truth assignment matching the bounded PLTL semantics
leads to the only truth assignment satisfying all constraints of the encoding. The new part in this proof compared to the future case is that
we also have to prove for all pairs of indexes $0 \leq i \leq k, 0 \leq d \leq \delta(\TSF)$
such that $d>0$ and $i < j$ the corresponding constraints
are satisfiable in a unique way.
This is the case because these constraints can be seen to form
Boolean circuits where all inputs are fixed and the output is not constrained in any way.
We thus obtain a unique satisfying truth assignment
for the full PLTL eventuality encoding in a similar manner as in the proof of Thm.~\ref{thm:ltl-eventuality}.

For a past formula $\TSF \in \mathit{cl}(\TLF)$ the proof starts by showing that
if the encoding is satisfiable, then
$\Trans{\TSF}^0_0$ corresponding to the first state of the bottom unrolling of Fig.~\ref{fig:counter3}
is true iff $\pi^{0} \models_k \TSF$. This can be easily checked by comparing the encoding
of $\Trans{\TSF}^0_0$ with the PLTL semantics of past formulas combined with our induction
hypothesis that the subformulas are correctly evaluated. 
Now it is also easy to check that
the rest of the encoding for all other pairs of indices $i,d$ in $0 \leq i \leq k, 0 \leq d \leq \delta(\TSF)$
such that $d=0$ or $i \geq j$ follows the one-step identities of the bounded PLTL semantics for $\TSF$
in a functional manner (proof by induction
following the straight black arcs of Fig.~\ref{fig:counter3} in the forward direction
jumping from one unrolling to the next as shown by the arcs)
and thus the truth assignment matching the bounded PLTL semantics for $\TSF$
leads to the only truth assignment satisfying all constraints of the encoding.
The new part in this proof compared to the future case is that
we also have to prove for all pairs of indexes $0 \leq i \leq k, 0 \leq d \leq \delta(\TSF)$
such that $d>0$ and $i < j$ the corresponding constraints
are satisfiable in a unique way. This is the case because these constraints
can be seen to form
Boolean circuits where all the inputs are fixed and the output is not constrained in any way.
We thus obtain a unique satisfying truth assignment
for the full PLTL eventuality encoding in a similar manner as in the proof of Thm.~\ref{thm:ltl-eventuality}.

Next consider the no-loop case (b): We first note that in the no-loop case
$\LoopExists$ is false and in this case the encoding for all indexes $d > 0$
can be seen to form Boolean circuits where all the inputs are fixed and the
output is not constrained in any way. Thus all of these constraints are satisfiable
in a unique way.

Therefore we need to only consider the case $d=0, 0 \leq i \leq k$. We proceed similarly to the proof of
Thm.~\ref{thm:ltl-eventuality} for future PLTL formulas, but due to the simplicity of the proof
we reproduce it here. Because $\LoopExists$ is false,
it is easy to see that
the $\Trans{\mathit{LastStateFormula}}_k$
constraints will force the proxy variables $\Trans{\TSF}^0_{k+1}$ to $\bot$, and
the encoding becomes exactly the same as in the
fixpoint encoding case and thus has a unique satisfying truth assignment.
Also the auxiliary encoding constraints will lead to a unique satisfying truth
assignment as as $\mathit{LoopExists}$ is false.

For a past PLTL formula $\TSF$ we first find that if the encoding is satisfiable,
$\Trans{\TSF}^0_0$ is true iff $\pi^{0} \modelsnl \TSF$. This can be easily checked by comparing the encoding
of $\Trans{\TSF}^0_0$ with the no-loop case PLTL semantics of past formulas combined with our induction
hypothesis that the subformulas are correctly evaluated. 
It is also easy to check that
the rest of the encoding for all other indices $0 < i \leq k$
follows the one-step identities of the no-loop case PLTL semantics for $\TSF$ in a functional manner,
and thus the truth assignment matching the no-loop case PLTL semantics for $\TSF$
leads to the only truth assignment satisfying all constraints of the encoding.
\qed
The size of the encoding is $O(\vert I\vert + k\cdot\vert T\vert 
+ k \cdot \vert\TLF\vert \cdot \delta(\TLF))$. 
The encoding for PLTL above also has the unique model property in the same sense
as in the LTL case. The unique model property allows us to read the exact bounded
semantics for all PLTL subformulas and all time indexes considered directly 
from the truth assignment given
by the SAT engine.
In fact, it also evaluates some value for the formula variables in
the light nodes of Fig.~\ref{fig:counter3}. These nodes could be easily detected
and forced to some fixed value (e.g., $\bot$) but that would make the encoding slightly larger.
For the BMC encoding we preferred not to do that, as the truth values of these nodes do not
matter because they cannot be referenced from $\Trans{\TLF}^0_0$ by either forward or
backward arcs. 

Similarly to the LTL case,
the PLTL eventuality encoding of this section (see Sect.~\ref{sec:ltl-eventuality} for the LTL version)
can alternatively be replaced with either a PLTL fixpoint evaluation encoding~\cite{LBHJ05}
(see Sect.~\ref{sec:ltl-fixe} for the LTL version) or the B{\"u}chi encoding (see Sect.~\ref{sec:ltl-buchi}
for the LTL case). Intuitively the main difference to the LTL case is evaluating the required
auxiliary encoding at such an unrolling depth $d=\delta(\TSF)$ that the evaluated formula $\TSF$
has stabilised according to Proposition~\ref{prop-depth}.

\subsubsection*{Partial Unrolling}
An interesting feature of the PLTL encoding 
is that a simple modification makes it sound even if we replace the function $\delta(\cdot)$
with a constant function that always returns $0$. 
In this case the size of the encoding will be linear in $\vert\TLF\vert$,
and a B{\"u}chi encoding variant of the PLTL encoding becomes essentially
a BMC encoding of~\cite{KPR98}, see also~\cite{SB05}.
In fact, we can limit the maximum virtual unrolling depth
of any subformula to any value $d_{max}$ between zero (minimal size encoding, potentially longer
counterexamples) and $\delta(\TSF)$ (minimal length counterexamples,
larger encoding).  Counterexamples will still be detected but
the bound required to do so will depend on the amount of unrolling done.

For the last unrolling we have to add the \emph{stabilisation forcing constraints} shown below
which  constrain the past formulas to also consider that the predecessor can be the last state 
of the last unrolling. 
\iffalse
Adding similar constraints to the encoding of~\cite{CiRoSh:FMCAD04} would also make that encoding work
correctly for formulas containing past operators.
\else
Such constraints are also required for the encoding of
\cite{CiRoSh:FMCAD04} to work correctly for formulas containing past
operators; \cite{CiRoSh:FMCAD04} does not state this explicitly.
\fi
The intuition for the stabilisation forcing constraints is that they ensure that 
past formulas in the loop
state of the last unrolling
evaluate to the same truth value, no matter whether it is seen as the successor
of the end state at current or the previous unrolling. In other words, 
all subformulas have stabilised. Proposition~\ref{prop-depth} will guarantee that when 
we have unrolled to the maximum depth
$\delta(\TSF)$, these constraints will not remove any satisfying models of the encoding
as the truth values of all formulas, in particular all the past formulas themselves,
have stabilised when the last unrolling has been reached.
{\encsize
\[
\begin{array}{c@{\quad}|@{\quad}c@{\quad}|}
\TSF & 1 \leq i \leq k, d = \delta(\TSF)\\
\hline&\\[-2ex]
\Since{\LSF}{\RSF} & \Trans{\Since{\LSF}{\RSF}}^d_{i}\Equiv
\Trans{\RSF}^d_{i}\vee\left(\Trans{\LSF}^d_i\wedge
\left(\left(l_i\wedge\Trans{\TSF}_k^{d}\right)\vee\left(\neg l_i\wedge\Trans{\TSF}^{d}_{i-1}\right)\right)\right) 
\\[1ex]
\Trigger{\LSF}{\RSF} & \Trans{\Trigger{\LSF}{\RSF}}^d_{i} \Equiv
\Trans{\RSF}^d_{i}\wedge\left(\Trans{\LSF}^d_i\vee
\left(\left(l_i\wedge\Trans{\TSF}_k^{d}\right)\vee\left(\neg l_i\wedge\Trans{\TSF}^{d}_{i-1}\right)\right)\right)
\\[1ex]
\Prev{\LSF} & \Trans{\Prev{\LSF}}^d_{i} \Equiv
\left(\mathit{l}_i\wedge\Trans{\LSF}^{d}_{k}\right)\vee
\left(\neg\mathit{l}_i\wedge\Trans{\LSF}^{d}_{i-1}\right)\\[1ex]
\PrevZ{\LSF} & \Trans{\PrevZ{\LSF}}^d_{i} \Equiv
\left(\mathit{l}_i\wedge\Trans{\LSF}^{d}_{k}\right)\vee
\left(\neg\mathit{l}_i\wedge\Trans{\LSF}^{d}_{i-1}\right)\\[1ex]
\end{array}
\]
}\smallskip%

The correctness of the stabilisation constraints is not difficult to see.
If we assume that for every past subformula $\TSF \in cl(\TLF)$ it holds that
$\pi^{j + (d_{max} \cdot p(\pi))} \models_k \TSF$ iff
$\pi^{k + 1 + (d_{max} \cdot p(\pi))} \models_k \TSF$ then we can easily prove that
the evaluated formula has stabilised for all subformulas at all indices in the unrolling $d_{max}$.

To prove soundness of the modified encoding we proceed as follows.
If the assumption of stabilisation at $d_{max}$ does not hold, we can find
a past time subformula $\TSF$ such that all its subformulas have stabilised at $d_{max}$
but $\pi^{j + (d_{max} \cdot p(\pi))} \models_k \TSF$ iff $\pi^{k + 1 + (d_{max} \cdot p(\pi))} \models_k \TSF$
does not hold. In this case it is easy to see that the original constraints force
$\Trans{\TSF}^{d_{max}}_j$ to true iff $\pi^{j + (d_{max} \cdot p(\pi))} \models_k \TSF$,
and it is easy to prove that the stabilisation forcing constraints force
$\Trans{\TSF}^{d_{max}}_j$ to true iff $\pi^{k + 1 + (d_{max} \cdot p(\pi))} \models_k \TSF$.
Therefore the whole encoding becomes unsatisfiable.

For completeness we note that Proposition~\ref{prop-depth} ensures that eventually
all PLTL formulas become periodic. This ensures that eventually 
all past subformulas will satisfy the stabilisation assumption above
with any value $0 \leq d_{max} \leq \delta(\TLF)$
when $k$ is increased large enough, for some value of $j$.
If the assumption about stabilisation holds, then by using
$min(d_{max}, \delta(\TSF))$ in the encoding and in the Proof of Thm.~\ref{thm:pltl-eventuality}
instead of $\delta(\TSF)$, we can prove the encoding to be satisfiable and
matching the bounded semantics of PLTL
using our assumption about
the stabilisation at unrolling $d_{max}$ instead of Proposition~\ref{prop-depth}.
The only new thing that needs to be proven is that the stabilisation enforcing constraints
are satisfiable, and this is immediate by the new constraints and our assumption of the stabilisation of
all past subformulas at the unrolling level $d_{max}$.

As a historical note, at the point of writing~\cite{LBHJ05} we were 
unfortunately not
aware of the symbolic PLTL B\"uchi automata translation of~\cite{KPR98}. 
Quite late in writing~\cite{HelJunLat:CAV05} we became aware of it by
stumbling on a bug --- stemming from the ambiguity mentioned above ---
in an unpublished prototype implementation of~\cite{CiRoSh:FMCAD04} kindly provided
to us by its authors. After that we discovered that~\cite{KPR98} did not 
have that problem, we quickly figured out how to use a similar optimisation in our context.

\subsection{Virtual Unrolling for B{\"u}chi Automata}\label{sec:pltl-buchi}

In this subsection we extend the idea of virtual unrolling to
B{\"u}chi automata. Starting from a B{\"u}chi automaton
$\tilde{B}^\TLF$ based on~\cite{KPR98}, which is tight only if $\TLF$
is a future time formula~\cite{SB05}, we obtain a B{\"u}chi automaton
$B^\TLF$ accepting the same language that is tight for all PLTL
formulae $\TLF$.

The situation is very similar to the BMC case: on a shortest witness,
the original B{\"u}chi automaton $\tilde{B}^\TLF$ needs some
additional unrollings of the transition relation of the model $M$ till
both have a loop of the same length. Note, that the intuition provided
by the example below does not rely on the fact that $\tilde{B}^\TLF$
is derived from~\cite{KPR98}. It only requires $\tilde{B}^\TLF$ to
have an accepting loop of the same length as the witness. Further
generalisation to arbitrary B{\"u}chi automata is possible but so far
of mostly theoretical interest~\cite{Schuppan06}.

For technical reasons we have to deviate from the convention that
$l_i$ is true at index $l$ of a $(k,l)$-loop and $\InLoop{}$ is true
from index $l$ through index $k$. Rather, both are shifted one state
towards the initial state, i.e., $l_i$ is true at index $l-1$ (which
could be regarded as being the loop start as well) and,
correspondingly, $\InLoop{}$ is true from $l-1$ through $k$.

The construction of the tight B{\"u}chi automaton is by and large the
same as in~\cite{SB05}. The presentation is changed to highlight
similarities with the encoding of PLTL for BMC in the previous
subsection.

\subsubsection*{Example}

We first walk through the steps of the construction using our running
example in Fig.~\ref{fig:counter1}. Figure
\ref{fig:pltlbuechiexample}a shows a run of a~\cite{KPR98}-like
B{\"u}chi automaton $\tilde{B}^\TLF$ on the path $(0 1) (2 3 4
5)^\omega$ --- remember, that we start the loop one state earlier in
this subsection. The model $M$ enters a loop of length 4 at time point
2 while $\tilde{B}^\TLF$ needs 6 more states until it enters a loop of
the same length. An accepting loop in the product $M \times
\tilde{B}^\TLF$ can be closed only at time point 12 (the last
occurrence of $x=4$ in Fig.~\ref{fig:pltlbuechiexample}a).

\def\tabpltlbuechiexampleline#1#2#3{%
\begin{minipage}[c]{0.23\linewidth}\includegraphics[scale=0.235]{figs/tableautight_legend#1}\end{minipage}&%
\begin{minipage}[c]{0.38\linewidth}\includegraphics[scale=0.235]{figs/tableautight_#2}\end{minipage}&%
\begin{minipage}[c]{0.38\linewidth}\includegraphics[scale=0.235]{figs/tableautight_#3}\end{minipage}\\%
\begin{minipage}{0.23\linewidth}\begin{center}\end{center}\end{minipage}&%
\begin{minipage}{0.38\linewidth}\begin{center}#2)\end{center}\end{minipage}&%
\begin{minipage}{0.38\linewidth}\begin{center}#3)\end{center}\end{minipage}%
}
\begin{figure}
\begin{tabular}{@{\hspace{0em}}ccc}
\tabpltlbuechiexampleline{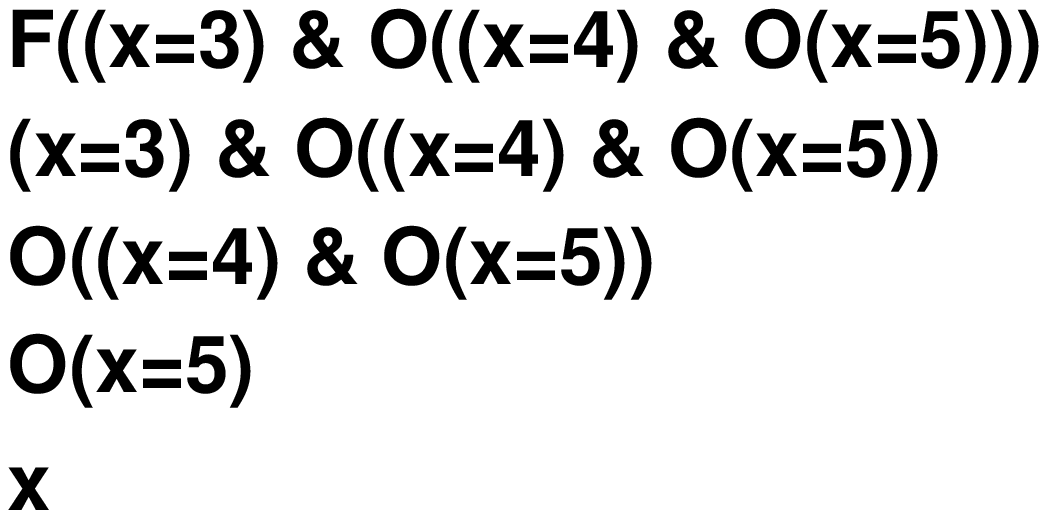}{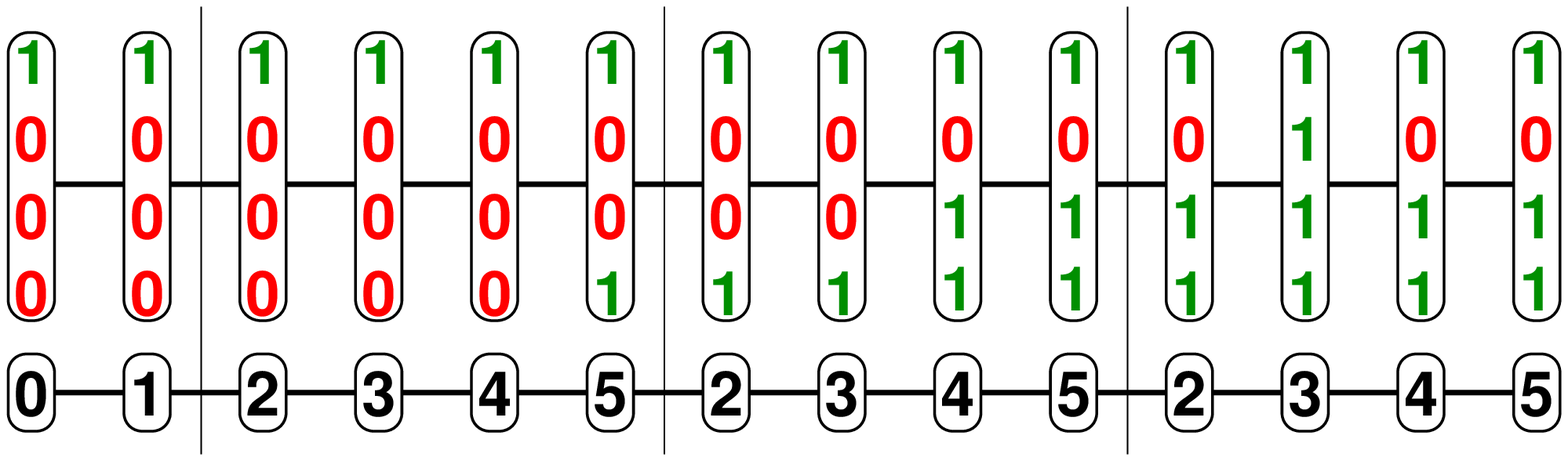}{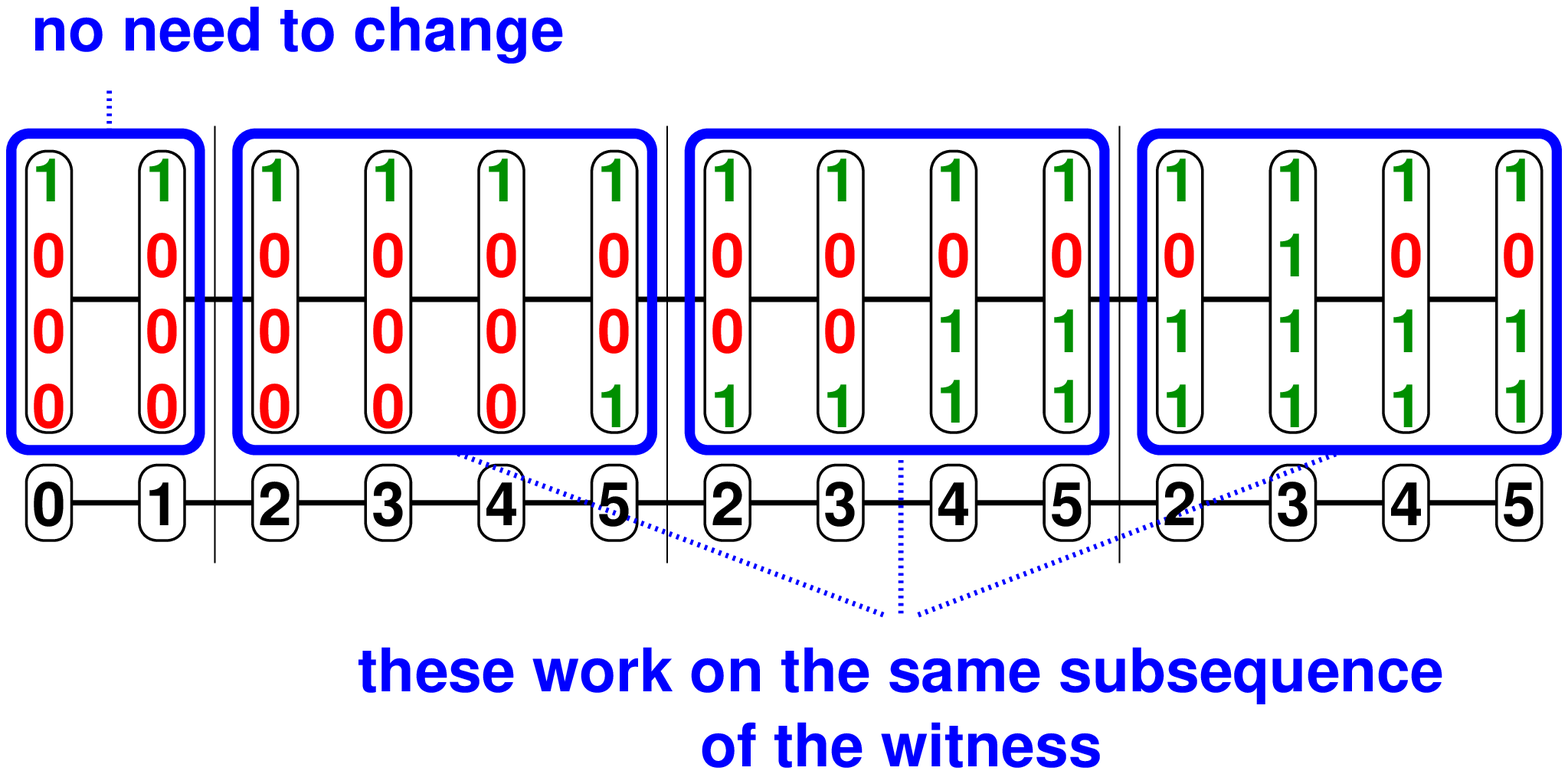}\\[1ex]
\tabpltlbuechiexampleline{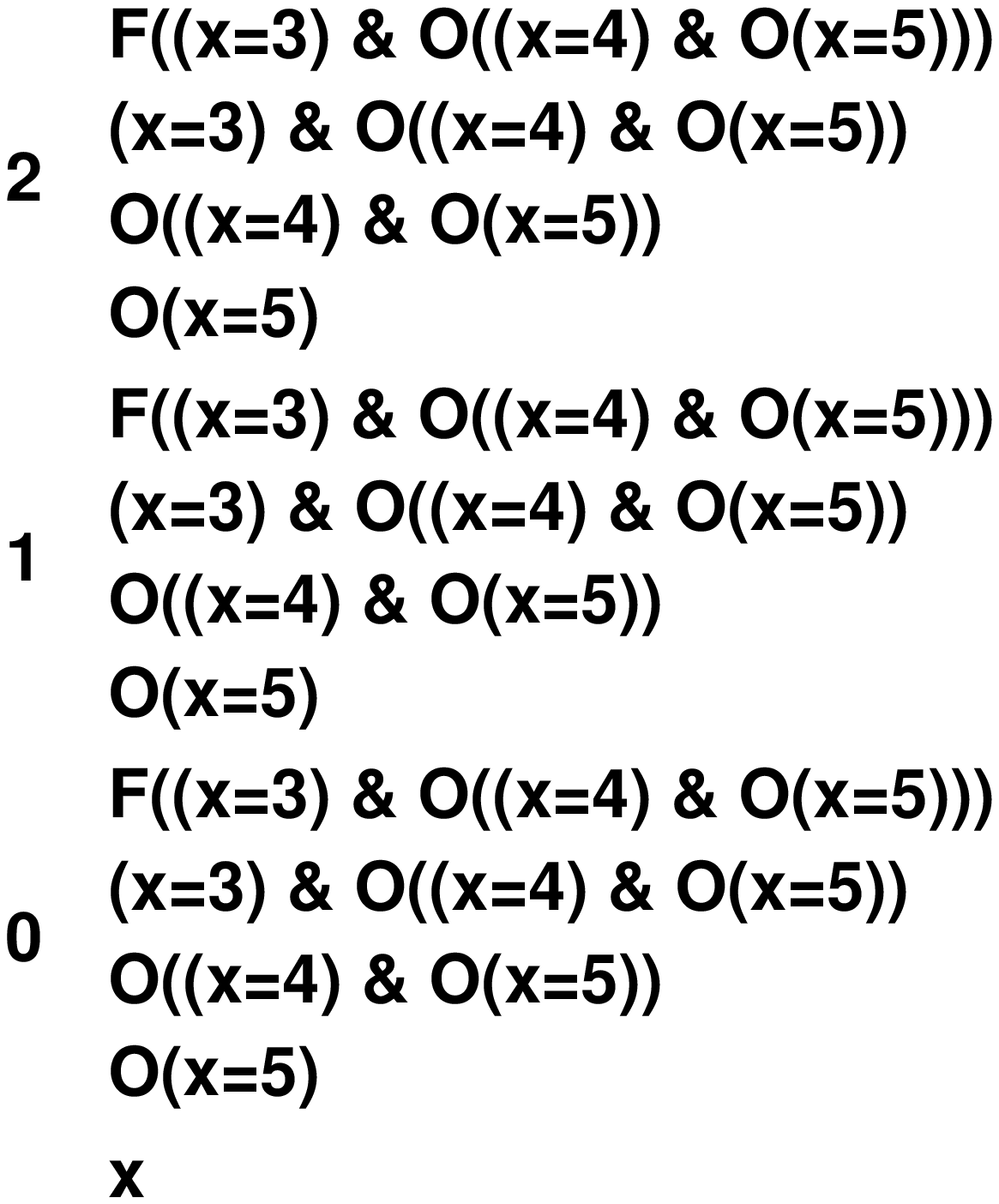}{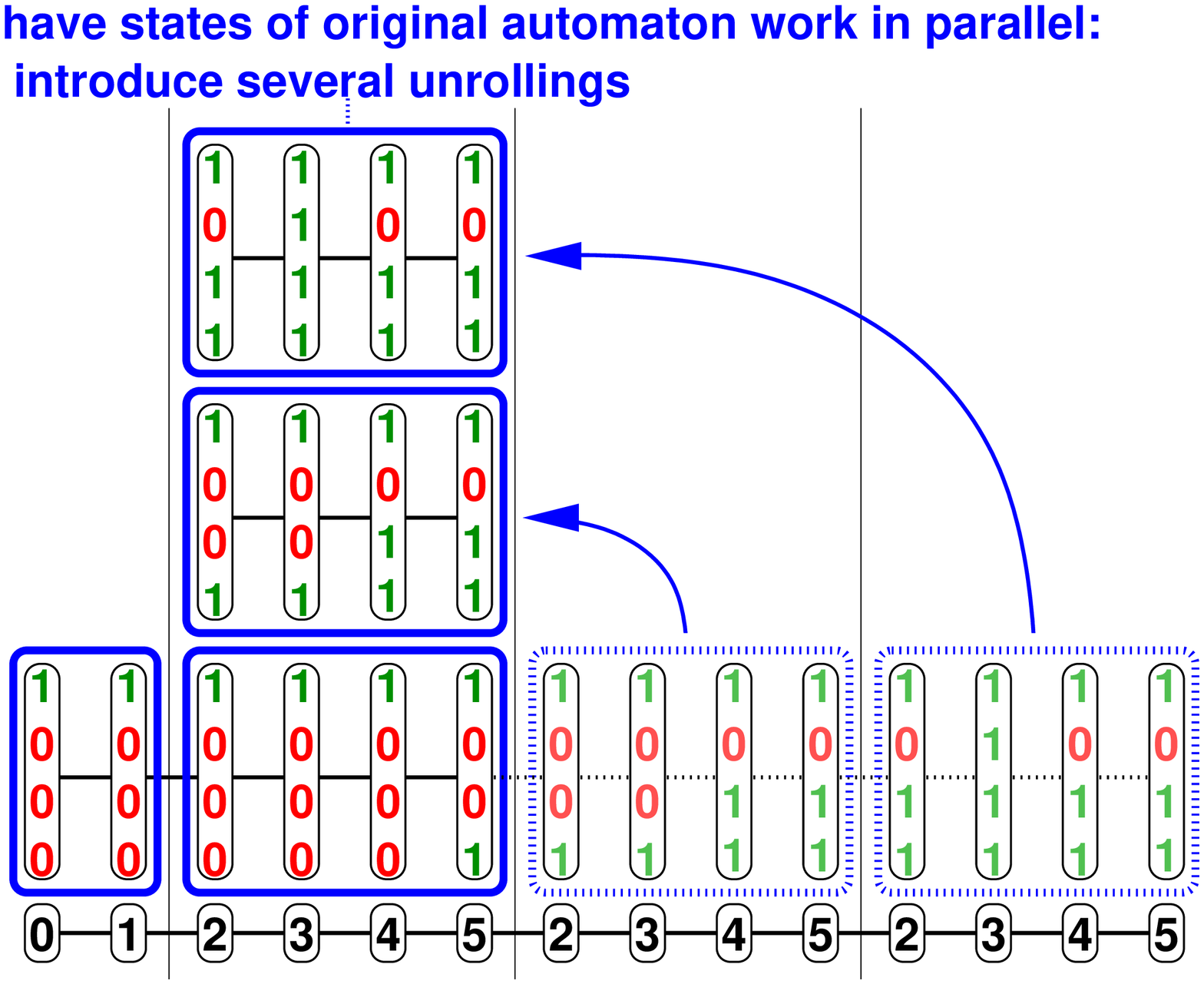}{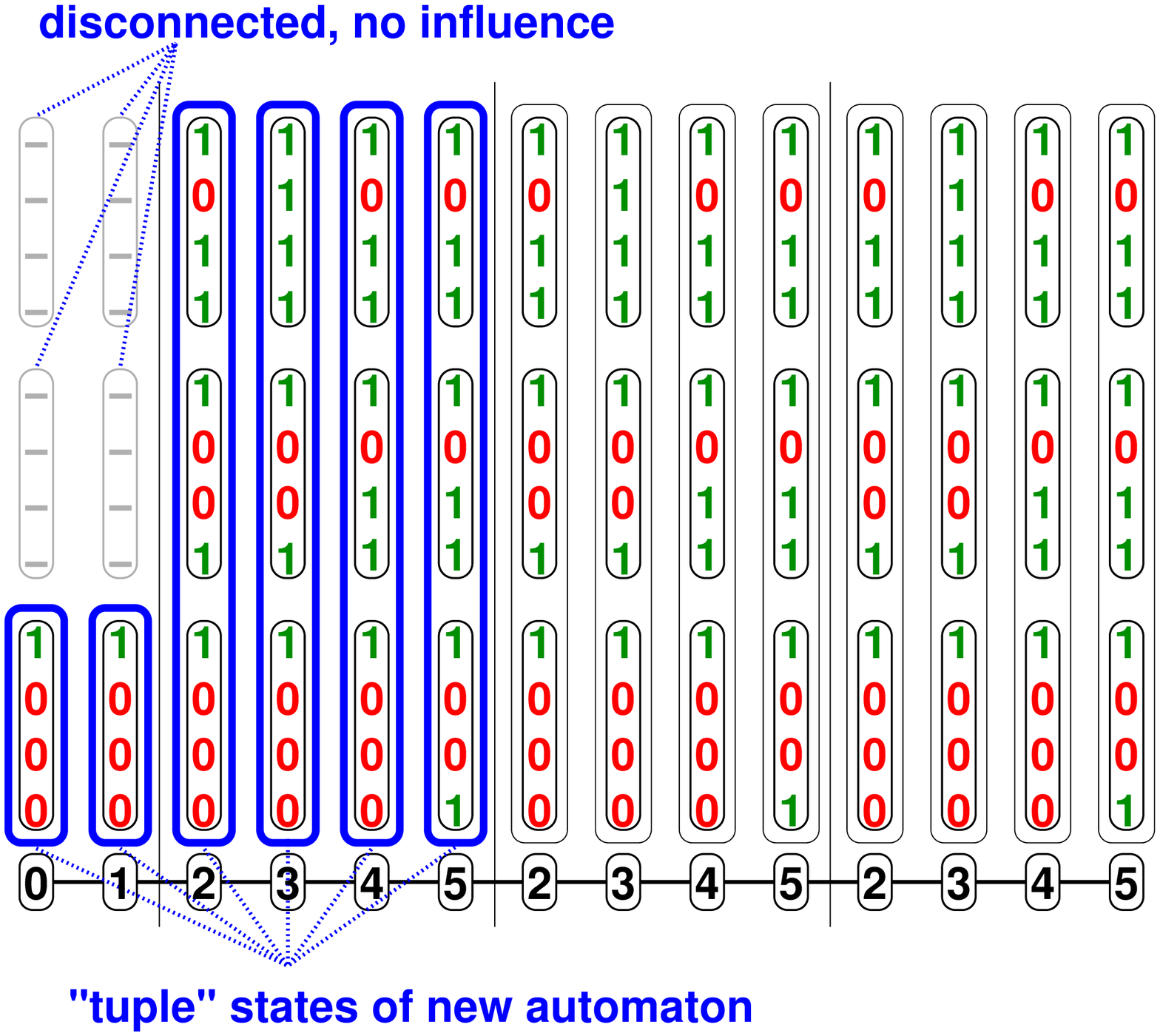}\\[1ex]
\tabpltlbuechiexampleline{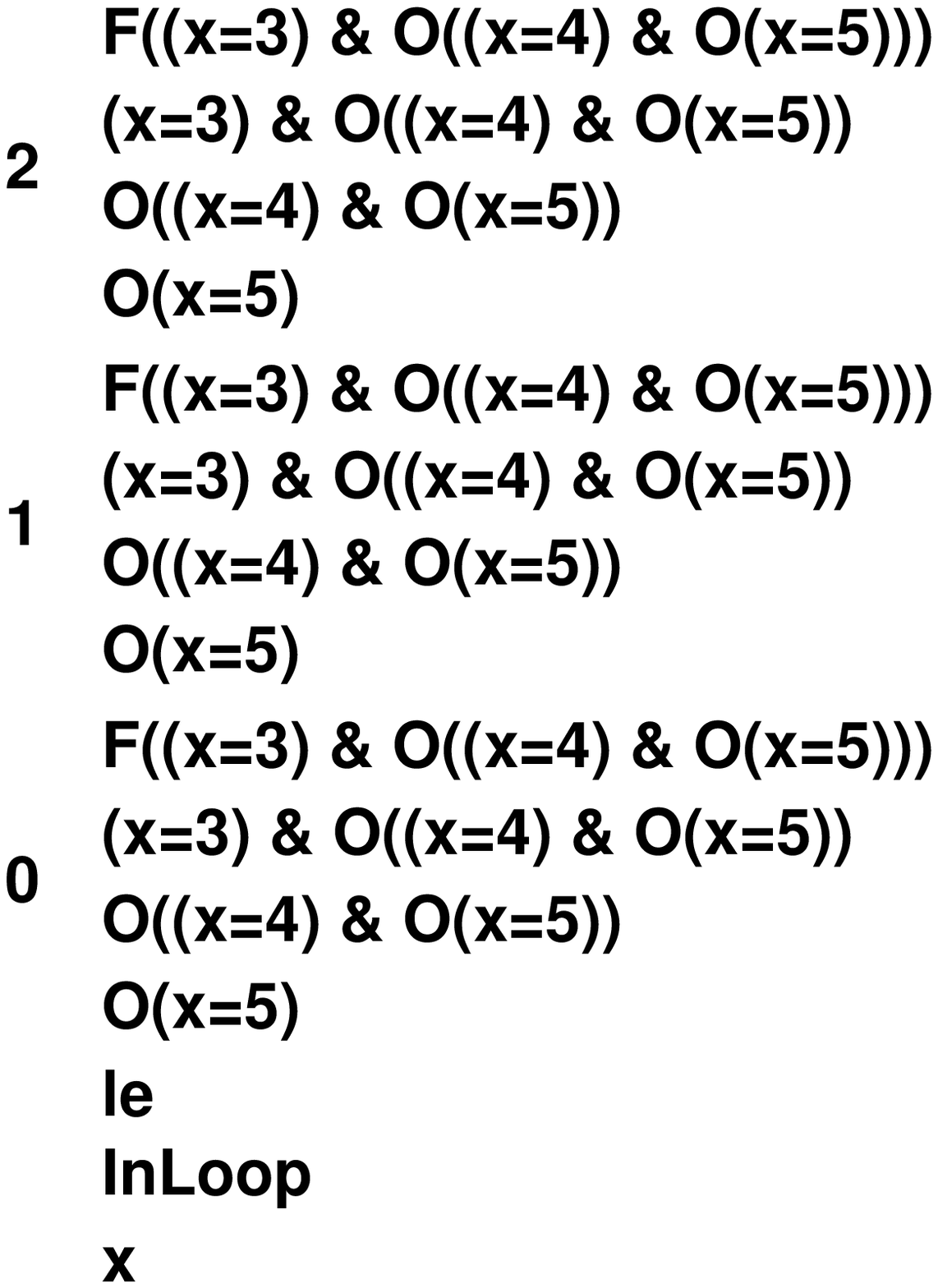}{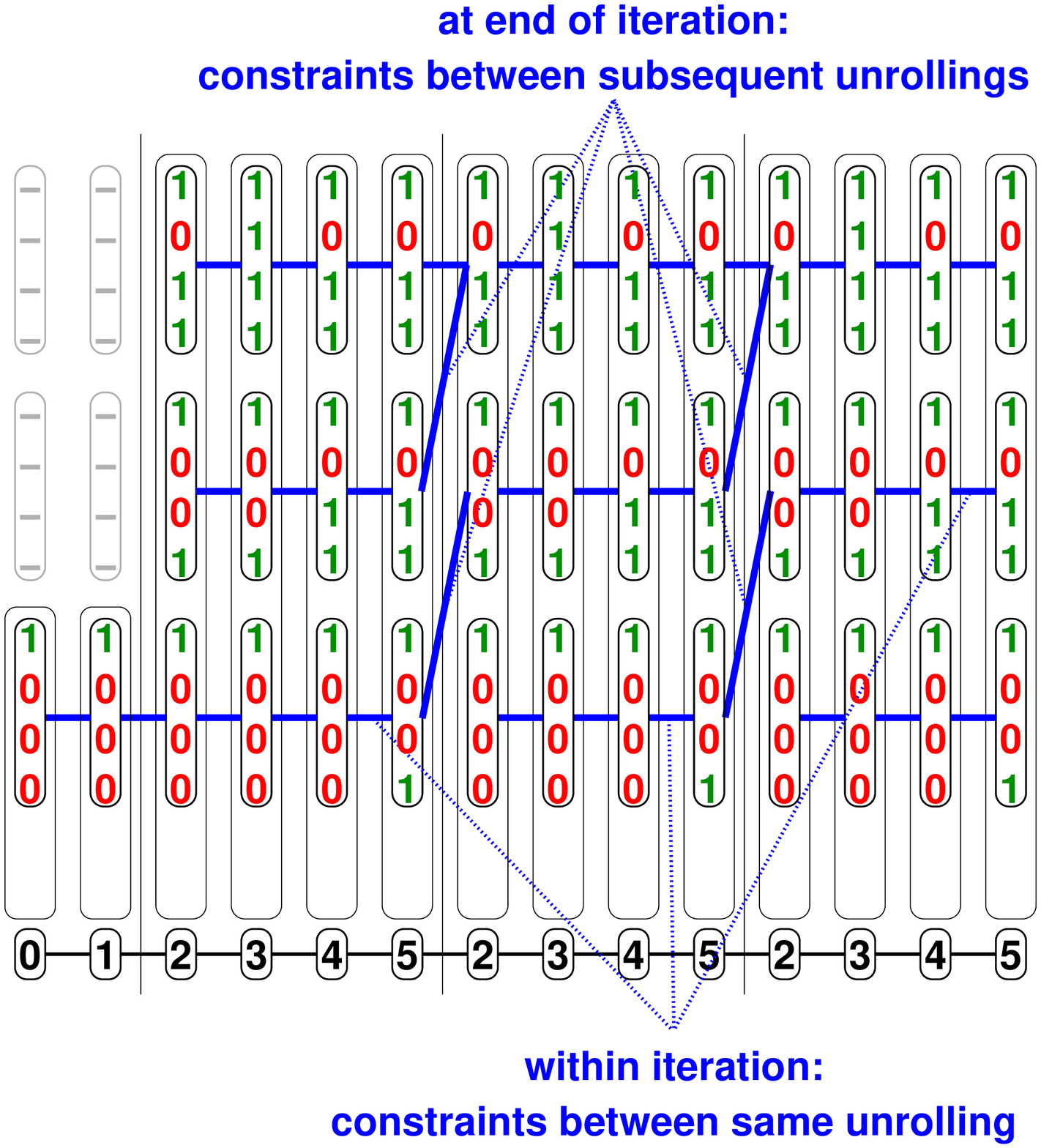}{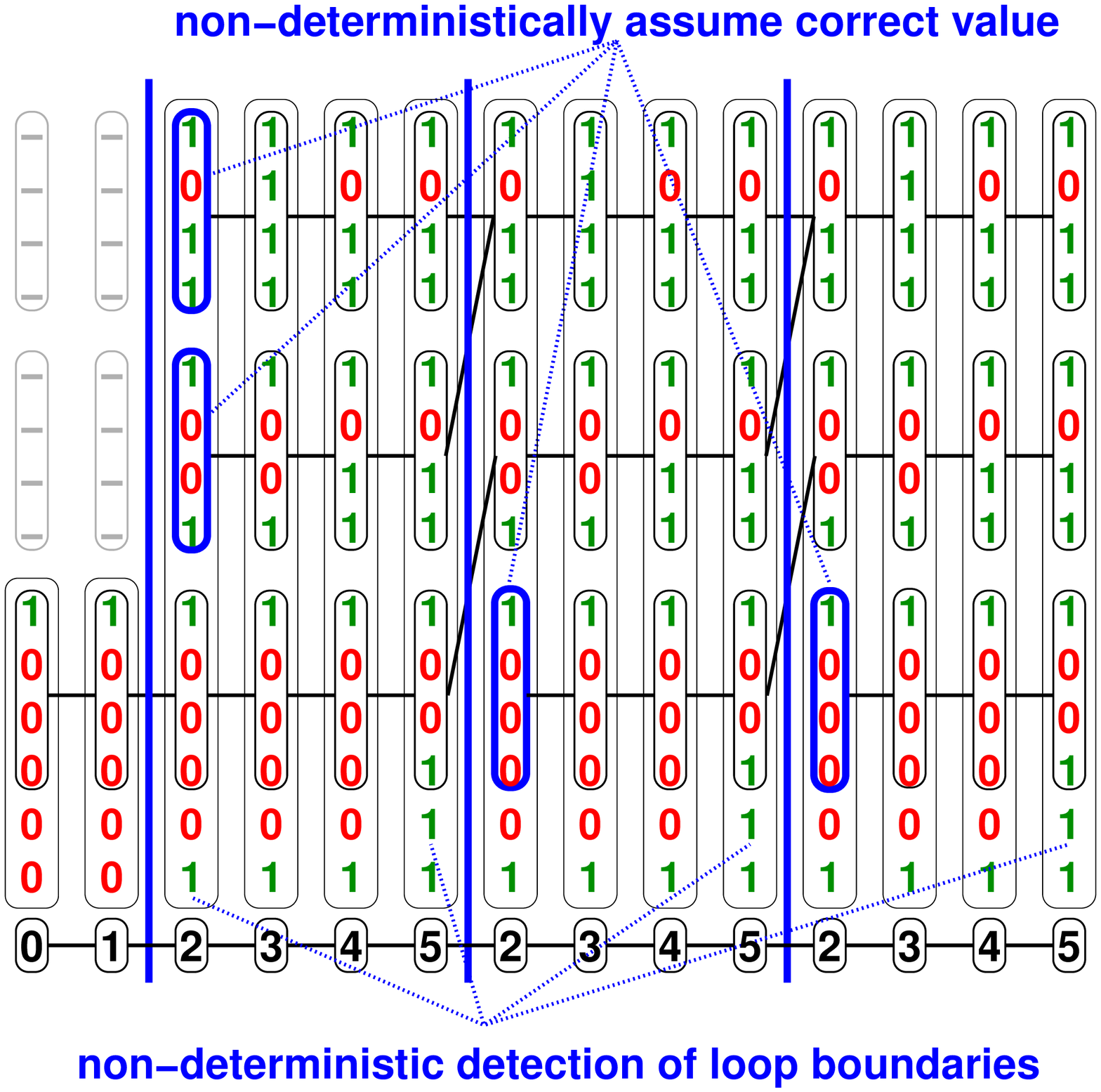}\\[1ex]
\tabpltlbuechiexampleline{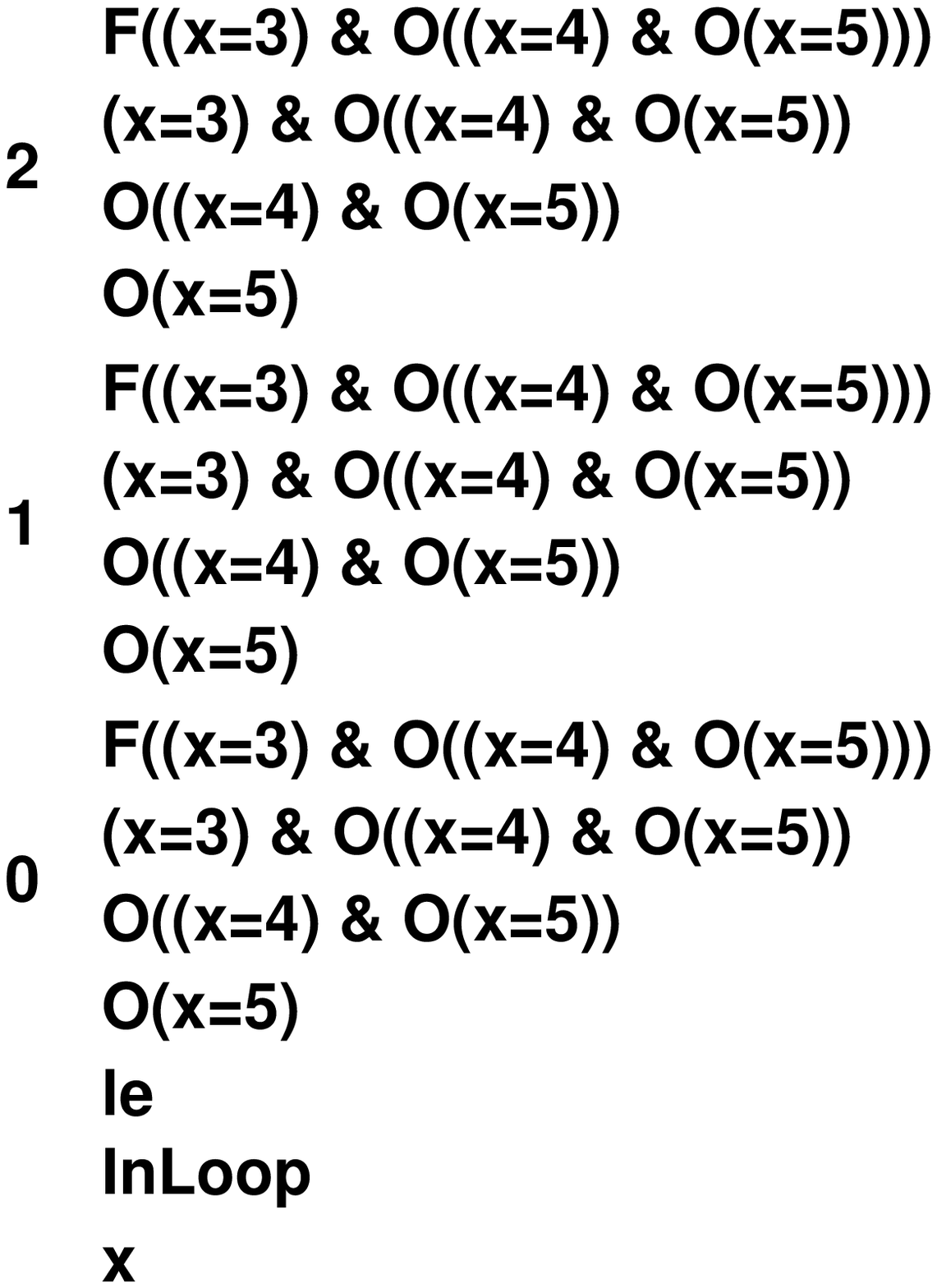}{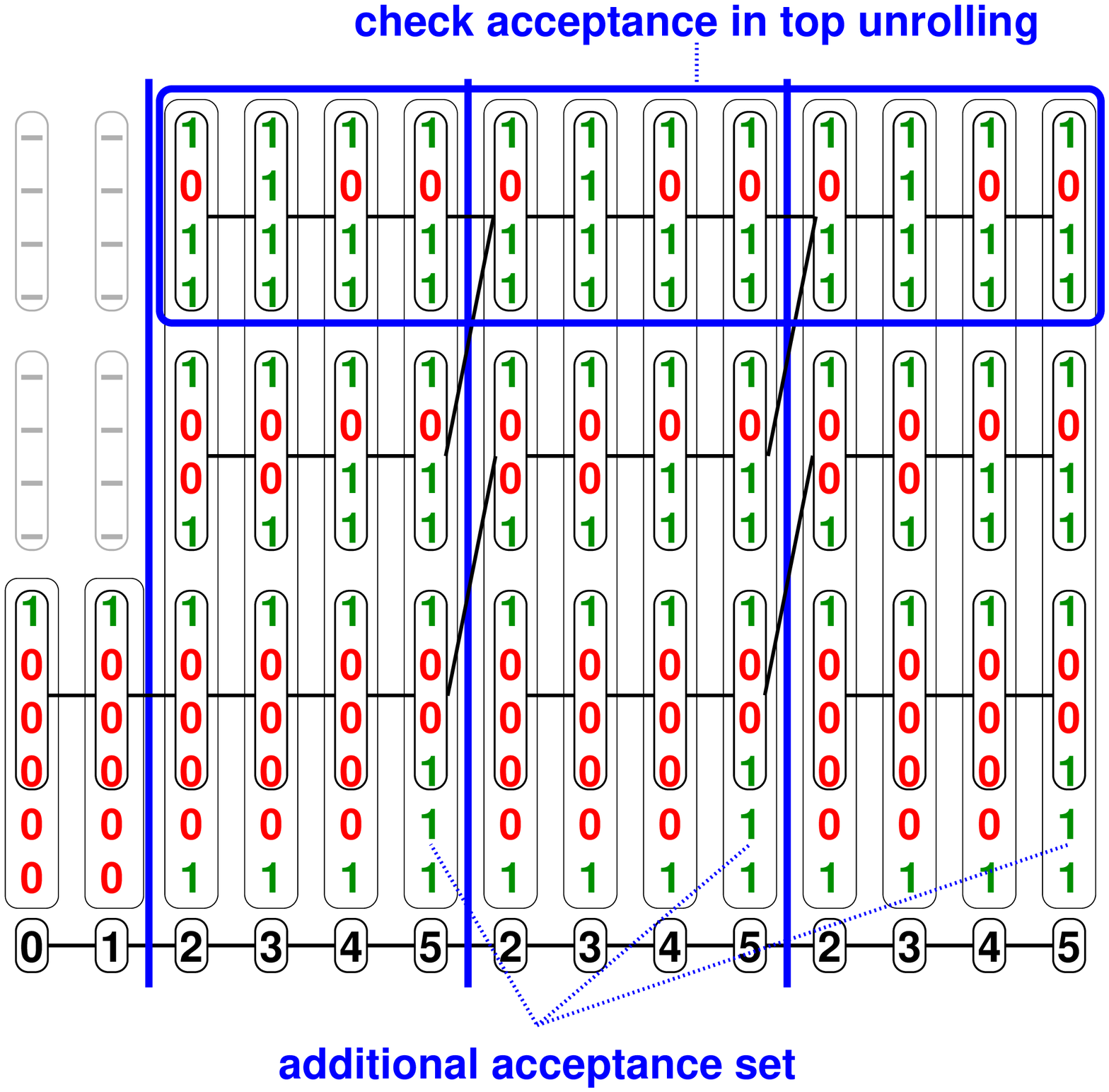}{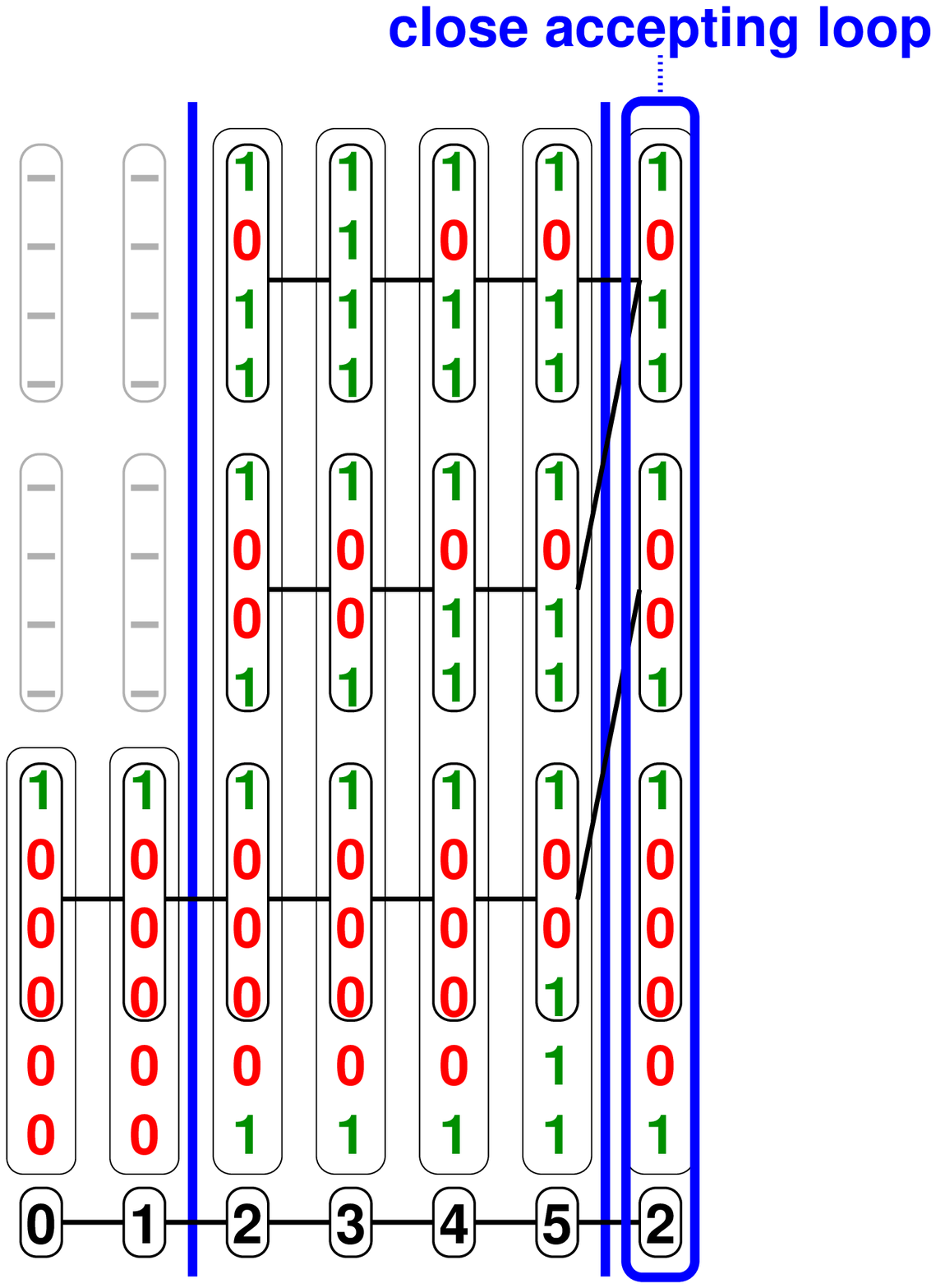}
\end{tabular}
\caption{\label{fig:pltlbuechiexample}Tightening~\cite{KPR98} by example}
\end{figure}

By virtually unrolling the transition relation of $M$ (or, in other
words, by folding in the transition relation of $\tilde{B}^\TLF$) some
parts of the run of $\tilde{B}^\TLF$ can take place in parallel to
reduce some or all of the excess length
(Fig.~\ref{fig:pltlbuechiexample}b,c). So far, there is no difference
to the BMC case. We now have to decide how to define states,
transition relation, and acceptance sets of the new automaton
$B^\TLF$.

The states of $B^\TLF$ consist of tuples of states of $\tilde{B}^\TLF$
(Fig.~\ref{fig:pltlbuechiexample}d). Before the loop starts the tuples
need only have size 1, i.e., they are identical to the states of
$\tilde{B}^\TLF$. After the loop start the tuples must be able to
accommodate the maximal excess length of an accepting run of
$\tilde{B}^\TLF$. If $\tilde{B}^\TLF$ is derived from~\cite{KPR98} we
can obtain a similar result as in Prop.~\ref{prop-depth} on the excess
length of accepting runs of $\tilde{B}^\TLF$~\cite{SB05}. Hence, the
maximum required size of the tuples is given by the past operator
depth of $\TLF$ plus one. In practice, the tuples before the loop
start also have that size but its constituent states at unrollings $>
0$ are disconnected from the rest of the automaton. Note that the
states of $B^\TLF$ at time points 6--9 and 10--13 are the same as
at time points 2--5.

Defining transitions not crossing a loop boundary is easy: there is a
transition from one tuple state to another in $B^\TLF$ iff each pair
of constituent states at the same unrolling has a transition in
$\tilde{B}^\TLF$ (Fig.~\ref{fig:pltlbuechiexample}e). When crossing a
loop boundary, a constituent state at unrolling $d-1$ in the pre-state
is connected to a constituent state at unrolling $d$ in the
post-state. In addition, there must be a transition in $\tilde{B}^\TLF$
between the constituent states at the highest unrolling of the pre-
and post-state to ensure that a loop exists in $\tilde{B}^\TLF$.

Clearly, we cannot know in which state $\tilde{B}^\TLF$ would be in an
unrolling $> 0$ when first entering the loop in $B^\TLF$ (time point 2
in Fig.~\ref{fig:pltlbuechiexample}f). Hence, the corresponding
constituent states in $B^\TLF$ are not constrained to the
past.\footnote{If $\tilde{B}^\TLF$ is derived from~\cite{KPR98} some
constraints similar to those in Sect.~5 of~\cite{HelJunLat:CAV05}
could be applied for monotonic operators.}
The constituent states at
unrolling 0 at time points 6 and 10 could in principle be forced to be
identical to their predecessor at time point 2; however, it turns out
that this is not required for correctness of the construction. The
loop boundaries are ``detected'' non-deterministically using oracle
variables $\InLoop{}$ with the same meaning as before and $\lend$
indicating the last state of a loop iteration.

As acceptance of a run in $\tilde{B}^\TLF$ is determined in its
looping part, each tuple state in $B^\TLF$ is in the acceptance set
$\tilde{F}_m$ of $B^\TLF$ iff its constituent state in the top
unrolling belongs to the corresponding acceptance set $F_m$ of
$\tilde{B}^\TLF$ (Fig.~\ref{fig:pltlbuechiexample}g). One additional
acceptance set is needed in $B^\TLF$ to guarantee that infinitely
often a loop boundary is guessed. Otherwise, there might not be a
connection between the bottom and top unrollings and, therefore,
acceptance might not be determined correctly. Finally, an accepting
loop can be closed (Fig.~\ref{fig:pltlbuechiexample}h).

\subsubsection*{Construction}

\def\TransTBA#1#2{\ensuremath{{{\vert [{#1}] \vert}^{#2}}}}

We symbolically construct a B{\"u}chi automaton $B^\TLF = (S,T,I,L,F)$
for a PLTL formula $\TLF$ as follows. For each $\TSF \in
\mathit{cl}(\TLF)$, $V$ contains state variables
$\TransTBA{\TSF}{0},\ldots,\TransTBA{\TSF}{\delta(\TSF)}$ meant to
represent the truth of $\TSF$ at unrollings $0 \le i \le
\delta(\TSF)$. Two oracles $\InLoop{}$ and $\lend$ signal the presumed
start of the loop and the end of each loop iteration. The rest of the
encoding is developed step by step below.
\newcounter{cnt:pltl-buechi-construction}
\def\tabpltlbuechiconstructionlinetwo#1#2{\stepcounter{cnt:pltl-buechi-construction}\arabic{cnt:pltl-buechi-construction} & {#1} & {#2}}
\def\tabpltlbuechiconstructionlinethree#1#2#3{\stepcounter{cnt:pltl-buechi-construction}\arabic{cnt:pltl-buechi-construction} & {#1} & {#2} & {#3}}
{\encsize
\[
\begin{array}{cc|c}
\mbox{line} & \mbox{constraint} & \mbox{applies to}\\
\hline
\tabpltlbuechiconstructionlinetwo{\InLoop{} \Rightarrow \InLoop{}'}{T}\\
\tabpltlbuechiconstructionlinetwo{\lend \Rightarrow \InLoop{}}{S}\\[1ex]
\end{array}
\]
}\smallskip%

\noindent
As in the BMC case we set $\TransTBA{\TSF}{d} \Leftrightarrow
\TransTBA{\TSF}{\delta(\TSF)}$ if $d > \delta(\TSF)$. The state
variables for atomic propositions are unconstrained; their valuations
are linked to the corresponding atomic propositions via $L$,
though.\footnote{Here we assume that the product of Kripke structures
is formed by demanding that product states match on shared atomic
propositions, see, e.g., \cite{Schuppan06}. In a symbolic setting
atomic propositions often correspond directly to valuations of state
variables and, hence, the product can be formed more directly by
sharing these state variables.} The valuation of the state variables
for Boolean operators is again the same as in the previous subsection:
{\encsize
\[
\begin{array}{cc|c|c}
\mbox{line} & \TSF & 0 \le d \le \delta(\TSF) & \mbox{applies to}\\
\hline
\tabpltlbuechiconstructionlinethree{p}{\top}{S}\\
\tabpltlbuechiconstructionlinethree{\neg p}{\top}{S}\\
\tabpltlbuechiconstructionlinethree{\LSF \wedge \RSF}{\TransTBA{\LSF \wedge \RSF}{d}\Leftrightarrow \TransTBA{\LSF}{d} \wedge \TransTBA{\RSF}{d}}{S}\\
\tabpltlbuechiconstructionlinethree{\LSF \vee \RSF}{\TransTBA{\LSF \vee \RSF}{d}\Leftrightarrow \TransTBA{\LSF}{d} \vee \TransTBA{\RSF}{d}}{S}\\[1ex]
\end{array}
\]
}\smallskip%

\noindent
Within a loop iteration and on the stem the valuation of the variables
for temporal operators directly follows their characterisation in
terms of current and next state values. Note that stem and loop are
disconnected at unrollings $> 0$. In the following tables we sometimes
use parentheses to disambiguate the scope of the next state operator
${}'$ and 'applies to' abbreviated with a.t.
{\tighterencsize
\[
\begin{array}{cc|c|c}
\mbox{line} &\TSF & 0 \le d \le \delta(\TSF) & \mbox{a.t.}\\
\hline
\tabpltlbuechiconstructionlinethree{\Next{\SF}}{\neg \lend \wedge \neg (\neg \InLoop{} \wedge \InLoop{}' \wedge d > 0) \Rightarrow (\TransTBA{\Next{\SF}}{d} \Leftrightarrow (\TransTBA{\SF}{d})')}{T}\\
\tabpltlbuechiconstructionlinethree{\Until{\LSF}{\RSF}}{\neg \lend \wedge \neg (\neg \InLoop{} \wedge \InLoop{}' \wedge d > 0) \Rightarrow (\TransTBA{\Until{\LSF}{\RSF}}{d} \Leftrightarrow \TransTBA{\RSF}{d} \vee (\TransTBA{\LSF}{d} \wedge (\TransTBA{\TSF}{d})'))}{T}\\
\tabpltlbuechiconstructionlinethree{\Release{\LSF}{\RSF}}{\neg \lend \wedge \neg (\neg \InLoop{} \wedge \InLoop{}' \wedge d > 0) \Rightarrow (\TransTBA{\Release{\LSF}{\RSF}}{d} \Leftrightarrow \TransTBA{\RSF}{d} \wedge (\TransTBA{\LSF}{d} \vee (\TransTBA{\TSF}{d})'))}{T}\\
\hline
\tabpltlbuechiconstructionlinethree{\Prev{\SF}}{\neg \lend \wedge \neg (\neg \InLoop{} \wedge \InLoop{}' \wedge d > 0) \Rightarrow ((\TransTBA{\Prev{\SF}}{d})' \Leftrightarrow \TransTBA{\SF}{d})}{T}\\
\tabpltlbuechiconstructionlinethree{\PrevZ{\SF}}{\neg \lend \wedge \neg (\neg \InLoop{} \wedge \InLoop{}' \wedge d > 0) \Rightarrow ((\TransTBA{\PrevZ{\SF}}{d})' \Leftrightarrow \TransTBA{\SF}{d})}{T}\\
\tabpltlbuechiconstructionlinethree{\Since{\LSF}{\RSF}}{\neg \lend \wedge \neg (\neg \InLoop{} \wedge \InLoop{}' \wedge d > 0) \Rightarrow ((\TransTBA{\Since{\LSF}{\RSF}}{d})' \Leftrightarrow (\TransTBA{\RSF}{d})' \vee ((\TransTBA{\LSF}{d})' \wedge \TransTBA{\TSF}{d}))}{T}\\
\tabpltlbuechiconstructionlinethree{\Trigger{\LSF}{\RSF}}{\neg \lend \wedge \neg (\neg \InLoop{} \wedge \InLoop{}' \wedge d > 0) \Rightarrow ((\TransTBA{\Trigger{\LSF}{\RSF}}{d})' \Leftrightarrow (\TransTBA{\RSF}{d})' \wedge ((\TransTBA{\LSF}{d})' \vee \TransTBA{\TSF}{d}))}{T}\\[1ex]
\end{array}
\]
}\smallskip

\noindent
When the end of a loop iteration is reached, subsequent unrollings
(other than the topmost) are linked by taking current state values
from unrolling $d$ and next state values from unrolling $d+1$. In the
topmost unrolling current and next state values are taken from the same
unrolling to ensure stabilisation of all variables. This case
corresponds to the loop-back case in BMC\@.
{\tighterencsize
\[
\begin{array}{cc|c|c}
\mbox{line} &\TSF & 0 \le d \le \delta(\TSF) & \mbox{a.t.}\\
\hline
\tabpltlbuechiconstructionlinethree{\Next{\SF}}{\lend \Rightarrow (\TransTBA{\Next{\SF}}{d} \Leftrightarrow (\TransTBA{\SF}{\min(d+1,\delta(\TSF))})')}{T}\\
\tabpltlbuechiconstructionlinethree{\Until{\LSF}{\RSF}}{\lend \Rightarrow (\TransTBA{\Until{\LSF}{\RSF}}{d} \Leftrightarrow \TransTBA{\RSF}{d} \vee (\TransTBA{\LSF}{d} \wedge (\TransTBA{\TSF}{\min(d+1,\delta(\TSF))})'))}{T}\\
\tabpltlbuechiconstructionlinethree{\Release{\LSF}{\RSF}}{\lend \Rightarrow (\TransTBA{\Release{\LSF}{\RSF}}{d} \Leftrightarrow \TransTBA{\RSF}{d} \wedge (\TransTBA{\LSF}{d} \vee (\TransTBA{\TSF}{\min(d+1,\delta(\TSF))})'))}{T}\\
\hline
\tabpltlbuechiconstructionlinethree{\Prev{\SF}}{\lend \Rightarrow ((\TransTBA{\Prev{\SF}}{\min(d+1,\delta(\TSF))})' \Leftrightarrow \TransTBA{\SF}{d})}{T}\\
\tabpltlbuechiconstructionlinethree{\PrevZ{\SF}}{\lend \Rightarrow ((\TransTBA{\PrevZ{\SF}}{\min(d+1,\delta(\TSF))})' \Leftrightarrow \TransTBA{\SF}{d})}{T}\\
\tabpltlbuechiconstructionlinethree{\Since{\LSF}{\RSF}}{\lend \Rightarrow ((\TransTBA{\Since{\LSF}{\RSF}}{\min(d+1,\delta(\TSF))})' \Leftrightarrow (\TransTBA{\RSF}{\min(d+1,\delta(\TSF))})' \vee ((\TransTBA{\LSF}{\min(d+1,\delta(\TSF))})' \wedge \TransTBA{\TSF}{d}))}{T}\\
\tabpltlbuechiconstructionlinethree{\Trigger{\LSF}{\RSF}}{\lend \Rightarrow ((\TransTBA{\Trigger{\LSF}{\RSF}}{\min(d+1,\delta(\TSF))})' \Leftrightarrow (\TransTBA{\RSF}{\min(d+1,\delta(\TSF))})' \wedge ((\TransTBA{\LSF}{\min(d+1,\delta(\TSF))})' \vee \TransTBA{\TSF}{d}))}{T}\\[1ex]
\end{array}
\]
}\smallskip

\noindent
Variables representing past operators are initialised in unrolling 0
as usual:
{\encsize
\[
\begin{array}{cc|c|c}
\mbox{line} &\TSF & & \mbox{applies to}\\
\hline
\tabpltlbuechiconstructionlinethree{\Prev{\SF}}{\TransTBA{\Prev{\SF}}{0} \Leftrightarrow \bot}{I}\\
\tabpltlbuechiconstructionlinethree{\PrevZ{\SF}}{\TransTBA{\PrevZ{\SF}}{0} \Leftrightarrow \top}{I}\\
\tabpltlbuechiconstructionlinethree{\Since{\LSF}{\RSF}}{\TransTBA{\Since{\LSF}{\RSF}}{0} \Leftrightarrow \TransTBA{\RSF}{0}}{I}\\
\tabpltlbuechiconstructionlinethree{\Trigger{\LSF}{\RSF}}{\TransTBA{\Trigger{\LSF}{\RSF}}{0} \Leftrightarrow \TransTBA{\RSF}{0}}{I}\\[1ex]
\end{array}
\]
}\smallskip%

\noindent
Acceptance for $\Until{}{}$- and $\Release{}{}$-formulae is defined in
their topmost unrolling but is otherwise standard:
{\encsize
\[
\begin{array}{cc|c|c}
\mbox{line} &\TSF & & \mbox{applies to}\\
\hline
\tabpltlbuechiconstructionlinethree{\Until{\LSF}{\RSF}}{\neg \TransTBA{\Until{\LSF}{\RSF}}{\delta(\TSF)} \vee \TransTBA{\RSF}{\delta(\TSF)}}{F}\\
\tabpltlbuechiconstructionlinethree{\Release{\LSF}{\RSF}}{\TransTBA{\Release{\LSF}{\RSF}}{\delta(\TSF)} \vee \neg \TransTBA{\RSF}{\delta(\TSF)}}{F}\\[1ex]
\end{array}
\]
}\smallskip%

\noindent
Finally, we add $\TransTBA{\TLF}{0}$ as an initial state constraint to
ensure the desired semantics and $\{\lend\}$ as acceptance set to
guarantee that ultimately all unrollings are linked. The labelling is
defined as $L(s) = \{p \in \mathit{AP}(\TLF) \mid \TransTBA{p}{0} \in
V \wedge \TransTBA{p}{0}(s) = \top\}$ where $\mathit{AP}(\TLF)$ is the
set of atomic propositions occurring in $\TLF$.

In the following we prove that $B^\TLF$ accepts the desired language
and is tight.

\begin{lemma}
\label{thm:pltl-buechi-correct}
$\Lang{B^\TLF} = \{\alpha \mid \alpha \models \TLF\}$
\end{lemma}

\begin{proof}
Let $\breve{B}^\TLF$ be defined as $B^\TLF$ without the initial state
constraint $\TransTBA{\TLF}{0}$.

(Correctness) We show that on every initialised fair path $\rho$ in
$\breve{B}^\TLF$ the values of $\TransTBA{\TSF}{d_i}(\rho_i)$
represent the validity of the subformula $\TSF$ at time point $i$, where
$d_i$ is either the number of $\lend$'s seen up to time point $i-1$ or
$\delta(\TSF)$, whichever is smaller. Formally, let $\rho$ be an initialised fair
path with $L(\rho) = \alpha$ in $\breve{B}^\TLF$. For each time point
$i$ in $\rho$, let $d_i = \min(|\{j \mid (j \le i-1) \wedge
(\mathit{\lend}(\rho_j) \Leftrightarrow
\top)\}|,\delta(\TSF))$.\footnote{In Fig.~\ref{fig:pltlbuechiexample}e
this corresponds to the thick sequence of transitions starting in
unrolling 0 at time point 0, jumping to unrolling 1 between time
points 5 and 6, and finally reaching unrolling 2 at time point 10.}
The initial, invariant, transition, and fairness constraints on
$\TransTBA{\TSF}{d_i}$ are identical to the constraints that a
B{\"u}chi automaton based on~\cite{KPR98} imposes on its state
variables representing the corresponding subformula. Hence, $\alpha^i
\models \TSF \Leftrightarrow \TransTBA{\TSF}{d_i}(\rho_i)$.

(Completeness) We show that there is an initialised fair path $\rho$
in $\breve{B}^\TLF$ with $L(\rho) = \alpha$ for each word
$\alpha$. Choose a set of indices $U = \{i_0,i_1,\ldots\}$ (for
``up'') such that $\lend(\rho_i) \Leftrightarrow i \in U$. Further,
choose $l \le i_0$ and set $\InLoop{}(\rho_j) \Leftrightarrow j \ge
l$. We inductively construct a valuation for
$\TransTBA{\TSF}{d}(\rho_i)$ for each subformula $\TSF$ of $\TLF$, $d
\le \delta(\TSF)$, and $i \ge 0$.
\begin{itemize}
\item If $\TSF$ is an atomic proposition $p$, set
$\TransTBA{p}{0}(\rho_i) \Equiv (\alpha^i \models p)$.

\item If the top level operator of $\TSF$ is Boolean, the valuation
follows directly from the semantics of the operator.

\item For $\Next{}$, each $\TransTBA{\Next{\SF}}{d}(\rho_i)$ appears
at most once in $\Next{}$'s defining constraint (line 7).

\item $\TSF = \Prev{\SF}$ is similar. Note that $\delta(\SF) =
\delta(\TSF) - 1$. Therefore, $\TransTBA{\Prev{\SF}}{\delta(\TSF)}'
\Leftrightarrow \TransTBA{\SF}{\delta(\TSF)}$ and
$\TransTBA{\Prev{\SF}}{\delta(\TSF)}' \Leftrightarrow
\TransTBA{\SF}{\delta(\TSF)-1}$ are
equivalent. $\TransTBA{\Prev{\SF}}{d}(\rho_i)$ is unconstrained if $d
= 0$ and $i - 1 \in U$ as well as if $d \ge 1$ and $i = l$.

\item For $\TSF = \Until{\LSF}{\RSF}$, start with the topmost
unrolling $\delta(\TSF)$. If $\TransTBA{\RSF}{\delta(\RSF)}$ remains
false from some $i_d$ on, assign $\forall i \ge i_d \;.\;
\TransTBA{\TSF}{\delta(\TSF)}(\rho_i) \Equiv \bot$. Now work towards
decreasing $i$ from each $i_n$ with
$\TransTBA{\RSF}{\delta(\RSF)}(i_n) \Equiv \top$, using line 8 in the
definition of $T$ for $\Until{}{}$. Continue with unrolling
$\delta(\TSF)-1$. Start at each $i \in U$ by obtaining
$\TransTBA{\TSF}{\delta(\TSF)-1}(\rho_i)$ from the previously assigned
$\TransTBA{\TSF}{\delta(\TSF)}(\rho_{i+1})$ via line 15. Then work
towards decreasing $i$ again using line 8 in the definition of $T$
until $\TransTBA{\TSF}{\delta(\TSF)-1}$ is assigned for all
$\rho_i$. This is repeated in decreasing order for each unrolling $0
\le d < \delta(\TSF) - 1$.

\item For $\Since{}{}$, start with $\TransTBA{\TSF}{0}(\rho_0)$ and
proceed towards increasing $i$, also increasing $d$ when $i \in U$
(lines 12, 19 in the definition of $T$ for $\Since{}{}$). When
$d=\delta(\TSF)$ is reached, assign
$\TransTBA{\TSF}{\delta(\TSF)}(\rho_i)$ for all $i$ using line 12 in
the definition of $T$. Then, similar to $\Until{}{}$, work towards
decreasing $d$ and $i$ from each $i \in U$.

\item $\PrevZ{}$, $\Release{}{}$, and $\Trigger{}{}$ are as their duals.
\end{itemize}
For state variables on the stem with $d > 0$ any assignment satisfying
the constraints in the definition of $B^\TLF$ can be chosen. It is
easy to verify that such assignment always exists. Fairness follows
from the definition of $U$, $l$, and the valuation chosen for
$\Until{}{}$ and $\Release{}{}$.

The claim is now immediate by the definition of $I$.
\end{proof}

\begin{lemma}
\label{thm:pltl-buechi-tight}
$B^\TLF$ is tight.
\end{lemma}

\begin{proof}
We show inductively that the valuations of the variables
$\TransTBA{\TSF}{d}(\rho_i)$ can be chosen such that the valuation at
a given relative index in a loop iteration is the same for each
iteration in an unrolling $d$. Formally, let $\alpha =
\beta\gamma^\omega$ with $\alpha \models \TLF$. There exists a run
$\rho$ on $\alpha$ such that for all subformulas $\TSF$ of $\TLF$
\[
\forall d \le \delta(\TSF) \;.\; \forall i_1,i_2 \ge |\beta| \;.\; ((\exists k \ge 0 \;.\; i_2 - i_1 = k |\gamma|) \Rightarrow (\TransTBA{\TSF}{d}(\rho_{i_1}) \Equiv \TransTBA{\TSF}{d}(\rho_{i_2})))
\]

Atomic propositions, Boolean connectives, and $\Next{}$ are
clear. $\Prev{}$ is also easy, we only have to assign the appropriate
value from other iterations when $\TransTBA{\TSF}{d}(i)$ is
unconstrained. For $\TSF = \Until{\LSF}{\RSF}$, by the induction
hypothesis, $\TransTBA{\RSF}{\delta(\RSF)}$ is either always false (in
which case we assign $\TransTBA{\TSF}{\delta(\TSF)}(\rho_i)$ to false
according to the proof of Lemma~\ref{thm:pltl-buechi-correct}) or
becomes true at the same time in each loop iteration. Hence, the claim
holds for unrolling $\delta(\TSF)$. From there we can proceed to lower
unrollings in the same manner as in the proof of Lemma~\ref{thm:pltl-buechi-correct}.
For $\Since{}{}$ we follow the order of
assignments from the proof of Lemma~\ref{thm:pltl-buechi-correct}. By
induction, the claim holds for unrolling $\delta(\TSF)$. From there,
we proceed towards decreasing $i$ and $d$. We use, by induction, the
same valuations of subformulas and the same equations (though in
reverse direction) as we used to get from $\TransTBA{\TSF}{0}(\rho_0)$
to unrolling $\delta(\TSF)$. $\PrevZ{}$, $\Release{}{}$, and
$\Trigger{}{}$ are as their duals.
\end{proof}

\begin{theorem}
\label{thm:pltl-buechi}
Let $\TLF$ be a PLTL formula, let $B^\TLF$ be defined as above. Then,
$\Lang{B^\TLF} = \{\alpha \mid \alpha \models \TLF\}$ and $B^\TLF$ is
tight.
\end{theorem}

\begin{proof}
By Lemma~\ref{thm:pltl-buechi-correct} and \ref{thm:pltl-buechi-tight}.
\end{proof}

As an optimisation, state variables representing atomic propositions,
Boolean operators, and values of subformulas $\TSF$ at unrollings $d >
\delta(\TSF)$ can be replaced with macros. If the automaton is used
with the liveness-to-safety transformation (with appropriate changes
to shift $\lstart$ and $\InLoop{}$ back one state), $\InLoop{}$ can be taken
directly from the transformation and $\lend$ can be defined as
$\LoopClosed'$.

\section{Incremental SAT and BMC}\label{sec:incbmc}
We now present an incremental eventuality encoding for PLTL
(see Sect.~\ref{sec:pltl-eventuality} for the non-incremental version).
The encoding has been first published in~\cite{HelJunLat:CAV05} and
is based on an earlier PLTL fixpoint evaluation encoding published in~\cite{LBHJ05}.

A promising technique for improving the performance of BMC is using
\emph{incremental} SAT solving.
When a solver is faced with a sequence of related problems,
learned clauses (see e.g., \cite{ZhangMMM01}) from the previous problems can drastically improve
the solution time for the next problem and thus for the whole sequence.
BMC is a natural candidate for incremental solving
as two BMC instances for bounds $k$ and $k+1$ are very similar. 
Strichman~\cite{Str01} and Whittemore et al.~\cite{WKS01} were among the 
first to consider incremental BMC\@.
Both papers presented frameworks for transforming a SAT problem to
the next in the sequence by adding and 
removing clauses from the current problem instance. 
E\'en and S\"orensson~\cite{ES03} consider incremental BMC combined with the
inductive scheme presented in~\cite{SSS00}.
Their approach is based on using the special syntactic structure of
the BMC encoding for invariants to forward all learned clauses,
and therefore they do not need to perform any potentially expensive
conflict analysis for learned clauses between two sequential problem instances.
%
%
Jin and Somenzi~\cite{DBLP:journals/entcs/JinS05} present efficient ways of filtering learned
clauses when creating the next problem instance. 
In~\cite{DBLP:conf/sat/BenedettiB04} a framework for incremental SAT solving based on incremental
compilation of the encoding to SAT is presented, however, their PLTL
encoding is based on the original and inefficient (for past formulas)
encoding of~\cite{BC03}.

The incremental encoding has been designed to allow easy separation of constraints
that remain active over all instances and constraints that should
be removed when the bound is increased.
In addition, we have tried to minimise the number of constraints that must be
removed in order to allow maximal learning in a solver independent fashion.
Both of these are achieved while maintaining the efficiency 
of the original encoding~\cite{LBHJ05}. 

There are a few considerations that need
to be taken into account for a good incremental encoding. First of all,
the encoding needs to be formulated so that it is easy to derive the case $k=i+1$
from $k=i$.
This is done by separating the encoding to a \emph{$k$-invariant} part
and a \emph{$k$-dependent} part. The information learned from the
$k$-invariant constraints can be reused when the bound is increased while
the information learned from the $k$-dependent constraints needs to be discarded.
Thus we try to minimise the use of $k$-dependent constraints in our
encoding. The so-called \emph{Base constraints} are also $k$-invariant, but they are
conditions that are constant for all values of $k$.

Keeping the number of $k$-dependent constraints small is achieved
largely by the introduction of \emph{proxy states}, which serve as
placeholders for the endpoint of a path. The disentanglement of the
constraints at index $k$ from the fixpoint encoding to the eventuality
encoding by introducing formula variables also for index $k+1$ can be
seen as a first step in that direction. This is the reason we chose
the eventuality encoding of Sect.~\ref{sec:pltl-eventuality} as the
base of our incremental encoding. Below only the differences needed to
obtain incrementality are given. All of the non-modified parts of the
encoding are $k$-invariant.

The loop constraints $\Trans{\mathit{LoopConstraints}}_k$ are modified to (changes are shown in blue boxes):
{\encsize
\begin{displaymath}
\begin{array}{l|rcl}
\hline
\mathrm{Base} 
& \lvar{0} & \Equiv & \bot 
\\[1ex]
&\InLoop{0} & \Leftrightarrow & \bot\\[1ex]
\hline
k\mathrm{-invariant} & \lvar{i} & \Rightarrow & (s_{i-1}=s_{\modification{E}{\scriptstyle}}) \\[1ex]
1\leq i\leq k&\InLoop{i} & \Leftrightarrow & \InLoop{i-1}\vee \lvar{i},\\[1ex]
&  \InLoop{i-1} & \Implies & \neg\lvar{i}\\[1ex]  
\hline
k\mathrm{-dependent} &\LoopExists & \Equiv & \InLoop{k}\\[1ex]
&\multicolumn{3}{c}{\hspace*{-1.5em}\modification{s_E \;\;\;\;=\;\;\;\; s_k}{\textstyle}} \\[1ex] 
\end{array}
\end{displaymath}
}\smallskip%

\noindent
Many $k$-dependent constraints of the non-incremental encoding of Sect.~\ref{sec:pltl-eventuality} have been
eliminated by introducing a new special system state $s_E$
with fresh (unconstrained) state variables acting as a proxy state for the endpoint $k$
of the path. In the $k$-dependent part the proxy state $s_E$ is constrained to be equivalent to $s_k$.
The constraint defining the variable $\LoopExists$ is $k$-dependent as it is defined in terms of 
$\InLoop{k}$.

The $\Trans{\mathit{LastStateFormula}}_k$ constraints
are modified to (changes are shown in blue boxes):
{\encsize
\[
\begin{array}{c@{\quad}|@{\quad}c@{\quad}|}
  & 0\leq d\leq \delta(\TSF) \\
\hline& \\[-2ex]
\mathrm{Base} & \neg\mathit{LoopExists}\Rightarrow\left(\Trans{\TSF}^d_{\modification{L}{\scriptstyle}}\Leftrightarrow\bot\right)\\[1ex]
\hline& \\[-2ex]
k\mathrm{-invariant}, 1\leq i \leq k & \lvar{i}\Implies\left(\Trans{\TSF}^d_{\modification{L}{\scriptstyle}}\Equiv\Trans{\TSF}^d_i\right) \\[1ex]
\hline &\\[-2ex]
k\mathrm{-dependent} & \modification{\Trans{\TSF}^d_E\Equiv\Trans{\TSF}^d_k}{\textstyle} \\[1ex]
 & \modification{\Trans{\TSF}^d_{k+1}\Equiv\Trans{\TSF}^{\mathit{min}(d+1,\delta(\TSF))}_L}{\textstyle}\\[1ex]
\end{array}
\]
}\smallskip%

\noindent
The proxy state $s_E$ has the corresponding new formula variables $\Trans{\TSF}^d_E$ which
have been introduced to make the encodings for the past formulas $k$-invariant.
For the future formulas another proxy state with index $L$ has been introduced.
This loop proxy state introduces new formula variables $\Trans{\TSF}^d_L$.
All the formulas at the proxy states are
bound to their corresponding states at the same time point, implementing jumping
from one unrolling to another as shown in Fig.~\ref{fig:counter3}.

We need to extend the first rule of the PLTL encoding also to indices $E$ and $L$,
for each subformula $\TSF \in \mathit{cl}(\TLF)$:
{
\begin{eqnarray*}
\encsize
&& \Trans{\TSF}_E^d = \Trans{\TSF}_E^{\delta(\TSF)}, \ \mathrm{when} \ d> \delta(\TSF)\mbox{; and}\\
&& \Trans{\TSF}_L^d = \Trans{\TSF}_L^{\delta(\TSF)}, \ \mathrm{when} \ d> \delta(\TSF).
\end{eqnarray*}
}\smallskip%

The auxiliary formula encoding is modified to (as before, changes are shown in blue boxes):
{\encsize
\[
\begin{array}{c@{\quad}|@{\quad}c@{\quad}|@{\quad}c|}
&\TSF &   \\[1ex]
\hline & &\\[-2ex]
\mathrm{Base} & \Until{\LSF}{\RSF} &\mathit{LoopExists}\Rightarrow\left(
\Trans{\Until{\LSF}{\RSF}}^{\delta(\TSF)}_{\modification{E}{\scriptstyle}}\Rightarrow\ATrans{\Finally{\RSF}}_{\modification{E}{\scriptstyle}}^{\delta(\RSF)}\right)\\[1ex] 
& \Release{\LSF}{\RSF} &\mathit{LoopExists}\Rightarrow\left(
\Trans{\Release{\LSF}{\RSF}}^{\delta(\TSF)}_{\modification{E}{\scriptstyle}}\Leftarrow\ATrans{\Globally{\RSF}}_{\modification{E}{\scriptstyle}}^{\delta(\RSF)}\right)\\[1ex]
& \Until{\LSF}{\RSF} & \ATrans{\Finally{\RSF}}_0^{\delta(\RSF)} \Leftrightarrow \bot\\[1ex] 
& \Release{\LSF}{\RSF} & \ATrans{\Globally{\RSF}}_0^{\delta(\RSF)} \Leftrightarrow \top \\[1ex]
\hline & &\\[-2ex]
k\mathrm{-invariant}
& \Until{\LSF}{\RSF} & \ATrans{\Finally{\RSF}}_i^{\delta(\RSF)}\Leftrightarrow 
\ATrans{\Finally{\RSF}}_{i-1}^{\delta(\RSF)}\vee \left(\InLoop{i}\wedge\Trans{\RSF}^{\delta(\RSF)}_i\right)\\[1ex]  
 1\leq i\leq k 
& \Release{\LSF}{\RSF} &  \ATrans{\Globally{\RSF}}_i^{\delta(\RSF)} \Leftrightarrow 
\ATrans{\Globally{\RSF}}_{i-1}^{\delta(\RSF)}\wedge \left(\neg\InLoop{i}\vee\Trans{\RSF}^{\delta(\RSF)}_i\right)\\[1ex]
\hline & &\\[-2ex]
k\mathrm{-dependent} 
& \Until{\LSF}{\RSF} & \modification{\ATrans{\Finally{\RSF}}_E^{\delta(\RSF)} \Leftrightarrow \ATrans{\Finally{\RSF}}_{k}^{\delta(\RSF)}}{\textstyle}\\[1ex] 
& \Release{\LSF}{\RSF} & \modification{\ATrans{\Globally{\RSF}}_E^{\delta(\RSF)} \Leftrightarrow\ATrans{\Globally{\RSF}}_{k}^{\delta(\RSF)}}{\textstyle}\\[1ex]
\end{array}
\]
}\smallskip%

\noindent
Basically all references to the index $k$ have been removed in the $k$-invariant parts by
references to $E$. The new $k$-dependent constraints constrain the auxiliary encodings
at $E$ to get their values from the state at the current bound $k$.

We also have to modify the encoding of past formulas slightly, as they explicitly
mention the bound $k$ used. The change is to replace the index $k$ with the proxy
end index $E$, and after this the encoding becomes $k$-invariant.
The case $d=0$ does not have to be changed and is therefore omitted. The indexing changes
required are again shown in blue boxes. The table below includes also the 
(optional) stabilisation forcing constraints.
{\tighterencsize
\[
\begin{array}{c@{\quad}|@{\quad}c@{\quad}}
\TSF & 1 \leq i \leq k, 1\leq d \leq \delta(\TSF)\\
\hline&\\[-2ex]
\Since{\LSF}{\RSF} & 
\Trans{\Since{\LSF}{\RSF}}^d_{i} \Equiv \Trans{\RSF}^d_{i}\vee\left(\Trans{\LSF}^d_i\wedge
\left(\left(l_i\wedge\Trans{\TSF}_{\modification{E}{\scriptstyle}}^{d-1}\right)\vee\left(\neg l_i\wedge\Trans{\TSF}^{d}_{i-1}\right)\right)\right) 
\\[1ex]
\Trigger{\LSF}{\RSF} & 
\Trans{\Trigger{\LSF}{\RSF}}^d_{i} \Equiv \Trans{\RSF}^d_{i}\wedge\left(\Trans{\LSF}^d_i\vee
\left(\left(l_i\wedge\Trans{\TSF}_{\modification{E}{\scriptstyle}}^{d-1}\right)\vee\left(\neg l_i\wedge\Trans{\TSF}^{d}_{i-1}\right)\right)\right)
\\[1ex]
\Prev{\LSF} & 
\Trans{\Prev{\LSF}}^d_{i} \Equiv \left(\mathit{l}_i\wedge\Trans{\LSF}^{d-1}_{\modification{E}{\scriptstyle}}\right)\vee
\left(\neg\mathit{l}_i\wedge\Trans{\LSF}^{d}_{i-1}\right)\\[1ex]
\PrevZ{\LSF} & 
\Trans{\PrevZ{\LSF}}^d_{i} \Equiv \left(\mathit{l}_i\wedge\Trans{\LSF}^{d-1}_{\modification{E}{\scriptstyle}}\right)\vee
\left(\neg\mathit{l}_i\wedge\Trans{\LSF}^{d}_{i-1}\right)\\[1ex]
\hline&\\[-2ex]
\Since{\LSF}{\RSF} &  \Trans{\Since{\LSF}{\RSF}}^{\delta(\TSF)}_{i}\Equiv
\Trans{\RSF}^{\delta(\TSF)}_{i}\vee\left(\Trans{\LSF}^{\delta(\TSF)}_i\wedge
\left(\left(l_i\wedge\Trans{\TSF}_{\modification{E}{\scriptstyle}}^{\delta(\TSF)}\right)\vee\left(\neg l_i\wedge\Trans{\TSF}^{\delta(\TSF)}_{i-1}\right)\right)\right) 
\\[1ex]
\Trigger{\LSF}{\RSF} & \Trans{\Trigger{\LSF}{\RSF}}^{\delta(\TSF)}_{i} \Equiv
\Trans{\RSF}^{\delta(\TSF)}_{i}\wedge\left(\Trans{\LSF}^{\delta(\TSF)}_i\vee
\left(\left(l_i\wedge\Trans{\TSF}_{\modification{E}{\scriptstyle}}^{\delta(\TSF)}\right)\vee\left(\neg l_i\wedge\Trans{\TSF}^{\delta(\TSF)}_{i-1}\right)\right)\right)
\\[1ex]
\Prev{\LSF} & \Trans{\Prev{\LSF}}^{\delta(\TSF)}_{i}\Equiv
\left(\mathit{l}_i\wedge\Trans{\LSF}^{\delta(\TSF)}_{\modification{E}{\scriptstyle}}\right)\vee
\left(\neg\mathit{l}_i\wedge\Trans{\LSF}^{\delta(\TSF)}_{i-1}\right)\\[1ex]
\PrevZ{\LSF} & \Trans{\PrevZ{\LSF}}^{\delta(\TSF)}_{i} \Equiv
\left(\mathit{l}_i\wedge\Trans{\LSF}^{\delta(\TSF)}_{\modification{E}{\scriptstyle}}\right)\vee
\left(\neg\mathit{l}_i\wedge\Trans{\LSF}^{\delta(\TSF)}_{i-1}\right)\\[1ex]
\end{array}
\]
}\smallskip%

Combining the tables above we get the full incremental PLTL encoding $\Trans{\mathit{IncPLTL}}_k$ 
for $\TLF$. Given a Kripke structure $M$, a PLTL formula $\TLF$, and a bound $k$, the
\emph{incremental PLTL eventuality encoding} as a propositional formula is given by: 
{\encsize
\[
\Trans{M,\TLF,k}=\Trans{M}_k\wedge\Trans{\mathit{LoopConstraints}}_k\wedge\Trans{\mathit{LastStateFormula}}_k\wedge\Trans{\mathit{IncPLTL}}_k\wedge\Trans{\TLF}^0_0.
\]
}\smallskip%

The correctness of our encoding is established by the following theorem. 
\begin{theorem}\label{thm:inc-pltl-eventuality}
Given a Kripke structure $M$ and a PLTL formula $\TLF$, 
$M$ has an initialised path $\pi$ such that $\pi\models\TLF$ iff there exists a $k\in\mathbb{N}$ such
that the incremental PLTL eventuality encoding $\Trans{M,\TLF,k}$ is satisfiable. In particular, if
$\pi\models_k\TLF$ then the incremental PLTL eventuality encoding $\Trans{M,\TLF,k}$ is satisfiable.
\end{theorem}
\proof
Note, that the fact that the encoding is used incrementally does not
influence correctness of the claim. Hence, we show that the
incremental PLTL eventuality encoding is satisfiable iff the
(non-incremental) PLTL eventuality encoding presented in Sect.~\ref{sec:pltl-bmc}
is satisfiable. Correctness then follows from
Thm.~\ref{thm:pltl-eventuality}.

It's not hard to verify that, essentially by applying substitutions to
the proxy variables, the incremental encoding can be transformed into
the non-incremental version plus
\iftrue
the following set of constraints
\[
\begin{array}{l}
s_E = s_k\\[0.5ex]
\neg \LoopExists{} \Rightarrow (\Trans{\TSF}^0_L \Leftrightarrow \bot)\\[0.5ex]
\forall 1 \le i \le k : l_i \Rightarrow (\Trans{\TSF}^0_L \Leftrightarrow \Trans{\TSF}^0_i)\\[0.5ex]
\forall 0 \le d \le \delta(\TSF) : \Trans{\TSF}^d_E \Leftrightarrow \Trans{\TSF}^d_k\\[0.5ex]
\forall 0 \le d \le \delta(\TSF) : \Trans{\TSF}^d_{k+1} \Leftrightarrow \Trans{\TSF}^{\min(d+1,\delta(\TSF))}_L\\[0.5ex]
\Trans{\TSF}^d_E \Leftrightarrow \Trans{\TSF}^{\delta(\TSF)}_E \mbox{ if } d > \delta(\TSF)\\[0.5ex]
\Trans{\TSF}^d_L \Leftrightarrow \Trans{\TSF}^{\delta(\TSF)}_L \mbox{ if } d > \delta(\TSF)\\[0.5ex]
\ATrans{\Finally{\RSF}}^{\delta(\RSF)}_E \Leftrightarrow \ATrans{\Finally{\RSF}}^{\delta(\RSF)}_k\\[0.5ex]
\ATrans{\Globally{\RSF}}^{\delta(\RSF)}_E \Leftrightarrow \ATrans{\Globally{\RSF}}^{\delta(\RSF)}_k\\[0.5ex]
\end{array}
\]
without changing the set of satisfying truth assignments. 
It is easy to see that with $l_i \Rightarrow
(\Trans{\TSF}^{\delta(\TSF)-1}_{k+1} \Leftrightarrow
\Trans{\TSF}^{\delta(\TSF)}_i \Leftrightarrow
\Trans{\TSF}^{\delta(\TSF)}_{k+1})$ the set of
constraints is a conflict-free assignment of the proxy variables.
\else
a set of constraints without changing the set of satisfying assignments. That set of
constraints turns out to be a conflict-free assignment of the proxy
variables.
\fi
\qed

The incrementality of the encoding works as follows. The encoding
$\Trans{M,\TLF,k+1}$ for bound $k+1$ is obtained from the encoding $\Trans{M,\TLF,k}$ for bound $k$.
First, all the $k$-dependent rules, and everything learned from them by the SAT solver have to be dropped.
After this the encoding must be extended by all the constraints needed for encoding the new time step $k+1$.

We have taken care to keep most of the encoding rules $k$-independent, and to make
all of the $k$-dependent constraints as simple as possible (they are all just equivalences
between two variables) in order to make the size of the $k$-dependent part as small as
possible. This was made in order to make the overhead to a non-incremental version
as small as possible. The experimental results of Sect.~\ref{sec:experiments} and~\cite{HelJunLat:CAV05}
confirm that the incremental approach does lead to performance benefits.

\section{Completeness: Proving Properties}\label{sec:completeness}

In its basic form bounded model checking only finds counterexamples
and does not prove systems to be correct. To prove that a system
has no counterexamples for a given property with BMC, we must prove that 
no counterexample can be longer than a certain bound, the \emph{completeness
threshold}, and prove that there are no shorter counterexamples.
The obvious upper bound for the completeness threshold is exponential
in the number of state bits in the system. We could thus obtain a complete BMC
procedure by always doing BMC until reaching this upper bound, but clearly such an approach
is unacceptable and we actually want a procedure that will in many
practical cases terminate with a much smaller bound.
There are several approaches to making BMC complete in a more practical sense,
i.e., which are able to prove properties by more precisely approximating
the required completeness threshold.

A complete method for proving invariant
properties is $k$-induction originally developed by Sheeran et al.~\cite{SSS00}.
They give several different variants for proving invariant properties.
The variant closest to our approach is the following:
If the invariant holds in every state in each initialised
path of length~$k$, and there is no initialised loop-free path, 
which does not visit an initial state, of length $k+1$;
then we can conclude that the invariant holds for the system.
The longest initialised loop-free path in the state graph is called the \emph{recurrence
diameter}, which can be used as an upper bound for the completeness
threshold when proving invariants. Clearly the number of reachable states of the system
gives a worst case upper bound for the recurrence diameter.
For a bound $k$ a straightforward encoding of
this loop-free path predicate is of the size $O(k^2)$. 
Kroening and Strichman~\cite{KS03} show that the size of this 
loop-free predicate can be optimised to $O(k\log^2 k)$ using sorting networks. They also suggest ways to
leave out state bits from the loop-free predicate to improve efficiency while maintaining 
completeness. The benefits of having a smaller predicate are two-fold: a smaller predicate
is easier to manage for the SAT solver and with fewer state variables we can prove
properties at shallower depths because the system loops earlier.

It is now easy to see that by combining the B{\"u}chi automata-based BMC encoding of Sect.~\ref{sec:pltl-buchi} for PLTL
and the liveness to safety reduction of Sect.~\ref{sec:l2s} with $k$-induction
we get a complete BMC method. The method can also be made incremental as shown in~\cite{ES03}.

In this section we show a more refined approach
to completeness based on the incremental
BMC encoding presented in Sect.~\ref{sec:incbmc}.
The approach has been first published in~\cite{HelJunLat:CAV05}.
It is based on similar ideas as~\cite{ES03} but due to the increased flexibility of the BMC encoding,
like the ability to refer to arbitrary states in the run, it is able to avoid the doubling of
the number of state bits as required by the liveness-to-safety transformation.
This doubling would increase the size of the already large loop-free predicate
needed by the approach.
Our method also works in the forward direction only (we always have the initial state predicate present),
unlike some other approaches to obtaining completeness such as~\cite{ES03,AS06}.

   
Practical experience seems to indicate that already model checking general 
safety properties using induction is challenging~\cite{DBLP:journals/entcs/ArmoniFFHPV05}.
Simply synchronising a finite state automaton (FSA) representing a safety 
property with the system to model check safety properties from does not scale well, and
forces model checkers to go deeper than the current capacity of SAT solvers. 
One reason is the non-determinism in the FSA representing the property~\cite{DBLP:journals/entcs/ArmoniFFHPV05}.
It seems that specifications using deterministic FSAs can be treated more efficiently~\cite{AEFKV05,Lat03a}.
Our BMC encodings follow this line of reasoning by trying to be as deterministic as possible.

Two papers that consider strengthening of induction without always doing deeper BMC
queries, which is expensive, are~\cite{MRS03,DBLP:journals/entcs/ArmoniFFHPV05}. In~\cite{MRS03} the inductive method 
of~\cite{SSS00} is generalised to an induction scheme based on simulations. Inductive 
invariants are automatically strengthened from failed induction proofs using a procedure 
based on existential quantification. Since existential quantification is resource intensive,
a method for quantifying on demand is developed. Another approach is presented in~\cite{DBLP:journals/entcs/ArmoniFFHPV05}.
They develop a methodology for flexible manual strengthening of induction. 
The key idea is to make the induction scheme part of the specification to allow a high 
degree of control of the induction process. Counterexamples produced by the model checker 
aid the designer in choosing new invariants.

Finding a completeness threshold for general LTL properties has proven fairly challenging.
Clarke et al.~\cite{ClarkeKroeningOuaknineStrichman05} show how the completeness threshold can be computed
for general LTL properties by computing the recurrence diameter of the product
of the system and a B\"uchi automaton representing the negation of the property. 
Awedh and Somenzi~\cite{AS06} apply the
same approach, but they use a refined method for calculating
the completeness threshold. Both papers have the problem that they use an 
explicit representation of B\"uchi automata in their implementations.  Thus, they
potentially use an exponential number of state bits in the size of the formula
to represent the B\"uchi automaton. 
Additionally, our encoding is able to find counterexamples for
full PLTL with smaller bounds than previous methods for LTL~\cite{CKOS04,AS04},
as these papers employ a method for translating generalised B\"uchi automata
to standard (non-generalised) B\"uchi automata in a way 
(called the counter method in~\cite{AS06}) which does not
preserve the minimal length of counterexamples.
Recently, the authors of~\cite{AS04} have refined their approach in~\cite{AS06}
to also in effect use generalised B{\"u}chi automata directly (called the flag method in~\cite{AS06}).

A different approach to proving completeness is taken by
McMillan~\cite{McM03}. He uses interpolants derived from unsatisfiability
proofs of BMC counterexample queries to over-approximate symbolic 
reachability. The deeper the BMC query is, the more exact the over-approximation 
is. The method is complete and can be extended to LTL model
checking through the liveness-to-safety transformation discussed in Sect.~\ref{sec:l2s}.
Although the method can in many cases converge more quickly 
than the recurrence diameter, which is the relevant completeness threshold for most other
methods, the unsatisfiability proofs can be of exponential size and cause a blow-up.

\subsubsection*{Suggested BMC Procedure for Completeness}
The incremental encoding of Sect.~\ref{sec:incbmc} can easily be extended to
also prove properties. 
The basic ideas used are similar to the variant of $k$-induction of~\cite{SSS00}
discussed above. However, the approach of~\cite{SSS00} is restricted to
proving invariants, while our approach can handle proving of all PLTL properties.

The procedure starts with bound $k=0$.
First we create a \emph{completeness formula},
denoted by $\CTrans{M,\TLF,k}$, which is satisfied
only for the initialised finite paths of length $k$
which one might be able to extend to a bounded witness
of formula $\TLF$ (of length $k$ or longer).
The completeness formula $\CTrans{M,\TLF,k}$
we use consists of exactly the incremental
translation $\Trans{M,\TLF,k}$ of Sect.~\ref{sec:incbmc}
with all $k$-dependent constraints removed.
Because these constraints are a subset of the constraints
$\Trans{M,\TLF,k'}$ for every $k' \geq k$, if
$\CTrans{M,\TLF,k}$ is unsatisfiable, so will also
$\Trans{M,\TLF,k'}$ be.

Now, similarly to the $k$-induction method, we want to conjunct
the completeness formula $\CTrans{M,\TLF,k}$ with a \emph{simple path}
formula which is satisfied for only initialised loop-free paths.
This formula is needed in order to guarantee termination of the procedure.
However, we use a certain product automaton instead of the Kripke
structure itself.
The states of this product automaton 
at time point $i$ consist of tuples of:
(a) system state $s_i$,
(b) a bit vector of values of all formula variables $\Trans{\TSF}^d_i$, denoted $\Trans{\ASF}_i$,
(c) a bit vector of values of all auxiliary formula variables $\ATrans{\TSF}^d_i$, denoted $\ATrans{\ASF}_i$, and
(d) value of the $\InLoop{i}$ predicate.
As an optimisation we disregard any differences in unrollings
$d > 0$ between two indices where $InLoop_i$ is false, as these bits
are not constrained by the top-level formula, and thus are always
satisfiable (these are the light nodes of Fig.~\ref{fig:counter3}).
To do so, we use $\Trans{\ASF}^0_i$ to denote
$\Trans{\ASF}_i$ restricted to the bits $\Trans{\TSF}^0_i$.
The simple path formula we use is the following:
{\small
\begin{displaymath}\label{eq:simplepath}
\begin{array}{lll}
\Trans{SimplePath}_k \Leftrightarrow \bigwedge_{0\leq i<j\leq k} & \left(s_i \neq s_j\vee 
\InLoop{i}\neq\InLoop{j}\vee
\Trans{\ASF}^0_i\neq\Trans{\ASF}^0_j\vee\right.\\
& \left(\InLoop{i}\wedge\InLoop{j}\wedge
\left.\left(\Trans{\ASF}_i\neq\Trans{\ASF}_j 
\vee\ATrans{\ASF}_i\neq\ATrans{\ASF}_j\right)\right)\right).
\end{array}
\end{displaymath}
}
If at bound $k$
the conjunction of the completeness $\CTrans{M,\TLF,k}$ and the simple path formula is
unsatisfiable the model checked formula $\neg \TLF$ holds in the system
and the procedure can be terminated.
Otherwise the \emph{witness formula} $\Trans{M,\TLF,k}$ is created
(and optionally conjuncted with the simple path formula) and the result is
satisfiable for bounded witnesses of length $k$ to the formula $\TLF$ (see Thm.~\ref{thm:inc-pltl-eventuality}).
If the witness formula is satisfiable, the model checked formula $\neg \TLF$ does not hold,
and the procedure can terminate.
Otherwise, the procedure is repeated after incrementing $k$ by one.

The $\Trans{SimplePath}_k$ constraint above is obviously quadratic in $k$.
We could use the standard simple path constraint used in other works employing $k$-induction
by slight modifications to the encoding, e.g., forcing the light nodes of Fig.~\ref{fig:counter3}
to $\bot$ in the encoding. This would enable, e.g., using the optimisations of~\cite{KS03}.

The procedure above has been designed to be easily implemented using one incremental
SAT solver only, and this is what our implementation does. The only place
where constraints have to be dropped is moving from a witness formula $\Trans{M,\TLF,k}$ for bound $k$
to the completeness formula $\CTrans{M,\TLF,k+1}$ for bound $k+1$, at which point all $k$-dependent constraints
of $\Trans{M,\TLF,k}$ and everything learned from them by the SAT solver have to be dropped. We use
implementation techniques similar to those of~\cite{ES03} to implement this.

We have the following result:
\begin{theorem}\label{thm:completeness}
Given a Kripke structure $M$ and a PLTL formula $\TLF$,
$M\models\TLF$ iff
for some $k \geq 0$:
$\CTrans{M,\neg\TLF,k}\wedge\Trans{SimplePath}_k$ is unsatisfiable
and $\Trans{M,\neg\TLF,i}\wedge\Trans{SimplePath}_i$ is unsatisfiable
for all $0\leq i < k$.
\end{theorem}

The proof requires the following Lemma:

\begin{lemma}\label{thm:simplepath}
Given a Kripke structure $M$ and a PLTL formula $\TLF$, if
$\Trans{M,\neg\TLF,k}$ is satisfiable for some $k$, there is
$\tilde{k} \le k$ such that $\Trans{M,\neg\TLF,\tilde{k}} \wedge
\Trans{SimplePath}_{\tilde{k}}$ is satisfiable.
\end{lemma}
\begin{proof}
We are given that $\Trans{M,\neg\TLF,k}$ is satisfiable.
If $\Trans{M,\neg\TLF,k} \wedge \Trans{SimplePath}_k$ is already
satisfiable for $k$ we are done.
Otherwise, the proof strategy is to show that for some $\tilde{k} < k$ the encoding
$\Trans{M,\neg\TLF,\tilde{k}}$ is satisfiable, and repeating the process.
By the finiteness of $k$,
this process can only be repeated a limited number of times.
The base case is proved by the fact that
$\Trans{SimplePath}_0$ is an empty set of constraints, thus proving termination
at some $\tilde{k}$ where
$\Trans{M,\neg\TLF,\tilde{k}} \wedge \Trans{SimplePath}_{\tilde{k}}$ is satisfiable.

Consider the induction step where $\Trans{M,\neg\TLF,k}$ is satisfiable
but $\Trans{SimplePath}_k$ is not satisfiable. Hence, there are $0 \le i
< j \le k$ such that $s_i = s_j$, $\InLoop{i} \Leftrightarrow
\InLoop{j}$ and either (a): $\InLoop{i} \wedge \InLoop{j} \wedge \Trans{\ASF}_i = \Trans{\ASF}_j \wedge \ATrans{\ASF}_i = \ATrans{\ASF}_j$,
or (b): $\neg \InLoop{i} \wedge \neg \InLoop{j} \wedge \Trans{\ASF}^0_i = \Trans{\ASF}^0_j$.
In the following we show that also
$\Trans{M,\neg\TLF,\tilde{k}}$ is satisfiable for
$\tilde{k} = k - j + i$, i.e., $\tilde{k} < k$.
Intuitively, we construct a satisfying truth
assignment by
``cutting out'' the part of the encoding between indices $i+1$ and $j$ (both inclusive)
of the satisfying truth assignment of $\Trans{M,\neg\TLF,k}$
and ``pasting together'' the remaining parts by reducing all variable
indices to the right of the cut point by $j-i$,
obtaining a satisfying truth assignment for $\Trans{M,\neg\TLF,\tilde{k}}$.

An exception to the above rule are the formula variables
with unrolling index $d>0$ such that $\InLoop{i}$ is false,
i.e., the light nodes of Fig.~\ref{fig:counter3}.
By similar reasoning as used in the proof of Thm.~\ref{thm:pltl-eventuality}
their constraints can never lead to the unsatisfiability of the encoding, and
they can therefore be ignored in constructing the (now actually partial)
truth assignment below. In other words a satisfying truth assignment
for them always exists, and will be fully determined by the partial truth
assignment for all the other variables to be constructed below.

Note, that in the case a loop exists: either $i < j < l$
or $l \le i < j$ as $\InLoop{i} \Leftrightarrow \InLoop{j}$.
Hence, the loop start at index $l$ is never cut out.
For ease of notation we define a function
$f$ mapping indices from the new to the old assignment:
\[
f(n) = \mbox{if } n \le i \mbox{ then } n \mbox{ else } n + j - i
\]
With that we define:
\[
\begin{array}{lrcl}
\forall 0 \le n \le \tilde{k} : & \tilde{s}_n & = & s_{f(n)}\\
\forall 0 \le n \le \tilde{k} : & \tilde{l}_n & \Leftrightarrow & l_{f(n)}\\
\forall 0 \le n \le \tilde{k} : & \widetilde{\InLoop{}}_n & \Leftrightarrow & \InLoop{f(n)}\\
& \widetilde{\LoopExists} & \Leftrightarrow & \LoopExists\\
& \tilde{s}_E & = & s_E\\
\end{array}
\]
%
%
We start with the model constraints.
Let $\pi$ be an initialised path in $M$ induced by a satisfying
truth assignment of $\Trans{M,\neg\TLF,k}$. Because $s_i = s_j$ the path
$\tilde{\pi}$ constructed from $\pi$ by cutting out indices $i+1$ and $j$ (both inclusive)
is still an initialised path of $M$. Hence, the model constraints are satisfied.

For the loop constraints note first that $\tilde{s}_{\tilde{k}} = s_k$
both in the case $j < k$ and in the case $j=k$. Furthermore, a loop
point is never cut out. Hence, if some $l_l$ was true in the original
assignment, there is $\tilde{l}$ such that $l_{\tilde{l}}$ is true in
the new assignment. In this case we also have $\tilde{s}_{\tilde{l}-1}
= s_{l-1}$. Thus by simple case analysis of the loop constraints
we get that they are satisfiable also in $\Trans{M,\neg\TLF,\tilde{k}}$.

What remains to be done is to prove that the formula encoding
$\widetilde{\Trans{\TSF}^0_0}$ is still satisfiable in $\Trans{M,\neg\TLF,\tilde{k}}$.
We do this by analysing the structure of the encoding rules for temporal
formulas. 
Below, in each case we consider the mapped pairs of indices $i,d$ such that
$d=0$ or $\InLoop{i}$ is true. For simplicity all indices below refer
to the original encoding $\Trans{M,\neg\TLF,k}$.

For all future formulas in $\Trans{M,\neg\TLF,k}$ at
index $i$ the references to formula values at $i+1$ have been replaced
in the encoding $\Trans{M,\neg\TLF,\tilde{k}}$  
with references to formula variables at index $j+1$ (note that potentially $j+1 = k+1$). Now because
both the formula values at $i$
and $j$ are identical and the future formula constraints at $i$ and $j$
are identical modulo index changes,
the constraints at $i$ will still be satisfiable
with the same truth assignment when all references to $i+1$ have been replaced
with references to $j+1$.

For all past formulas in $\Trans{M,\neg\TLF,k}$ at index $j+1$
(at the loop index $l$, when $j = k$)
the references to formula values at $j$ have been replaced
in the encoding $\Trans{M,\neg\TLF,\tilde{k}}$
with references to formula variables at index $i$. 
Now because the formula values at $i$
and $j$ are identical, 
the constraints at $j+1$
(at the loop index $l$, when $j = k$)
will still be satisfiable
with the same truth assignment when all references to $j$ have been replaced
with references to $i$.

For the auxiliary encoding all the constraints are also satisfied by replacing
all references to index $j$ in $\Trans{M,\neg\TLF,k}$ with references
to index $i$ in $\Trans{M,\neg\TLF,\tilde{k}}$. This
is the case because $\ATrans{\ASF}_i = \ATrans{\ASF}_j$ holds in case (a)
due to the simple path constraint $\ATrans{\ASF}_i = \ATrans{\ASF}_j$,
and in case (b) because the encoding for auxiliary variables keeps
them constant for all indices $0\leq i < j < l$.

Now combining all the cases above we were able to ``cut out'' a part of the
encoding $\Trans{M,\neg\TLF,k}$ while still retaining its satisfiability.
Thus $\Trans{M,\neg\TLF,\tilde{k}}$ will also be satisfiable.
\end{proof}

We can now continue with the proof of Thm.~\ref{thm:completeness}:

\begin{proof}
``$\Rightarrow$'' We only deal with finite models $M$ and finite
formulas $\neg\TLF$. $\Trans{SimplePath}_k$ must therefore become and
remain unsatisfiable from some $k$ onward. From correctness of the
incremental PLTL eventuality encoding (Thm.~\ref{thm:inc-pltl-eventuality}) we have that $\Trans{M,\neg\TLF,i}$ is
unsatisfiable for all $i \ge 0$ if $M \models \TLF$.

\begin{sloppypar}
``$\Leftarrow$'' Assume that
$\CTrans{M,\neg\TLF,k}\wedge\Trans{SimplePath}_k$ is unsatisfiable for
some $k \ge 0$ and $\Trans{M,\neg\TLF,i}\wedge\Trans{SimplePath}_i$ is
unsatisfiable for all $0\leq i < k$. As noted above, unsatisfiability
of $\CTrans{M,\neg\TLF,k}$ implies unsatisfiability of
$\Trans{M,\neg\TLF,k'}$ for all $k' \ge k$. Similarly, if
$\Trans{SimplePath}_k$ is unsatisfiable, so is
$\Trans{SimplePath}_{k'}$ for all $k' \ge k$. Hence, we have that
$\Trans{M,\neg\TLF,i} \wedge \Trans{SimplePath}_i$ is unsatisfiable
for all $i \ge 0$. Using Thm.~\ref{thm:inc-pltl-eventuality} together
with Lemma~\ref{thm:simplepath} in the reverse
direction we can conclude $M \models \TLF$.
\end{sloppypar}
\end{proof}

We could also increase the bound $k$ by more than one at a time if the
witness formula is not conjuncted with the simple path formula.
The proof requires the fact that if $\Trans{M,\neg\psi,k}$ is
satisfiable for some $k$, it is satisfiable for all $k' \ge k$.

\begin{lemma}\label{thm:satk-implies-satlargerthank}
Given a Kripke structure $M$ and a PLTL formula $\TLF$, if
$\Trans{M,\neg\TLF,k}$ is satisfiable for some $k$, then
$\Trans{M,\neg\TLF,k'}$ is satisfiable for all $k' \ge k$.
\end{lemma}

\begin{proof}
Assume $\pi = s_0 \ldots s_k$ is a bounded witness for $\neg\TLF$. We
show below that $\pi$ can be extended by one state so that the
result is again a bounded witness for $\neg\TLF$. By 
Thm.~\ref{thm:inc-pltl-eventuality}, $\Trans{M,\neg\TLF,k+1}$ is then also
satisfiable. Repeated application gives satisfiability of
$\Trans{M,\neg\TLF,k'}$ for any $k' \ge k$.

Consider the no-loop case first. By definition of $\modelsnl$,
$\pi$ extended with an arbitrary successor of $s_k$, $s_{k+1}$, is
also a bounded no-loop witness for $\neg\TLF$. If $\pi$ is a
$(k,l)$-loop, we rewrite $\pi$ into a $(k+1,l+1)$-loop by delaying the
loop start by one state: $\pi' = s_0 \ldots s_l s_{l+1} \ldots s_k
s_{k+1} = s_l$. Clearly, $\pi'$ interpreted as $(k+1,l+1)$-loop
represents the same infinite path as $\pi$ interpreted as $(k,l)$-loop
and, hence, also satisfies $\neg \TLF$.
\end{proof}

\begin{theorem}\label{thm:completeness-incmore}
Given a Kripke structure $M$ and a PLTL formula $\TLF$,
$M\models\TLF$ iff
for some $k \geq 0$:
$\CTrans{M,\neg\TLF,k}\wedge\Trans{SimplePath}_k$ is
unsatisfiable and either $k = 0$ or $\Trans{M,\neg\TLF,k-1}$
is unsatisfiable.
\end{theorem}
\begin{proof}
The ``$\Rightarrow$'' direction is exactly as in the proof of Thm.~\ref{thm:completeness}.
For ``$\Leftarrow$'' assume that for some $k
\geq 0$ we have that $\CTrans{M,\neg\TLF,k}\wedge\Trans{SimplePath}_k$ is
unsatisfiable and either $k=0$ or $\Trans{M,\neg\TLF,k-1}$ is
unsatisfiable. In the case $k=0$ the result follows directly from Thm.~\ref{thm:completeness}.
Now consider the case $k>0$.
By Lemma~\ref{thm:satk-implies-satlargerthank}, we have that
$\Trans{M,\neg\TLF,k'}$ is unsatisfiable for all $0 \leq k' < k$.
Therefore also obviously 
$\Trans{M,\neg\TLF,k'} \wedge\Trans{SimplePath}_{k'}$ is unsatisfiable for all $0 \leq k' < k$
and the result follows from Thm.~\ref{thm:completeness}.
\end{proof}

\section{Experiments and Comparisons}\label{sec:experiments}

In this section we experimentally evaluate and compare the approaches
presented in this paper. The benchmarks, implementations, and scripts
are available at
\begin{center}
\url{http://www.tcs.hut.fi/Software/benchmarks/LMCS-2006}
\end{center}

\subsection{Benchmark Instances}

We mostly use examples of nontrivial complexity. The majority are taken
from the NuSMV distribution~\cite{NuSMV}, one is from the examples of
the Rebeca tool \cite{SirjaniMovagharShalideBoer04}, and two are from
previous work of the authors~\cite{SchuppanBiere03,LBHJ05}.
Table~\ref{tab:modeldescriptions} provides a brief description of the
models. For ``1394'' and ``dme'' we use instances of different sizes
(indicated by the numerical parameters). For ``1394'' a buggy variant
is used as well (denoted ``1394b'').

\def\tabmodeldescriptionline#1#2#3#4{\begin{minipage}[t]{0.17\linewidth}{#1}\end{minipage}&\begin{minipage}[t]{0.07\linewidth}{\hfill#2}\end{minipage}&\begin{minipage}[t]{0.57\linewidth}{#3}\end{minipage}&\begin{minipage}[t]{0.095\linewidth}{#4}\end{minipage}}
\begin{table}
\begin{center}
{\footnotesize
\begin{tabular}{|l|r|l|l|}
\hline
\tabmodeldescriptionline{model}{state\-bits}{description}{source}\\[2.5ex]
\hline
\tabmodeldescriptionline{1394\{b\}-[345]-[23]}{97--197}{IEEE 1394 FireWire tree identify protocol with 3--5 nodes and 2 or 3 ports per node}{\cite{SchuppanBiere03}}\\[3.25ex]
\tabmodeldescriptionline{abp4}{30}{alternating bit protocol for 4 bits}{\cite{NuSMV}}\\[0.5ex]
\tabmodeldescriptionline{brp}{45}{bounded retransmission protocol}{\cite{NuSMV}}\\[0.5ex]
\tabmodeldescriptionline{counter}{3}{3-bit counter}{\cite{NuSMV}}\\[0.5ex]
\tabmodeldescriptionline{csmacd}{126}{MAC sublayer of CSMA/CD protocol}{\cite{SirjaniMovagharShalideBoer04}}\\[0.5ex]
\tabmodeldescriptionline{dme[35]}{54, 90}{asynchronous distributed mutual exclusion circuit with 3 or 5 nodes}{\cite{NuSMV}}\\[0.5ex]
\tabmodeldescriptionline{mutex}{5}{mutual exclusion with 2 participants}{\cite{NuSMV}}\\[0.5ex]
\tabmodeldescriptionline{pci}{64}{PCI Bus protocol}{\cite{NuSMV}}\\[0.5ex]
\tabmodeldescriptionline{prod-cons}{26}{producer consumer}{\cite{NuSMV}}\\[0.5ex]
\tabmodeldescriptionline{production-cell}{54}{production cell control model}{\cite{NuSMV}}\\[0.5ex]
\tabmodeldescriptionline{bc57-sensors}{78}{reactor system model}{\cite{NuSMV}}\\[0.5ex]
\tabmodeldescriptionline{ring}{3}{3 inverters forming a cycle}{\cite{NuSMV}}\\[0.5ex]
\tabmodeldescriptionline{short}{2}{simple request handler}{\cite{NuSMV}}\\[0.5ex]
\tabmodeldescriptionline{srg5}{8}{5 bit shift register}{\cite{LBHJ05}}\\
\hline
\end{tabular}
}
\end{center}
\caption{\label{tab:modeldescriptions}Models used in the experiments}
\end{table}

Table~\ref{tab:propertytemplates} gives templates of the properties
used. The first column states the name of the model. Columns 2--4
indicate names and truth of the properties. To save space we combine a
property and its negated version in a single line. The negation of
property ``p'' is later referred to as ``$\neg$p''. Truth is indicated
by ``t'' for true, ``f'' for false, ``?'' for unknown (if none of our
approaches terminated successfully), and ``--'' for not
used. Sometimes we make the resulting witnesses more interesting by
requiring that the request of a request-response property holds
infinitely often (marked ``nv''). We also prefix a property with
``$\Finally{}$'' to turn a safety property into a liveness
property. For ``1394'' the first entry in column 3 refers to the
correct, the second to the buggy version. The last two columns give
past operator depth and the template of the property.

\def\TempOp#1{{\mathrm{\bf#1}}}
\def\TempConst#1{{\mathrm{#1}}}
\def\X{\TempOp{X}}
\def\U{\mathrel{\TempOp{U}}}
\def\R{\mathrel{\TempOp{R}}}
\def\F{\TempOp{F}}
\def\G{\TempOp{G}}
\def\Y{\TempOp{Y}}
\def\Z{\TempOp{Z}}
\def\S{\mathrel{\TempOp{S}}}
\def\T{\mathrel{\TempOp{T}}}
\def\O{\TempOp{O}}
\def\H{\TempOp{H}}
\def\tabpropertytemplateline#1#2#3#4#5#6{%
\begin{minipage}[t]{0.075\linewidth}{#1}\end{minipage}&%
\begin{minipage}[t]{0.06\linewidth}{\hfill#2}\end{minipage}&%
\begin{minipage}[t]{0.015\linewidth}{\hfill#3}\end{minipage}&%
\begin{minipage}[t]{0.010\linewidth}{\hfill#4}\end{minipage}&%
\begin{minipage}[t]{0.02\linewidth}{\hfill#5}\end{minipage}&%
\begin{minipage}[t]{0.65\linewidth}{#6}\end{minipage}%
\\%
&&&&&\\[-2.25ex]%
}
\def\tabpropertytemplatelinelast#1#2#3#4#5#6{%
\begin{minipage}[t]{0.075\linewidth}{#1}\end{minipage}&%
\begin{minipage}[t]{0.06\linewidth}{\hfill#2}\end{minipage}&%
\begin{minipage}[t]{0.015\linewidth}{\hfill#3}\end{minipage}&%
\begin{minipage}[t]{0.010\linewidth}{\hfill#4}\end{minipage}&%
\begin{minipage}[t]{0.02\linewidth}{\hfill#5}\end{minipage}&%
\begin{minipage}[t]{0.65\linewidth}{#6}\end{minipage}%
\\[2.5ex]%
}
\begin{table}
\begin{center}
{\fontsize{6}{6}\selectfont 
\begin{tabular}{|l|l|l|l|l|l|}
\hline
&&\multicolumn{2}{l|}{}&&\\[-2.25ex]model & property & \multicolumn{2}{l|}{truth} & $\delta(\TLF)$ & template\\
\cline{3-4}
 & & $p$ & $\neg p$ & &\\&&&&&\\[-2.25ex]
\hline\\[-2.25ex]
\tabpropertytemplateline{1394\{b\}}{1}{t}{f}{0}{$\F((p)\vee((q\vee(r)))$}
\tabpropertytemplateline{-[345]-[23]}{2}{t/f}{--}{4}{$\G((\O((p)\wedge(\O((\neg(p))\wedge(\O((p)\wedge(\O(\neg(p)))))))))\rightarrow(\F(\G(\X(\neg(p))))))$}
\tabpropertytemplateline{}{3}{t/f}{--}{6}{$\G((\O((p)\wedge(\O((\neg(p))\wedge(\O((p)\wedge(\O((\neg(p))\wedge(\O((p)\wedge(\O(\neg(p)))))))))))))\rightarrow(\F(\G(\X(\neg(p))))))$}
\tabpropertytemplateline{}{4}{t/f}{--}{8}{$\G((\O((p)\wedge(\O((\neg(p))\wedge(\O((p)\wedge(\O((\neg(p))\wedge(\O((p)\wedge(\O((\neg(p))\wedge(\O((p)\wedge\\\hspace*{0.5em}(\O(\neg(p)))))))))))))))))\rightarrow(\F(\G(\X(\neg(p))))))$}
\tabpropertytemplateline{}{5}{t}{f}{0}{$(\G(p))\vee((q)\U(\G((r)\vee(s))))$}
\hline\\[-2.25ex]
\tabpropertytemplateline{abp4}{0}{f}{t}{2}{$\G((p)\rightarrow(\Y(\H(q))))$}
\tabpropertytemplateline{}{1}{t}{--}{0}{$\G(\F(p))$}
\tabpropertytemplateline{}{2}{f}{--}{0}{$\G((p)\rightarrow(\X((\neg(p))\U((q)\wedge(((\neg(r))\wedge(s))\vee((r)\wedge(t)))))))$}
\tabpropertytemplateline{}{3}{t}{--}{0}{$\G((p)\rightarrow(\X(((p)\U(\neg(p)))\U((q)\wedge(((\neg(r))\wedge(s))\vee((r)\wedge(t)))))))$}
\hline\\[-2.25ex]
\tabpropertytemplateline{brp}{0}{t}{f}{2}{$\F(\G((p)\rightarrow(\O((q)\rightarrow(\O(r))))))$}
\tabpropertytemplateline{}{$\neg$ 0, nv}{f}{--}{2}{$\neg((\F(\G((p)\rightarrow(\O((q)\rightarrow(\O(r)))))))\wedge((\G(\F(p)))\wedge(\G(\F(q)))))$}
\tabpropertytemplateline{}{1}{t}{f}{0}{$(\G((p)\rightarrow((\X((q)\vee((r)\vee(s))))\R(p))))\wedge((\G((q)\rightarrow((\X((t)\vee(s)))\R(q))))\wedge((\G((t)\rightarrow\\\hspace*{0.5em}((\X((p)\vee(s)))\R(t))))\wedge((\G((q)\rightarrow((\X((p)\vee(s)))\R(q))))\wedge(\G((s)\rightarrow((\X(p))\R(s)))))))$}
\hline\\[-2.25ex]
\tabpropertytemplateline{counter}{0}{t}{f}{0}{$\F(\G(p))$}
\hline\\[-2.25ex]
\tabpropertytemplateline{csmacd}{0}{f}{f}{0}{$\G((p)\rightarrow(\F(q)))$}
\tabpropertytemplateline{}{1}{?}{f}{0}{$(p)\wedge((\F(q))\rightarrow((((((r)\U(s))\U(t))\U(u))\U(v))\U(q)))$}
\hline\\[-2.25ex]
\tabpropertytemplateline{dme[35]}{0}{f}{f}{2}{$\G((p)\rightarrow((p)\T((\neg(p))\T(\neg(q))))$}
\tabpropertytemplateline{}{$\neg$ 0, nv}{f}{--}{2}{$\neg ((\G((p)\rightarrow((p)\T((\neg(p))\T(\neg(q))))))\wedge(\G(\F(p))))$}
\tabpropertytemplateline{}{1}{t}{f}{0}{$\G(((p)\wedge(\X(\neg(p))))\rightarrow(\X((\G(\neg(p)))\vee(((\neg(p))\U(q))\U(r)))))$}
\hline\\[-2.25ex]
\tabpropertytemplateline{mutex}{0}{t}{f}{0}{$\G((p)\rightarrow(\F(q)))$}
\hline\\[-2.25ex]
\tabpropertytemplateline{pci}{0}{f}{f}{4}{$\G((p)\rightarrow(\G(((q)\wedge(\Y((r)\wedge(\O((s)\wedge(\O((t)\wedge(\O(u)))))))))\rightarrow$\\\hspace*{0.5em}$(\O((v)\wedge(\O((w)\wedge(\neg(\O(x))))))))))$}
\tabpropertytemplateline{}{$\F$ 0}{f}{--}{4}{$\F(\G((p)\rightarrow(\G(((q)\wedge(\Y((r)\wedge(\O((s)\wedge(\O((t)\wedge(\O(u)))))))))\rightarrow$\\\hspace*{0.5em}$(\O((v)\wedge(\O((w)\wedge(\neg(\O(x)))))))))))$}
\tabpropertytemplateline{}{1}{?}{f}{0}{$(((((\G((o)\rightarrow((o)\U(p))))\wedge(\G((q)\rightarrow((q)\U((r)\vee(o))))))\wedge(\G((s)\rightarrow((s)\U((t)\vee((q)\vee\\\hspace*{0.5em}(o)))))))\wedge(\G((u)\rightarrow((u)\U((v)\vee((s)\vee((q)\vee(o))))))))\wedge(\G((w)\rightarrow((w)\U((x)\vee((w)\vee((u)\vee\\\hspace*{0.5em}((q)\vee(o)))))))))\wedge(\G((y)\rightarrow((y)\U((z)\vee((y)\vee((w)\vee((u)\vee((q)\vee(o)))))))))$}
\hline\\[-2.25ex]
\tabpropertytemplateline{prod-cons}{0}{f}{f}{1}{$((\G(\neg(p)))\wedge(\G(\F((q)\wedge((q)\S(r))))))\wedge(\G(\F(((q)\wedge((q)\S(r)))\rightarrow((s)\S(t)))))$}
\tabpropertytemplateline{}{1}{t}{--}{4}{$\G((p)\rightarrow((p)\S((q)\S((r)\S((s)\S(t))))))$}
\tabpropertytemplateline{}{$\neg$ 1, nv}{f}{--}{4}{$\neg((\G((p)\rightarrow((p)\S((q)\S((r)\S((s)\S(t)))))))\wedge(\G(\F(p))))$}
\tabpropertytemplateline{}{2}{f}{--}{0}{$\G((p)\rightarrow(\F(((q)\wedge(r))\wedge(s))))$}
\tabpropertytemplateline{}{3}{f}{--}{0}{$\G((p)\rightarrow(\F(q)))$}
\tabpropertytemplateline{}{4}{t}{--}{0}{$\G((p)\rightarrow(\F(q)))$}
\tabpropertytemplateline{}{5}{t}{f}{0}{$(\X(((\X(((\X(p))\R(q))\wedge(r)))\R(s))\wedge(t)))\R(u)$}
\hline\\[-2.25ex]
\tabpropertytemplateline{production-cell}{0}{t}{f}{6}{$\G(\F(((p)\vee(q))\wedge(\O((r)\wedge(\O(((s)\vee(t))\wedge(\O((u)\wedge$\\\hspace*{0.5em}$(\O(((s)\vee(t))\wedge(\O(((v)\vee(w))\wedge(\O(x))))))))))))))$}
\tabpropertytemplateline{}{1}{t}{f}{12}{$\G(\F(((p)\vee(q))\wedge(\Y(\O((r)\wedge(\Y(\O(((s)\vee(t))\wedge(\Y(\O((u)\wedge$\\\hspace*{0.5em}$(\Y(\O(((s)\vee(t))\wedge(\Y(\O(((v)\vee(w))\wedge(\Y(\O(x))))))))))))))))))))$}
\tabpropertytemplateline{}{2}{t}{f}{10}{$\G(\F(((\neg(p))\vee(\neg(q)))\wedge(\O((\neg(r))\wedge(\Y(\O(((\neg(s))\vee(\neg(t)))\wedge(\O((\neg(u))\wedge\\\hspace*{0.5em}(\Y(\O(((\neg(s))\vee(\neg(t)))\wedge(\Y(\O(((\neg(v))\vee(\neg(w)))\wedge(\Y(\O(x))))))))))))))))))$}
\tabpropertytemplateline{}{3}{t}{f}{0}{$(((((((((((((1)\U((e)\wedge(\neg(f))))\U((e)\wedge(f)))\U((g)\wedge(h\wedge((i)\wedge(j)))))\U((k)\wedge((l)\wedge((i)\wedge\\\hspace*{0.5em}(j)))))\U((k)\wedge((l)\wedge(m))))\U((n)\wedge((l)\wedge(o))))\U((p)\wedge(q)))\U((r)\wedge(q)))\U((s)\wedge(q)))\U\\\hspace*{0.5em}((t)\wedge((u)\wedge(v))))\U((w)\wedge((u)\wedge(x))))\U((y)\wedge(\neg(z))))\U((y)\wedge(z))$}
\tabpropertytemplateline{}{4}{t}{f}{0}{$((((((((((((((((1)\U((a)\wedge(\neg(b))))\U((a)\wedge(b)))\U((c)\wedge((d)\wedge((e)\wedge(f)))))\U((g)\wedge((h)\wedge((e)\wedge\\\hspace*{0.5em}(f)))))\U((i)\wedge((j)\wedge(k))))\U((l)\wedge((j)\wedge(m))))\U((n)\wedge(o)))\U((p)\wedge(o)))\U((q)\wedge(o)))\U\\\hspace*{0.5em}((r)\wedge((s)\wedge(t))))\U((u)\wedge((s)\wedge(v))))\U((w)\wedge(\neg(x))))\U((w)\wedge(x)))\U((y)\wedge((z)\wedge((aa)\wedge\\\hspace*{0.5em}((ab)\wedge((ac)\wedge(ad)))))))\U((ae)\wedge((af)\wedge(ag))))\U((a)\wedge(\neg(b)))$}
\hline\\[-2.25ex]
\tabpropertytemplateline{bc57}{0}{t}{f}{2}{$\G(\F((p)\wedge(\O((q)\wedge(\F((r)\wedge(\O(s))))))))$}
\tabpropertytemplateline{}{1}{t}{f}{0}{$(((((\G((a)\rightarrow(((b)\wedge((c)\wedge(d)))\R((e)\wedge(f)))))\wedge(\G((g)\rightarrow(((b)\wedge((h)\wedge(i)))\R((j)\wedge(k))))))\wedge\\\hspace*{0.5em}(\G((l)\rightarrow(((b)\wedge((m)\wedge(n)))\R((o)\wedge(p))))))\wedge(\G((q)\rightarrow(((b)\wedge((r)\wedge(s)))\R((t)\wedge(u))))))\wedge\\\hspace*{0.5em}(\G((v)\rightarrow(((b)\wedge((w)\wedge(\neg(x))))\R((e)\wedge(y))))))\wedge(\G((z)\rightarrow(((b)\wedge((aa)\wedge(\neg(ab))))\R((ac)\wedge\\\hspace*{0.5em}(ad)))))$}
\tabpropertytemplateline{}{2}{t}{f}{0}{$(((((\G((a)\rightarrow(((b)\wedge((c)\wedge(d)))\U((e)\wedge(f)))))\wedge(\G((g)\rightarrow(((b)\wedge((h)\wedge(i)))\U((j)\wedge(k))))))\wedge\\\hspace*{0.5em}(\G((l)\rightarrow(((b)\wedge((m)\wedge(n)))\U((o)\wedge(p))))))\wedge(\G((q)\rightarrow(((b)\wedge((r)\wedge(s)))\U((t)\wedge(u))))))\wedge\\\hspace*{0.5em}(\G((v)\rightarrow(((b)\wedge((w)\wedge(\neg(x))))\U((e)\wedge(y))))))\wedge(\G((z)\rightarrow(((b)\wedge((aa)\wedge(\neg(ab))))\U((ac)\wedge\\\hspace*{0.5em}(ad)))))$}
\tabpropertytemplateline{}{3}{f}{--}{0}{$(((((\G(\F(p)))\vee(\G(\F(q))))\vee(\G(\F(r))))\vee(\G(\F(s))))\vee(\G(\F(t))))\vee(\G(\F(u)))$}
\hline\\[-2.25ex]
\tabpropertytemplateline{ring}{0}{t}{f}{0}{$(\G(\F(p)))\wedge(\G(\F(\neg(p))))$}
\hline\\[-2.25ex]
\tabpropertytemplateline{short}{0}{t}{f}{0}{$\G((p)\rightarrow(\F(q)))$}
\hline\\[-2.25ex]
\tabpropertytemplateline{srg5}{0}{t}{f}{4}{$(((\F(\G(\neg(p))))\wedge(\G(\F(q))))\wedge(\G(\F(r))))\rightarrow(\F((s)\S((t)\S((u)\S((v)\S(w))))))$}
\tabpropertytemplatelinelast{}{$\neg$ 0, nv}{f}{--}{4}{$\neg(((((\F(\G(\neg(p))))\wedge(\G(\F(q))))\wedge(\G(\F(r))))\rightarrow$\\\hspace*{0.5em}$(\F((s)\S((t)\S((u)\S((v)\S(w)))))))\wedge(((\F(\G(\neg(p))))\wedge(\G(\F(q))))\wedge(\G(\F(r)))))$}
\hline
\end{tabular}
}
\end{center}
\caption{\label{tab:propertytemplates}Templates of the properties used in the experiments}
\end{table}

\subsection{Implementations}

Following the automata-theoretic approach to LTL \cite{VW86b}, a mo\-del
checking procedure consists of encoding the property and subsequent
fair cycle detection. As a special case, the second step can be
performed by applying the liveness-to-safety translation and doing
invariant checking. Where available we use off-the-shelf model
checking procedures that include all steps to evaluate a particular
approach. We make the following exceptions to that rule. Our
implementation of the liveness-to-safety translation has the encoding
of the property included but needs to be complemented with an
algorithm to check invariants. To determine whether the effort of a
dedicated implementation of a BMC encoding with the corresponding
opportunities for optimisation is worthwhile we also combine our BMC
encodings with the liveness-to-safety translation and with separately
generated B{\"u}chi automata (Fig.~\ref{exp:boundplots}(e),
(f)). Finally, when comparing a tight with a non-tight B{\"u}chi
automaton in BDD-based symbolic model checking we invoke the
conversion from LTL to a B{\"u}chi automaton externally for both
variants to minimise the influence of different variable orders
(Fig.~\ref{exp:boundplots}(j)).

Encoding of PLTL properties for model checking has been widely
researched (for references see Sect.~\ref{sec:ltl-buchi}). However, in
symbolic model checking, the dominating encodings are still more or
less close to a symbolic implementation of the tableau construction
\cite{LP85} in \cite{BurchClarkeMcMillanDillHwang92,CGH97}. This
shifts a potential exponential blow-up from generation of the
B{\"u}chi automaton to the search for a fair cycle. All encodings
presented in this paper fall in this category. The question whether
optimised B{\"u}chi automata constructions actually yield better
overall performance in symbolic model checking algorithms is still
open: while actual search for a fair cycle seems to benefit from
optimised B{\"u}chi automata, there are cases where those benefits are
more than offset by generating the B{\"u}chi automaton
\cite{SebastianiTonettaVardi05,CimattiRoveriSempriniTonetta06}. Note,
finally, that we currently don't have a construction that yields an
explicit Büchi automaton that is both, small and tight. In
Fig.~\ref{exp:boundplots}(f) below we evaluate whether there is any
overhead in forming the product of the model and the Büchi automaton
for the property first and have the conversion to SAT only encode the
search for a fair cycle compared to encoding the property in a way
very similar to such Büchi automaton as part of the conversion to SAT
(which our encodings do). Therefore, the translation of a PLTL formula
into a Büchi automaton in Fig.~\ref{exp:boundplots}(f) is chosen to be
similar to our BMC encoding. Comparing the performance of optimised
translations from PLTL into Büchi automata with SAT-based approaches
is out of the scope of this work. Similar reservations apply to our
other experiments.

The details of the approaches and implementations used in the
experiments are listed below. The first four approaches include all
steps while the latter three are partial.
\begin{description}
\item[CAV2005\textnormal{($c$,$u$,$o$)}] means an implementation of a
linear, incremental BMC procedure for PLTL on top of NuSMV 2.2.3. The
exact BMC encoding is described in Sect.~\ref{sec:incbmc} and is
essentially the encoding given in~\cite{HelJunLat:CAV05}. The
parameters describe
\begin{itemize}
\item
   whether the completeness check of Sect.~\ref{sec:completeness}
   is enabled ($c=\textup{compl}$) or not ($c=\textup{nocompl}$),
  \item
   whether full virtual unrolling is applied ($u=\textup{unroll}$)
   or not ($u=\textup{nounroll}$),
   and
  \item
   whether the optimisations described in Sect.~5 of~\cite{HelJunLat:CAV05}
   are active ($o=\textup{opt}$) or not ($o=\textup{noopt}$).
 \end{itemize}

\item[VMCAI2005] stands for an implementation of a linear,
non-incremental BMC procedure for PLTL on top of NuSMV version
2.2.3. The exact BMC encoding is described in~\cite{LBHJ05}; it is
very similar to the non-incremental PLTL encoding given in
Sect.~\ref{sec:pltl-eventuality} except that it uses the fixed point
encoding similar to that in Sect.~\ref{sec:ltl-fixe} instead of the
eventuality encoding.

\item[NuSMV\textnormal{(BMCLTL)}] is an example of a non-linear,
non-incremental BMC encoding. It is the standard way to perform
SAT-based bounded model checking for PLTL in NuSMV \cite{NuSMV},
version 2.2.3. For a description see \cite{BC03}.

\item[NuSMV\textnormal{(BDDLTL)}] is the standard method for BDD-based
PLTL model checking in NuSMV \cite{NuSMV}, version 2.2.3. The property
is translated into a symbolic B{\"u}chi automaton with
\cite{KPR98}. Cycle detection is performed with the backward version
of the Emerson-Lei algorithm \cite{EmersonLei86}; we always enabled
the restriction to the set of reachable states. Neither dynamic
reordering nor model-specific variable orders are used.

\item[L2S\textnormal{($t$,$o$)}] is the liveness-to-safety
transformation. The implementation of the transformation is based on
previous work~\cite{SB04,Schuppan06} rather than on the formulation in
Sect.~\ref{sec:l2s}. The encoding of the automaton representing the
property is based on the construction outlined in
Sect.~\ref{sec:pltl-buchi} but is slightly modified for a tighter
integration with the liveness-to-safety transformation. As an example,
the signals indicating the start of the looping part and the end of a
loop iteration are provided directly by the reduction rather than
being separate input variables. The result is close to \cite{LBHJ05}
--- in fact, \cite{LBHJ05} was the starting point of our construction
of a tight B{\"u}chi automaton.

The first parameter states which degree of virtual unrolling is used
in the encoding of the property:
\begin{itemize}
\item $t = \textup{tight}$ means full virtual unrolling up to the past
operator depth of the property, and
\item $t = \textup{notight}$ performs no virtual unrolling at all.
\end{itemize}
The second parameter, $o$, indicates whether variable optimisation
(see Sect.~\ref{sec:l2svaropt}) is
\begin{itemize}
\item enabled ($o = \textup{ic}$), or
\item not ($o = \textup{none}$).
\end{itemize}
Identification of input and constant variables is largely based on
(conservative) syntactic criteria: a variable $v$ in an SMV model
clearly is an input variable if $v$ appears only on the right-hand
side of \texttt{next} assignments such that $v$ is not itself in the
scope of a \texttt{next} operator. Sometimes knowledge of a model was
used to conclude that a variable is either a constant or an input
variable.

A full model checking procedure is obtained in combination with any of
the previous four or with \textbf{NuSMV}(BDDINVAR) below. We only use
\textbf{CAV2005} and \textbf{NuSMV}(BDDINVAR).

\item[B\textnormal{($t$)}] stands for a symbolic implementation of a
B{\"u}chi automaton. The parameter $t$ indicates whether a tight
($t=\textup{tight}$) version is used or not
($t=\textup{notight}$). The former corresponds to
Sect.~\ref{sec:pltl-buchi} while the latter is produced by NuSMV's
ltl2smv tool, which implements~\cite{KPR98}.\footnote{Note, that the
$\textup{notight}$ version still accepts shortest counterexamples for
the future fragment of LTL. This is in contrast to the notation used
in \cite{AS06} where ``tight'' refers to an automaton based on
\cite{CGH97} (i.e., \cite{KPR98} restricted to the future fragment of
PLTL) and ``non-tight'' to one based on \cite{SB00}.} To obtain a
model checking procedure we combine \textbf{B}($t$) either with
\textbf{NuSMV}(BDDLTL) or with \textbf{CAV2005}.

\item[NuSMV\textnormal{(BDDINVAR)}] is BDD-based forward invariant
checking with version 2.2.3 of NuSMV \cite{NuSMV}. We use this to
perform BDD-based model checking with the liveness-to-safety
transformation. State variables of the original model and their second
instances are interleaved, but neither dynamic reordering nor a
model-specific variable order are employed.

\end{description}

\subsection{Results and Comparisons}

\subsubsection{Setting and notation.}
We ran the benchmarks on Linux PC machines with a AMD Athlon(tm) 64 3200+
processor 
and 2~GB of memory.
The memory limit for each run was set to 1.5~GB and the time limit to 1 hour
by using the Linux \texttt{ulimit} command.
For all SAT-based BMC procedures we used zChaff~\cite{zChaff},
version 2004.11.15, as the SAT solver.

Tables~\ref{exp:array1} and~\ref{exp:array2} show the results for
selected approaches.
The $a$ columns tell whether the property was found to be true (t) or false (f)
in the instance (model,property) by the approach in question.
The running times in $t$-columns are given in seconds except
that TO (MO) means that the instance was not solved because of
a timeout (running out of memory).
For \textbf{L2S}(t,o)+$X$ and \textbf{B}(t)+$X$ approaches the running
time does not include the liveness-to-safety transformation or
B\"uchi automaton generation time, but
only the solving time of $X$.\footnote{Note that both the liveness-to-safety transformation and the Büchi
automaton generation are performed symbolically and, therefore, can be
done in polynomial time in the length of the description of the model
and the formula.}

The BMC-based approaches were run in the usual way:
starting with the bound 0 and increasing it by one until
(i) a counterexample or a proof was found, or
(ii) the time or memory limit was reached.
In the incremental approaches (\textbf{CAV2005})
there is only one SAT instance that is updated and solved again
when the bound increases, while
in the non-incremental approaches (\textbf{VMCAI2005},\textbf{NuSMV}(BMCLTL)),
the SAT instance for each bound is independently generated and checked.
The $k$-columns give the bound that was reached.
In particular,
if the problem was solved (no TO or MO in the $t$-column),
the $k$ column gives the length of the counterexample or
the bound required to prove the property.

For BDD-based approaches,
the $|\textup{cex}|$-column gives the length of the produced counterexample.

Fig.~\ref{exp:scatters} shows scatter plots comparing the running times of
different approaches to solve the benchmark instances.
Red squares denote benchmark instances where the property is false
(i.e., they have a counterexample), and black diamonds denote instances
where the property holds.
As mentioned earlier, the time limit was set to 3600 seconds (1 hour):
timeouts are denoted by the time ``value'' 7200 and
running out of memory by the time ``value'' 14400 in the scatter plots.

\subsubsection{Evaluation of different approaches.}

Comparing the columns \textbf{NuSMV}(BMCLTL) and \textbf{VMCAI2005} of
Table~\ref{exp:array1},
plotted against each other also in Fig.~\ref{exp:scatters}(a),
we can see the positive effect of having a more compact
BMC encoding for PLTL formulae.
Most of the properties we use involve past operators and
for such formulae the encoding of \textbf{VMCAI2005}~\cite{LBHJ05} is linear
in $k$ while the encoding of \textbf{NuSMV}(BMCLTL)~\cite{BC03} is not.

From the columns \textbf{VMCAI2005} and \textbf{CAV2005}(nocompl,unroll,opt)
of Table~\ref{exp:array1} and from Fig.~\ref{exp:scatters}(b)
we see that by adapting the compact encoding of \textbf{VMCAI2005}
to exploit modern \emph{incremental} SAT solvers gives an additional major
performance boost.
Note that although \textbf{VMCAI2005} uses a fixed point encoding
while \textbf{CAV2005}(nocompl,unroll,opt) uses eventuality encoding,
we can claim that the major part of the observed performance boost
is due to incrementality
because of the results in Table~1 of~\cite{HelJunLat:CAV05}:
\textbf{VMCAI2005} and \textbf{CAV2005}(nocompl,unroll,opt) with
no incrementality seem to behave very similarly.

As we can see from Fig.~\ref{exp:scatters}(c), the effect of doing virtual
unrolling on the running times is not clear.
However, there are slightly more cases in which unrolling helped than
where it made things slower. This is due to the fact that unrolling can
shorten counterexamples for formulas with past operators.
Figure~\ref{exp:boundplots}(a) illustrates that removing virtual unrolling
may increase the counterexample length not only in theory but
in practice, too.
We also experimented with the option of not applying the optimisations
of~\cite[Sect.~5]{HelJunLat:CAV05} in \textbf{CAV2005} and found that
the optimisations don't seem to have noticeable effect in practice, except
that they sometimes reduce the bound required to prove a property
when virtual unrolling is applied.

Figure~\ref{exp:scatters}(d) shows the effect of adding the completeness check
described in Sect.~\ref{sec:completeness} to the incremental PLTL BMC procedure
\textbf{CAV2005}(nocompl,unroll,opt).
The results demonstrate that the completeness check
(i) enables one to sometimes also prove properties and
    not only find counterexamples,
and
(ii) generally slows down the BMC procedure by a factor of 2 or 3.
However, if we compare the incremental and complete
\textbf{CAV2005}(compl,unroll,opt) to the non-incremental and
incomplete \textbf{VMCAI2005},
we see that incremental SAT solving techniques allows us to have a complete
BMC procedure that almost always outperforms a non-incremental and incomplete
state-of-the-art BMC procedure on the benchmarks we ran.

In Fig.~\ref{exp:scatters}(e) and (f) we compare bounded model
checking with the specialised BMC encoding
\textbf{CAV2005}(nocompl,unroll,opt) with an encoding based on the
liveness-to-safety transformation (requiring only invariant checking
in the BMC procedure) and based on using a tight B{\"u}chi automaton
(requiring only fair loop detection in the BMC procedure). There is a
noticeable overhead when using the liveness-to-safety transformation
while, based on our set of experiments, we cannot conclude that a
specialised encoding improves performance over the B{\"u}chi
automaton. A main benefit of the specialised BMC encoding is, however,
that it can also capture no-loop counterexamples.

Figure \ref{exp:scatters}(g) contrasts finding shortest
counterexamples using a BMC-based and a BDD-based method.
While the former solves slightly more instances (that have a counterexample)
within the given resource bounds,
there are also some instances that it can not solve but the latter can.
With respect to running time there is no clear winner either.

If we compare standard BDD-based PLTL model checking (\textbf{NuSMV}(BDDLTL))
and a state-of-the-art complete BMC procedure
(\textbf{CAV2005}(compl,unroll,opt)), refer to Fig.~\ref{exp:scatters}(h),
we can see that they are quite incomparable.
Although the BDD-based approach seems to be better in proving properties
as it produces less timeouts and memouts,
there are instances that BMC proves much faster.
And vice versa for properties having counterexamples:
\textbf{NuSMV}(BDDLTL) solves (brp,$\neg 0$,nv) much faster but
it (or any of the BDD-based methods we experimented) cannot solve
the properties appearing in Tables~\ref{exp:array1} and~\ref{exp:array2}
on the 1394-5-2 and 1394b-6-4 models.

Using a tight B{\"u}chi automaton with standard BDD-based model
checking incurs a severe performance penalty (Fig.~\ref{exp:scatters}(i)).
The lengths of the counterexamples produced by the tight and non-tight variants
are very different with neither being consistently better
(Fig.~\ref{exp:boundplots}(b)).
Note that \textbf{B}(tight)+\textbf{NuSMV}(BDDLTL) does not necessarily
produce shortest possible counterexamples although it uses tight automata:
the fair path finding algorithm employed in \textbf{NuSMV}(BDDLTL)
does not produce shortest fair $(k,l)$-loops.

Tightness tends to come with a price in the liveness-to-safety and
BDD-based approach as seen in Fig.~\ref{exp:scatters}(j), though less
noticeable than in the standard BDD-based approach. While there is a
price that grows with increasing past operator depth for some examples
(1394-4-2,p2--4 and production-cell,p0/2/1), there is also the
opposite case (1394b-4-2,p2--4). For the production-cell examples the
partial unrolling optimisation proved valuable (not shown here): one
level of unrolling (i.e., treating the specification as having past
operator depth 1) gives shortest counterexamples as with a tight
encoding but takes time only as with a non-tight encoding.

BDD-based model checking using the liveness-to-safety transformation
is often faster than the standard approach of using BDDs
(Fig.~\ref{exp:scatters}(k)) when the property is false, while it is
typically slower for true properties. Further analysis indicates that
early termination might play a role in this behaviour. Another, yet
unexplored factor could be that
\textbf{L2S}$(t,o)$+\textbf{NuSMV}(BDDINVAR) uses a forward invariant
checking algorithm while \textbf{NuSMV}(BDDLTL) uses the backward version of
the Emerson-Lei algorithm. While not shown,
\textbf{L2S}$(t,o)$ + \textbf{NuSMV}(BDDINVAR) tends to use more memory
for both, false and true properties \cite{Schuppan06}. However, it
produces significantly shorter counterexamples
(Fig.~\ref{exp:boundplots}(c)) and is able to solve some examples
where the standard approach reaches the time or memory limit. Note
that the gain in counterexample length in
(Fig.~\ref{exp:boundplots}(c)) is the same when using \textbf{CAV2005}
with unrolling.

The plot in Fig.~\ref{exp:scatters}(l) illustrates that the variable
optimisation presented in Sect.~\ref{sec:l2svaropt} helps in BDD-based
model checking with the liveness-to-safety transformation as expected,
and it does not seem to have any adverse side effects.

We also experimented with a combination of the
liveness-to-safety transformation and the temporal induction of~\cite{ES03}.
That is, we use \textbf{L2S}($t$,ic) to transform the PLTL problem to
an invariant problem and then apply the temporal induction algorithm
implemented in NuSMV (the command \texttt{check\_invar\_bmc\_inc -a zigzag}).
We were surprised that the resulting approach could not prove any of
the true properties among the benchmarks we ran.
We have no explanation for this behaviour at the moment, but we suspect that
the liveness-to-safety transformation and the backwards working
completeness checking of~\cite{ES03} might not fit together well.

\iftrue
\def\expfont{\fontsize{5}{6}\selectfont}
\else
\def\expfont{\tiny}
\fi

\begin{table}[p]
\caption{Results of the experiments 1}
\label{exp:array1}
\centering
{\expfont
\begin{tabular}{|lr|r|r|r|r|r|r|r|r|r|r|r|r|r|r|}
\hline
 & & 
\multicolumn{3}{|c}{\textbf{NuSMV}} & 
\multicolumn{3}{|c}{\textbf{VMCAI2005}} & 
\multicolumn{3}{|c}{\textbf{CAV2005}} & 
\multicolumn{3}{|c|}{\textbf{CAV2005}}\\
 & & 
\multicolumn{3}{|c}{(\textbf{BMCLTL})} & 
\multicolumn{3}{|c}{} & 
\multicolumn{3}{|c}{(\textbf{nocompl,unroll,opt})} & 
\multicolumn{3}{|c|}{(\textbf{compl,unroll,opt})}\\
model & prop. &
a & k & t & 
a & k & t & 
a & k & t & 
a & k & t\\
\hline
1394-3-2 & $1$ & 
 & 45 & TO & 
 & 40 & TO & 
 & 1237 & MO & 
t & 16 & 45 \\
1394-3-2 & $\neg{}1$ & 
f & 11 & 17 & 
f & 11 & 31 & 
f & 11 & 9 & 
f & 11 & 12 \\
1394-3-2 & $5$ & 
 & 47 & TO & 
 & 49 & TO & 
 & 946 & TO & 
 & 65 & TO \\
1394-3-2 & $\neg{}5$ & 
f & 11 & 19 & 
f & 11 & 17 & 
f & 11 & 8 & 
f & 11 & 11 \\
1394-4-2 & $1$ & 
 & 17 & TO & 
 & 18 & TO & 
 & 24 & TO & 
 & 24 & TO \\
\hline
1394-4-2 & $\neg{}1$ & 
f & 16 & 1316 & 
f & 16 & 1676 & 
f & 16 & 424 & 
f & 16 & 679 \\
1394-4-2 & $2$ & 
 & 17 & TO & 
 & 22 & TO & 
 & 33 & TO & 
 & 29 & TO \\
1394-4-2 & $3$ & 
 & 8 & TO & 
 & 21 & TO & 
 & 31 & TO & 
 & 29 & TO \\
1394-4-2 & $4$ & 
 & 5 & TO & 
 & 22 & TO & 
 & 30 & TO & 
 & 29 & TO \\
1394-4-2 & $5$ & 
 & 26 & TO & 
 & 26 & TO & 
 & 43 & TO & 
 & 38 & TO \\
\hline
1394-4-2 & $\neg{}5$ & 
f & 16 & 1287 & 
f & 16 & 1371 & 
f & 16 & 338 & 
f & 16 & 551 \\
1394-5-2 & $1$ & 
 & 14 & TO & 
 & 14 & TO & 
 & 15 & TO & 
 & 16 & TO \\
1394-5-2 & $\neg{}1$ & 
f & 14 & 2937 & 
f & 14 & 2749 & 
f & 14 & 976 & 
f & 14 & 1295 \\
1394-5-2 & $5$ & 
 & 16 & TO & 
 & 16 & TO & 
 & 21 & TO & 
 & 20 & TO \\
1394-5-2 & $\neg{}5$ & 
f & 14 & 3360 & 
f & 14 & 3425 & 
f & 14 & 938 & 
f & 14 & 1200 \\
\hline
1394b-4-2 & $2$ & 
f & 11 & 172 & 
f & 11 & 44 & 
f & 11 & 12 & 
f & 11 & 24 \\
1394b-4-2 & $3$ & 
 & 8 & TO & 
f & 11 & 46 & 
f & 11 & 15 & 
f & 11 & 23 \\
1394b-4-2 & $4$ & 
 & 5 & TO & 
f & 11 & 43 & 
f & 11 & 17 & 
f & 11 & 25 \\
1394b-5-3 & $2$ & 
f & 11 & 832 & 
f & 11 & 654 & 
f & 11 & 223 & 
f & 11 & 505 \\
1394b-5-3 & $3$ & 
 & 8 & TO & 
f & 11 & 665 & 
f & 11 & 484 & 
f & 11 & 426 \\
\hline
1394b-5-3 & $4$ & 
 & 5 & TO & 
f & 11 & 775 & 
f & 11 & 385 & 
f & 11 & 259 \\
1394b-6-4 & $2$ & 
 & 10 & TO & 
 & 10 & TO & 
f & 11 & 1875 & 
f & 11 & 2241 \\
1394b-6-4 & $3$ & 
 & 8 & TO & 
 & 10 & TO & 
f & 11 & 1930 & 
f & 11 & 2070 \\
1394b-6-4 & $4$ & 
 & 5 & TO & 
 & 9 & TO & 
f & 11 & 2125 & 
f & 11 & 2738 \\
abp4 & $0$ & 
f & 16 & 62 & 
f & 16 & 46 & 
f & 16 & 27 & 
f & 16 & 20 \\
\hline
abp4 & $\neg{}0$ & 
 & 47 & TO & 
 & 52 & TO & 
 & 354 & TO & 
 & 46 & TO \\
abp4 & $1$ & 
 & 30 & TO & 
 & 29 & TO & 
 & 45 & TO & 
 & 38 & TO \\
abp4 & $2$ & 
f & 17 & 70 & 
f & 17 & 36 & 
f & 17 & 39 & 
f & 17 & 59 \\
abp4 & $3$ & 
 & 29 & TO & 
 & 30 & TO & 
 & 37 & TO & 
 & 36 & TO \\
brp & $0$ & 
 & 31 & TO & 
 & 241 & TO & 
 & 3040 & TO & 
 & 86 & TO \\
\hline
brp & $\neg{}0$ & 
f & 1 & 0 & 
f & 1 & 0 & 
f & 1 & 0 & 
f & 1 & 0 \\
brp & $\neg{}0,\textup{nv}$ & 
 & 22 & TO & 
 & 21 & TO & 
f & 24 & 600 & 
f & 24 & 573 \\
brp & $1$ & 
 & 25 & TO & 
 & 38 & TO & 
 & 196 & TO & 
 & 78 & TO \\
brp & $\neg{}1$ & 
f & 1 & 0 & 
f & 1 & 0 & 
f & 1 & 0 & 
f & 1 & 0 \\
counter & $0$ & 
 & 202 & TO & 
 & 1263 & MO & 
 & 11849 & TO & 
t & 23 & 0 \\
\hline
counter & $\neg{}0$ & 
f & 8 & 0 & 
f & 8 & 0 & 
f & 8 & 0 & 
f & 8 & 0 \\
csmacd & $0$ & 
 & 18 & TO & 
 & 18 & TO & 
 & 19 & TO & 
 & 19 & TO \\
csmacd & $\neg{}0$ & 
f & 6 & 3 & 
f & 7 & 5 & 
f & 6 & 2 & 
f & 6 & 5 \\
csmacd & $1$ & 
 & 22 & TO & 
 & 24 & TO & 
 & 33 & TO & 
 & 29 & TO \\
csmacd & $\neg{}1$ & 
f & 6 & 3 & 
f & 7 & 5 & 
f & 6 & 2 & 
f & 6 & 4 \\
\hline
dme3 & $0$ & 
 & 27 & MO & 
 & 49 & TO & 
 & 48 & TO & 
f & 62 & 2547 \\
dme3 & $\neg{}0$ & 
f & 1 & 0 & 
f & 1 & 0 & 
f & 1 & 0 & 
f & 1 & 0 \\
dme3 & $\neg{}0,\textup{nv}$ & 
 & 27 & MO & 
f & 59 & 2330 & 
f & 59 & 641 & 
f & 59 & 1136 \\
dme3 & $1$ & 
 & 42 & TO & 
 & 53 & TO & 
 & 58 & TO & 
 & 65 & TO \\
dme3 & $\neg{}1$ & 
f & 1 & 0 & 
f & 1 & 0 & 
f & 1 & 0 & 
f & 1 & 0 \\
\hline
dme5 & $0$ & 
 & 27 & MO & 
 & 57 & TO & 
 & 74 & TO & 
 & 67 & TO \\
dme5 & $\neg{}0$ & 
f & 1 & 0 & 
f & 1 & 0 & 
f & 1 & 0 & 
f & 1 & 0 \\
dme5 & $\neg{}0,\textup{nv}$ & 
 & 27 & MO & 
 & 58 & TO & 
 & 75 & TO & 
 & 69 & TO \\
dme5 & $1$ & 
 & 42 & TO & 
 & 44 & TO & 
 & 48 & TO & 
 & 68 & TO \\
dme5 & $\neg{}1$ & 
f & 1 & 0 & 
f & 1 & 0 & 
f & 1 & 0 & 
f & 1 & 0 \\
\hline
mutex & $0$ & 
 & 226 & TO & 
 & 950 & MO & 
 & 10624 & TO & 
t & 18 & 0 \\
mutex & $\neg{}0$ & 
f & 6 & 0 & 
f & 6 & 0 & 
f & 6 & 0 & 
f & 6 & 0 \\
pci & $0$ & 
 & 17 & TO & 
f & 18 & 3092 & 
f & 18 & 1339 & 
f & 18 & 1631 \\
pci & $\neg{}0$ & 
f & 0 & 0 & 
f & 0 & 0 & 
f & 0 & 0 & 
f & 0 & 0 \\
pci & $\textbf{F}0$ & 
 & 14 & TO & 
f & 18 & 1121 & 
f & 18 & 514 & 
f & 18 & 610 \\
\hline
pci & $1$ & 
 & 16 & TO & 
 & 18 & TO & 
 & 20 & TO & 
 & 20 & TO \\
pci & $\neg{}1$ & 
f & 1 & 0 & 
f & 1 & 0 & 
f & 1 & 0 & 
f & 1 & 0 \\
prod-cons & $0$ & 
f & 21 & 972 & 
f & 21 & 63 & 
f & 21 & 14 & 
f & 21 & 35 \\
prod-cons & $\neg{}0$ & 
 & 25 & TO & 
f & 26 & 390 & 
f & 26 & 96 & 
f & 26 & 233 \\
prod-cons & $1$ & 
 & 16 & TO & 
 & 68 & TO & 
 & 180 & TO & 
 & 72 & TO \\
\hline
prod-cons & $\neg{}1,\textup{nv}$ & 
 & 16 & TO & 
f & 21 & 49 & 
f & 21 & 16 & 
f & 21 & 29 \\
prod-cons & $2$ & 
f & 24 & 114 & 
f & 24 & 101 & 
f & 24 & 2 & 
f & 24 & 10 \\
prod-cons & $3$ & 
f & 24 & 145 & 
f & 24 & 140 & 
f & 24 & 15 & 
f & 24 & 47 \\
prod-cons & $4$ & 
 & 65 & TO & 
 & 63 & TO & 
 & 259 & TO & 
 & 91 & TO \\
prod-cons & $5$ & 
 & 44 & TO & 
 & 49 & TO & 
 & 233 & TO & 
 & 82 & TO \\
\hline
prod-cons & $\neg{}5$ & 
f & 21 & 79 & 
f & 21 & 58 & 
f & 21 & 13 & 
f & 21 & 32 \\
production-cell & $0$ & 
 & 6 & TO & 
 & 87 & TO & 
 & 1255 & TO & 
 & 149 & MO \\
production-cell & $\neg{}0$ & 
 & 6 & TO & 
 & 66 & TO & 
 & 73 & TO & 
f & 81 & 53 \\
production-cell & $1$ & 
 & 4 & TO & 
 & 82 & TO & 
 & 619 & TO & 
 & 105 & MO \\
production-cell & $\neg{}1$ & 
 & 4 & TO & 
 & 63 & TO & 
f & 81 & 301 & 
f & 81 & 104 \\
\hline
production-cell & $2$ & 
 & 4 & TO & 
 & 85 & TO & 
 & 1115 & MO & 
 & 115 & MO \\
production-cell & $\neg{}2$ & 
 & 4 & TO & 
 & 66 & TO & 
f & 81 & 104 & 
f & 81 & 85 \\
production-cell & $3$ & 
 & 5 & TO & 
 & 19 & TO & 
 & 2391 & MO & 
t & 110 & 103 \\
production-cell & $\neg{}3$ & 
 & 6 & TO & 
 & 19 & TO & 
f & 81 & 234 & 
f & 81 & 42 \\
production-cell & $4$ & 
 & 4 & TO & 
 & 15 & TO & 
 & 2295 & MO & 
t & 110 & 111 \\
\hline
production-cell & $\neg{}4$ & 
 & 5 & TO & 
 & 16 & TO & 
f & 81 & 73 & 
f & 81 & 46 \\
bc57-sensors & $0$ & 
 & 20 & TO & 
 & 300 & MO & 
 & 1845 & MO & 
 & 130 & MO \\
bc57-sensors & $\neg{}0$ & 
 & 20 & TO & 
 & 99 & TO & 
f & 103 & 863 & 
 & 101 & TO \\
bc57-sensors & $1$ & 
 & 57 & TO & 
 & 101 & TO & 
 & 101 & TO & 
 & 111 & TO \\
bc57-sensors & $\neg{}1$ & 
 & 56 & TO & 
 & 96 & TO & 
f & 103 & 1078 & 
 & 94 & TO \\
\hline
bc57-sensors & $2$ & 
 & 56 & TO & 
 & 130 & TO & 
 & 1709 & MO & 
 & 132 & MO \\
bc57-sensors & $\neg{}2$ & 
 & 57 & TO & 
 & 97 & TO & 
f & 103 & 1215 & 
 & 92 & TO \\
bc57-sensors & $3$ & 
 & 89 & TO & 
 & 99 & TO & 
f & 103 & 1173 & 
f & 103 & 3158 \\
ring & $0$ & 
 & 184 & TO & 
 & 312 & TO & 
 & 756 & MO & 
t & 65 & 1012 \\
ring & $\neg{}0$ & 
f & 7 & 0 & 
f & 7 & 0 & 
f & 7 & 0 & 
f & 7 & 0 \\
\hline
short & $0$ & 
 & 213 & TO & 
 & 1132 & MO & 
 & 3414 & TO & 
t & 10 & 0 \\
short & $\neg{}0$ & 
f & 1 & 0 & 
f & 1 & 0 & 
f & 1 & 0 & 
f & 1 & 0 \\
srg5 & $0$ & 
 & 13 & TO & 
 & 312 & TO & 
 & 805 & MO & 
 & 56 & TO \\
srg5 & $\neg{}0$ & 
f & 1 & 0 & 
f & 1 & 0 & 
f & 1 & 0 & 
f & 1 & 0 \\
srg5 & $\neg{}0,\textup{nv}$ & 
f & 6 & 8 & 
f & 6 & 0 & 
f & 6 & 0 & 
f & 6 & 0 \\
\hline
\hline
\end{tabular}

}
\end{table}

\begin{table}[p]
\caption{Results of the experiments 2}
\label{exp:array2}
\centering
{\expfont
\begin{tabular}{|lr|r|r|r|r|r|r|r|r|r|r|r|r|r|r|r|r|r|r|}
\hline
 & & 
\multicolumn{3}{|c}{\textbf{NuSMV}} & 
\multicolumn{3}{|c}{\textbf{L2S(notight,ic)}+} & 
\multicolumn{3}{|c}{\textbf{L2S(tight,ic)}+} & 
\multicolumn{3}{|c}{\textbf{L2S(tight,ic)}+} & 
\multicolumn{3}{|c|}{\textbf{B(tight)}+}\\
 & & 
\multicolumn{3}{|c}{(\textbf{BDDLTL})} & 
\multicolumn{3}{|c}{\textbf{NuSMV}} & 
\multicolumn{3}{|c}{\textbf{NuSMV}} & 
\multicolumn{3}{|c}{\textbf{CAV2005}} & 
\multicolumn{3}{|c|}{\textbf{CAV2005}}\\
 & &
\multicolumn{3}{|c}{} & 
\multicolumn{3}{|c}{\textbf{(BDDINVAR)}} & 
\multicolumn{3}{|c}{\textbf{(BDDINVAR)}} & 
\multicolumn{3}{|c}{\textbf{(nocompl,unroll,opt)}} & 
\multicolumn{3}{|c|}{\textbf{(nocompl,unroll,opt)}}\\
model & prop. &
a & $|\textup{cex}|$ & t  & 
a & $|\textup{cex}|$ & t  & 
a & $|\textup{cex}|$ & t  & 
a & k & t  & 
a & k & t \\
\hline
1394-3-2 & $1$ & 
t &  & 5 & 
t &  & 7 & 
t &  & 7 & 
 & 696 & MO & 
 & 1190 & MO \\
1394-3-2 & $\neg{}1$ & 
f & 12 & 15 & 
f & 11 & 5 & 
f & 11 & 5 & 
f & 11 & 18 & 
f & 11 & 8 \\
1394-3-2 & $5$ & 
t &  & 3 & 
t &  & 13 & 
t &  & 13 & 
 & 177 & MO & 
 & 1099 & TO \\
1394-3-2 & $\neg{}5$ & 
f & 11 & 25 & 
f & 11 & 7 & 
f & 11 & 7 & 
f & 11 & 25 & 
f & 11 & 8 \\
1394-4-2 & $1$ & 
t &  & 505 & 
t &  & 445 & 
t &  & 445 & 
 & 20 & TO & 
 & 26 & TO \\
\hline
1394-4-2 & $\neg{}1$ & 
f & 20 & 721 & 
f & 16 & 336 & 
f & 16 & 336 & 
f & 16 & 689 & 
f & 16 & 462 \\
1394-4-2 & $2$ & 
t &  & 271 & 
t &  & 470 & 
t &  & 615 & 
 & 24 & TO & 
 & 27 & TO \\
1394-4-2 & $3$ & 
t &  & 288 & 
t &  & 610 & 
t &  & 825 & 
 & 24 & TO & 
 & 28 & TO \\
1394-4-2 & $4$ & 
t &  & 300 & 
t &  & 726 & 
t &  & 1276 & 
 & 23 & TO & 
 & 28 & TO \\
1394-4-2 & $5$ & 
t &  & 318 & 
t &  & 811 & 
t &  & 811 & 
 & 27 & TO & 
 & 40 & TO \\
\hline
1394-4-2 & $\neg{}5$ & 
f & 19 & 1231 & 
f & 16 & 536 & 
f & 16 & 535 & 
f & 16 & 902 & 
f & 16 & 368 \\
1394-5-2 & $1$ & 
 &  & MO & 
 &  & MO & 
 &  & MO & 
 & 15 & TO & 
 & 15 & TO \\
1394-5-2 & $\neg{}1$ & 
 &  & MO & 
 &  & MO & 
 &  & MO & 
f & 14 & 1097 & 
f & 14 & 896 \\
1394-5-2 & $5$ & 
 &  & MO & 
 &  & MO & 
 &  & MO & 
 & 18 & TO & 
 & 20 & TO \\
1394-5-2 & $\neg{}5$ & 
 &  & MO & 
 &  & MO & 
 &  & MO & 
f & 14 & 1634 & 
f & 14 & 908 \\
\hline
1394b-4-2 & $2$ & 
f & 20 & 131 & 
f & 15 & 110 & 
f & 11 & 90 & 
f & 11 & 41 & 
f & 11 & 21 \\
1394b-4-2 & $3$ & 
f & 23 & 140 & 
f & 18 & 132 & 
f & 11 & 97 & 
f & 11 & 48 & 
f & 11 & 25 \\
1394b-4-2 & $4$ & 
f & 26 & 152 & 
f & 21 & 183 & 
f & 11 & 100 & 
f & 11 & 32 & 
f & 11 & 20 \\
1394b-5-3 & $2$ & 
 &  & MO & 
 &  & MO & 
 &  & MO & 
f & 11 & 1601 & 
f & 11 & 577 \\
1394b-5-3 & $3$ & 
 &  & MO & 
 &  & MO & 
 &  & MO & 
f & 11 & 850 & 
f & 11 & 202 \\
\hline
1394b-5-3 & $4$ & 
 &  & MO & 
 &  & MO & 
 &  & MO & 
f & 11 & 524 & 
f & 11 & 198 \\
1394b-6-4 & $2$ & 
 &  & TO & 
 &  & TO & 
 &  & TO & 
 & 10 & TO & 
f & 11 & 2839 \\
1394b-6-4 & $3$ & 
 &  & TO & 
 &  & TO & 
 &  & TO & 
f & 11 & 2857 & 
f & 11 & 1521 \\
1394b-6-4 & $4$ & 
 &  & TO & 
 &  & TO & 
 &  & TO & 
f & 11 & 2505 & 
f & 11 & 3415 \\
abp4 & $0$ & 
f & 37 & 1 & 
f & 19 & 11 & 
f & 16 & 6 & 
f & 16 & 33 & 
f & 16 & 17 \\
\hline
abp4 & $\neg{}0$ & 
t &  & 0 & 
t &  & 0 & 
t &  & 0 & 
 & 58 & TO & 
 & 51 & TO \\
abp4 & $1$ & 
t &  & 0 & 
t &  & 58 & 
t &  & 58 & 
 & 33 & TO & 
 & 38 & TO \\
abp4 & $2$ & 
f & 40 & 2 & 
f & 17 & 11 & 
f & 17 & 11 & 
f & 17 & 153 & 
f & 17 & 38 \\
abp4 & $3$ & 
t &  & 0 & 
t &  & 62 & 
t &  & 62 & 
 & 24 & TO & 
 & 38 & TO \\
brp & $0$ & 
t &  & 0 & 
 &  & TO & 
 &  & TO & 
 & 312 & TO & 
 & 106 & TO \\
\hline
brp & $\neg{}0$ & 
f & 6 & 4 & 
f & 1 & 0 & 
f & 1 & 0 & 
f & 1 & 0 & 
f & 1 & 0 \\
brp & $\neg{}0,\textup{nv}$ & 
f & 68 & 14 & 
f & 24 & 81 & 
f & 24 & 85 & 
f & 24 & 1485 & 
f & 24 & 579 \\
brp & $1$ & 
t &  & 4 & 
 &  & TO & 
 &  & TO & 
 & 26 & TO & 
 & 29 & TO \\
brp & $\neg{}1$ & 
f & 23 & 3 & 
f & 1 & 0 & 
f & 1 & 0 & 
f & 1 & 0 & 
f & 1 & 0 \\
counter & $0$ & 
t &  & 0 & 
t &  & 0 & 
t &  & 0 & 
 & 2277 & TO & 
 & 1598 & TO \\
\hline
counter & $\neg{}0$ & 
f & 8 & 0 & 
f & 8 & 0 & 
f & 8 & 0 & 
f & 8 & 0 & 
f & 8 & 0 \\
csmacd & $0$ & 
 &  & MO & 
f & 27 & 439 & 
f & 27 & 438 & 
 & 18 & TO & 
 & 18 & TO \\
csmacd & $\neg{}0$ & 
 &  & MO & 
f & 6 & 2 & 
f & 6 & 2 & 
f & 6 & 7 & 
f & 6 & 3 \\
csmacd & $1$ & 
 &  & TO & 
 &  & TO & 
 &  & TO & 
 & 30 & TO & 
 & 21 & TO \\
csmacd & $\neg{}1$ & 
 &  & MO & 
f & 6 & 2 & 
f & 6 & 2 & 
f & 6 & 10 & 
f & 6 & 3 \\
\hline
dme3 & $0$ & 
f & 215 & 42 & 
f & 63 & 8 & 
f & 63 & 8 & 
f & 63 & 1488 & 
f & 63 & 670 \\
dme3 & $\neg{}0$ & 
f & 1 & 0 & 
f & 1 & 0 & 
f & 1 & 0 & 
f & 1 & 0 & 
f & 1 & 0 \\
dme3 & $\neg{}0,\textup{nv}$ & 
f & 217 & 39 & 
f & 59 & 5 & 
f & 59 & 7 & 
f & 59 & 477 & 
f & 59 & 513 \\
dme3 & $1$ & 
t &  & 17 & 
t &  & 246 & 
t &  & 246 & 
 & 62 & TO & 
 & 64 & TO \\
dme3 & $\neg{}1$ & 
f & 68 & 56 & 
f & 1 & 0 & 
f & 1 & 0 & 
f & 1 & 0 & 
f & 1 & 0 \\
\hline
dme5 & $0$ & 
f & 343 & 1289 & 
f & 103 & 303 & 
f & 103 & 299 & 
 & 71 & TO & 
 & 76 & TO \\
dme5 & $\neg{}0$ & 
f & 1 & 10 & 
f & 1 & 1 & 
f & 1 & 1 & 
f & 1 & 0 & 
f & 1 & 0 \\
dme5 & $\neg{}0,\textup{nv}$ & 
f & 344 & 1371 & 
f & 99 & 271 & 
f & 99 & 309 & 
 & 74 & TO & 
 & 76 & TO \\
dme5 & $1$ & 
t &  & 483 & 
 &  & TO & 
 &  & TO & 
 & 46 & TO & 
 & 57 & TO \\
dme5 & $\neg{}1$ & 
f & 108 & 1465 & 
f & 1 & 1 & 
f & 1 & 1 & 
f & 1 & 0 & 
f & 1 & 0 \\
\hline
mutex & $0$ & 
t &  & 0 & 
t &  & 0 & 
t &  & 0 & 
 & 392 & MO & 
 & 1958 & TO \\
mutex & $\neg{}0$ & 
f & 7 & 0 & 
f & 6 & 0 & 
f & 6 & 0 & 
f & 6 & 0 & 
f & 6 & 0 \\
pci & $0$ & 
f & 23 & 214 & 
 &  & MO & 
 &  & MO & 
f & 18 & 2971 & 
 & 13 & TO \\
pci & $\neg{}0$ & 
f &  & 3593 & 
f & 1 & 0 & 
f & 1 & 1 & 
f & 1 & 0 & 
f & 1 & 0 \\
pci & $\textbf{F}0$ & 
f & 40 & 212 & 
 &  & MO & 
 &  & MO & 
f & 18 & 2450 & 
f & 18 & 893 \\
\hline
pci & $1$ & 
 &  & TO & 
 &  & MO & 
 &  & MO & 
 & 15 & TO & 
 & 20 & TO \\
pci & $\neg{}1$ & 
f & 6 & 374 & 
f & 1 & 0 & 
f & 1 & 0 & 
f & 1 & 0 & 
f & 1 & 0 \\
prod-cons & $0$ & 
f & 36 & 1200 & 
f & 21 & 78 & 
f & 21 & 85 & 
f & 21 & 49 & 
f & 21 & 14 \\
prod-cons & $\neg{}0$ & 
f & 69 & 42 & 
f & 26 & 188 & 
f & 26 & 204 & 
f & 26 & 210 & 
f & 26 & 122 \\
prod-cons & $1$ & 
t &  & 1 & 
 &  & MO & 
 &  & TO & 
 & 48 & TO & 
 & 30 & TO \\
\hline
prod-cons & $\neg{}1,\textup{nv}$ & 
f & 53 & 43 & 
f & 21 & 37 & 
f & 21 & 51 & 
f & 21 & 25 & 
f & 21 & 14 \\
prod-cons & $2$ & 
f & 57 & 308 & 
f & 24 & 50 & 
f & 24 & 50 & 
f & 24 & 7 & 
f & 24 & 16 \\
prod-cons & $3$ & 
f & 42 & 119 & 
f & 24 & 50 & 
f & 24 & 50 & 
f & 24 & 44 & 
f & 24 & 24 \\
prod-cons & $4$ & 
t &  & 15 & 
 &  & MO & 
 &  & MO & 
 & 72 & TO & 
 & 196 & TO \\
prod-cons & $5$ & 
t &  & 3 & 
 &  & TO & 
 &  & TO & 
 & 140 & TO & 
 & 169 & TO \\
\hline
prod-cons & $\neg{}5$ & 
f & 54 & 191 & 
f & 21 & 65 & 
f & 21 & 65 & 
f & 21 & 61 & 
f & 21 & 16 \\
production-cell & $0$ & 
t &  & 694 & 
t &  & 3 & 
t &  & 12 & 
 & 192 & MO & 
 & 498 & TO \\
production-cell & $\neg{}0$ & 
f & 85 & 1094 & 
f & 83 & 6 & 
f & 81 & 10 & 
f & 81 & 219 & 
f & 81 & 13 \\
production-cell & $1$ & 
t &  & 95 & 
t &  & 6 & 
 &  & TO & 
 & 168 & MO & 
 & 358 & TO \\
production-cell & $\neg{}1$ & 
f & 146 & 37 & 
f & 126 & 8 & 
 &  & TO & 
 & 66 & TO & 
f & 81 & 95 \\
\hline
production-cell & $2$ & 
t &  & 81 & 
t &  & 6 & 
t &  & 517 & 
 & 171 & MO & 
 & 390 & TO \\
production-cell & $\neg{}2$ & 
f & 126 & 34 & 
f & 125 & 11 & 
f & 81 & 256 & 
 & 70 & TO & 
f & 81 & 17 \\
production-cell & $3$ & 
t &  & 1657 & 
t &  & 5 & 
t &  & 5 & 
 & 1188 & MO & 
 & 1726 & MO \\
production-cell & $\neg{}3$ & 
f & 271 & 3050 & 
f & 81 & 11 & 
f & 81 & 11 & 
f & 81 & 457 & 
f & 81 & 19 \\
production-cell & $4$ & 
 &  & MO & 
t &  & 8 & 
t &  & 8 & 
 & 1130 & MO & 
 & 1584 & MO \\
\hline
production-cell & $\neg{}4$ & 
 &  & MO & 
f & 81 & 26 & 
f & 81 & 26 & 
f & 81 & 551 & 
f & 81 & 35 \\
bc57-sensors & $0$ & 
t &  & 65 & 
t &  & 2430 & 
t &  & 2192 & 
 & 101 & TO & 
 & 116 & TO \\
bc57-sensors & $\neg{}0$ & 
f & 112 & 206 & 
f & 103 & 109 & 
f & 103 & 102 & 
f & 103 & 2793 & 
f & 103 & 1198 \\
bc57-sensors & $1$ & 
t &  & 152 & 
t &  & 3388 & 
t &  & 3391 & 
 & 93 & TO & 
 & 111 & TO \\
bc57-sensors & $\neg{}1$ & 
f & 104 & 909 & 
f & 103 & 99 & 
f & 103 & 99 & 
f & 103 & 2819 & 
f & 103 & 1084 \\
\hline
bc57-sensors & $2$ & 
t &  & 39 & 
t &  & 2347 & 
t &  & 2352 & 
 & 82 & TO & 
 & 120 & TO \\
bc57-sensors & $\neg{}2$ & 
f & 104 & 866 & 
f & 103 & 101 & 
f & 103 & 101 & 
f & 103 & 3501 & 
f & 103 & 1317 \\
bc57-sensors & $3$ & 
 &  & TO & 
f & 103 & 96 & 
f & 103 & 95 & 
f & 103 & 2973 & 
f & 103 & 1109 \\
ring & $0$ & 
t &  & 0 & 
t &  & 0 & 
t &  & 0 & 
 & 90 & TO & 
 & 639 & MO \\
ring & $\neg{}0$ & 
f & 13 & 0 & 
f & 7 & 0 & 
f & 7 & 0 & 
f & 7 & 0 & 
f & 7 & 0 \\
\hline
short & $0$ & 
t &  & 0 & 
t &  & 0 & 
t &  & 0 & 
 & 336 & MO & 
 & 3288 & TO \\
short & $\neg{}0$ & 
f & 3 & 0 & 
f & 1 & 0 & 
f & 1 & 0 & 
f & 1 & 0 & 
f & 1 & 0 \\
srg5 & $0$ & 
t &  & 0 & 
t &  & 0 & 
t &  & 0 & 
 & 117 & TO & 
 & 241 & MO \\
srg5 & $\neg{}0$ & 
f & 16 & 0 & 
f & 1 & 0 & 
f & 1 & 0 & 
f & 1 & 0 & 
f & 1 & 0 \\
srg5 & $\neg{}0,\textup{nv}$ & 
f & 15 & 0 & 
f & 6 & 0 & 
f & 6 & 1 & 
f & 6 & 0 & 
f & 6 & 0 \\
\hline
\hline
\end{tabular}

}
\end{table}

%
\newcommand{\incscatter}[1]{\includegraphics[width=4.7cm]{#1}}
\begin{figure}
  \centering
  \begin{tabular}{ccc}
    \incscatter{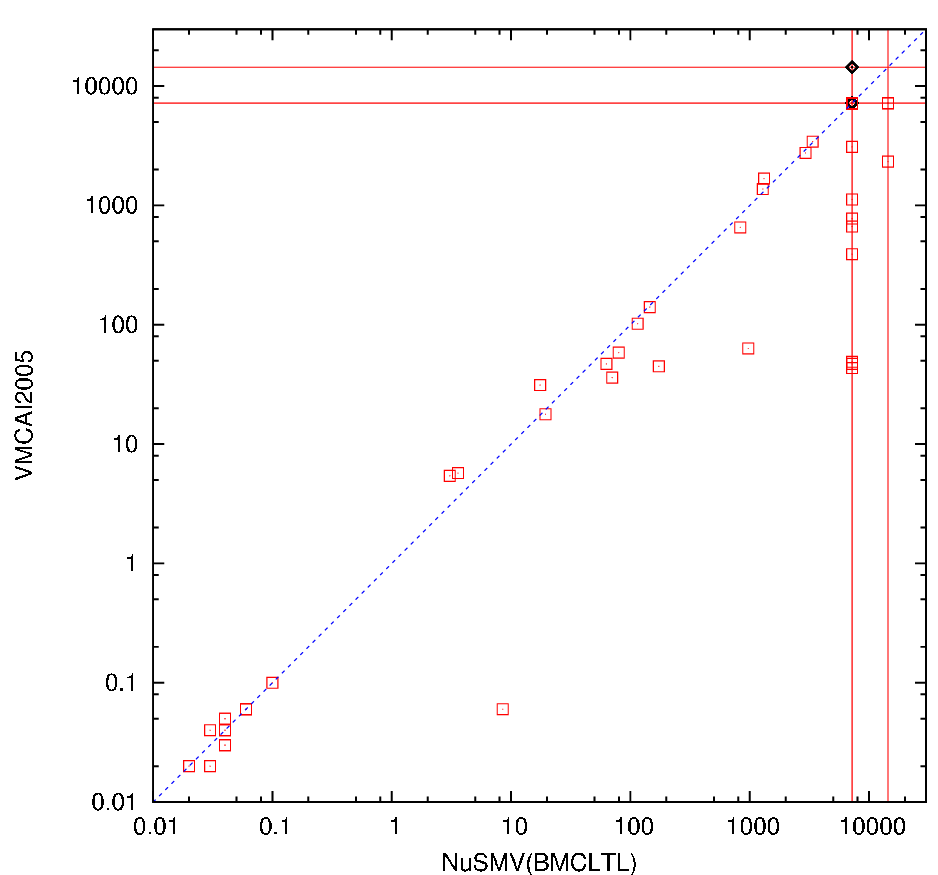}
    &
    \incscatter{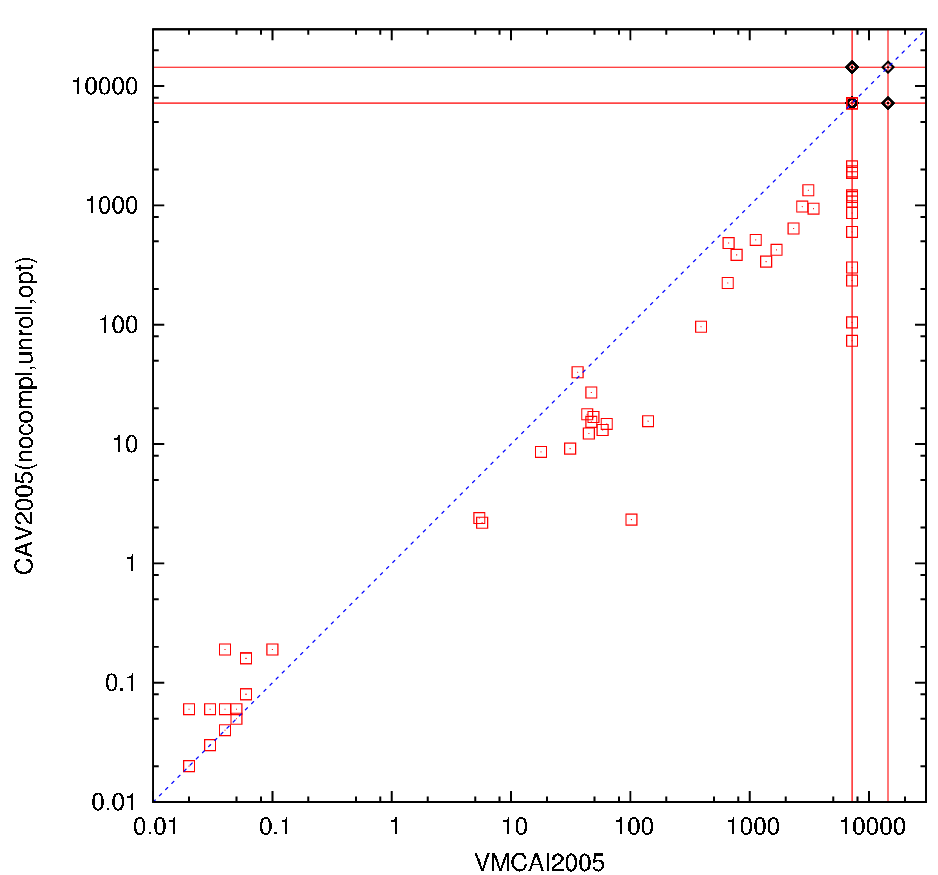}
    &
    \incscatter{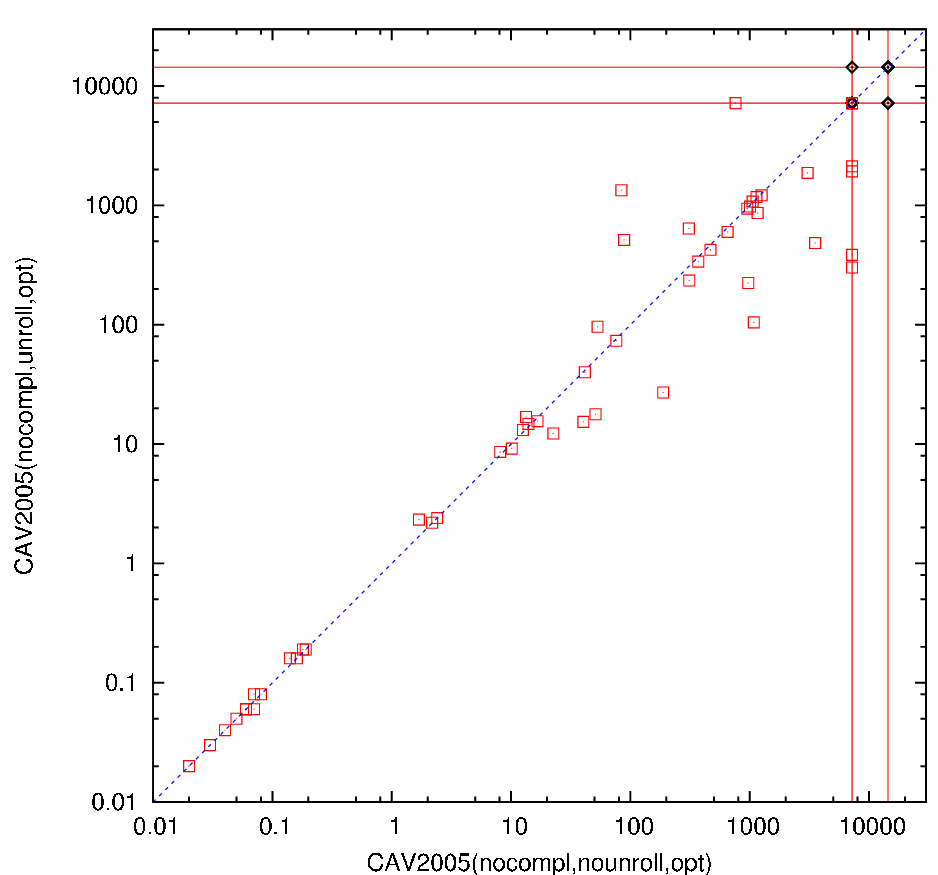}
    \\
    (a) & (b) & (c) \\
    \incscatter{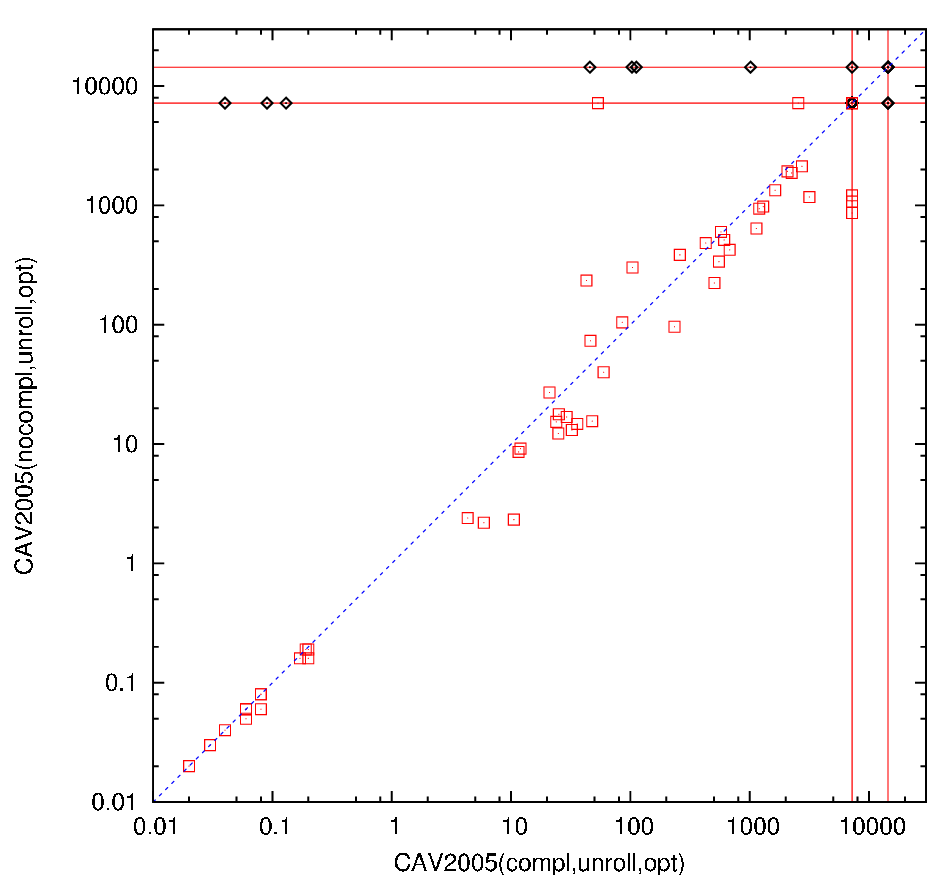}
    &
    \incscatter{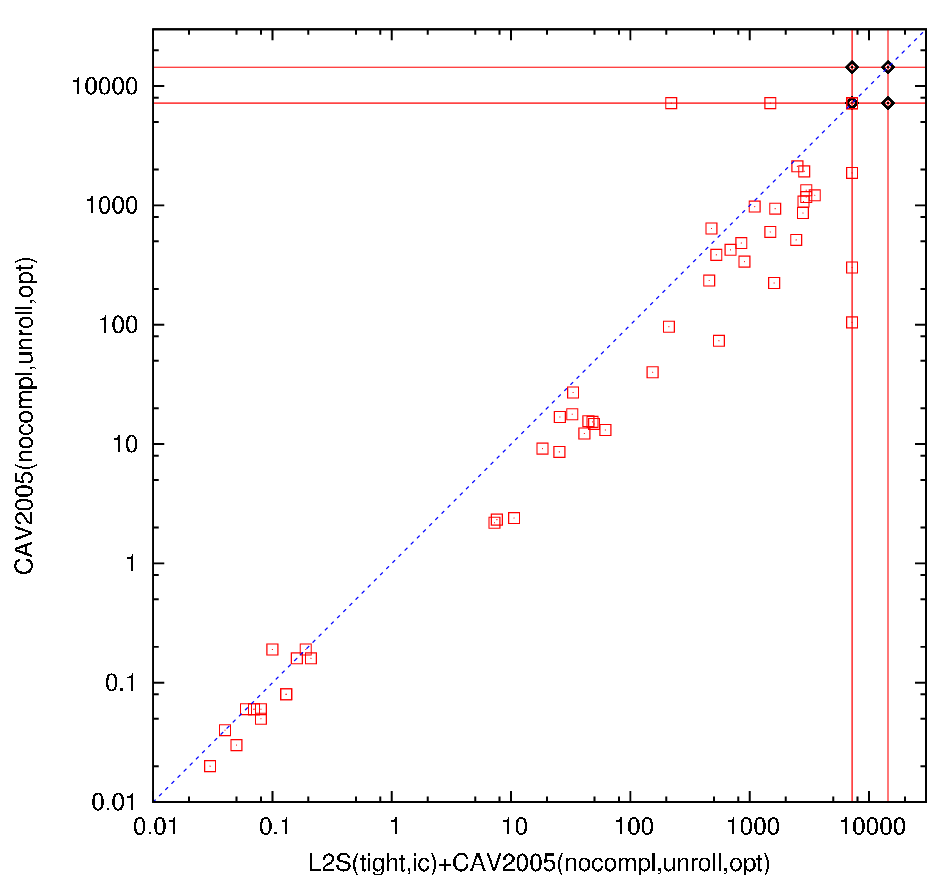}
    &
    \incscatter{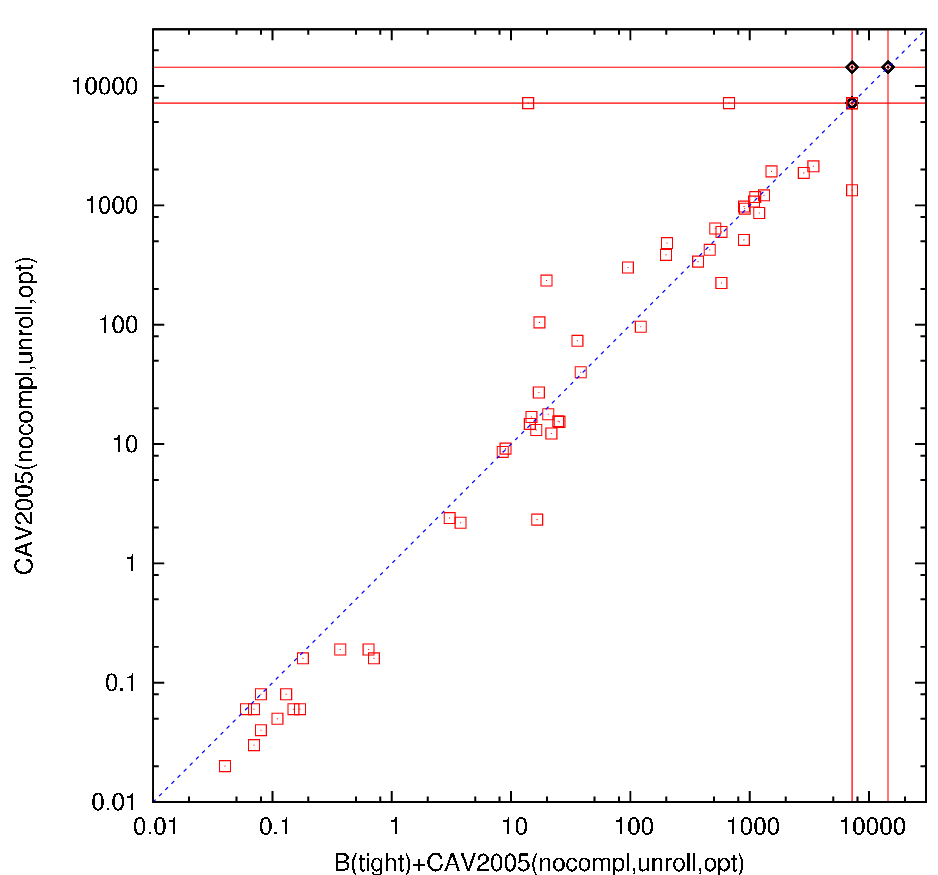}
    \\
    (d) & (e) & (f) \\
    \incscatter{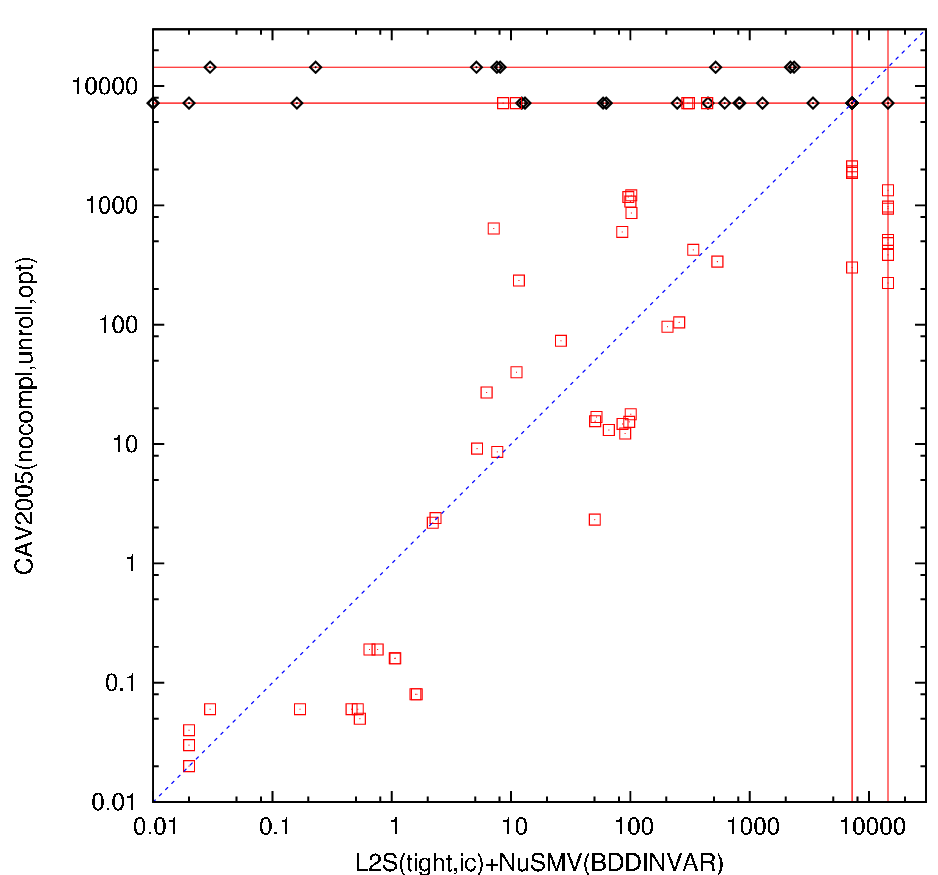}
    &
    \incscatter{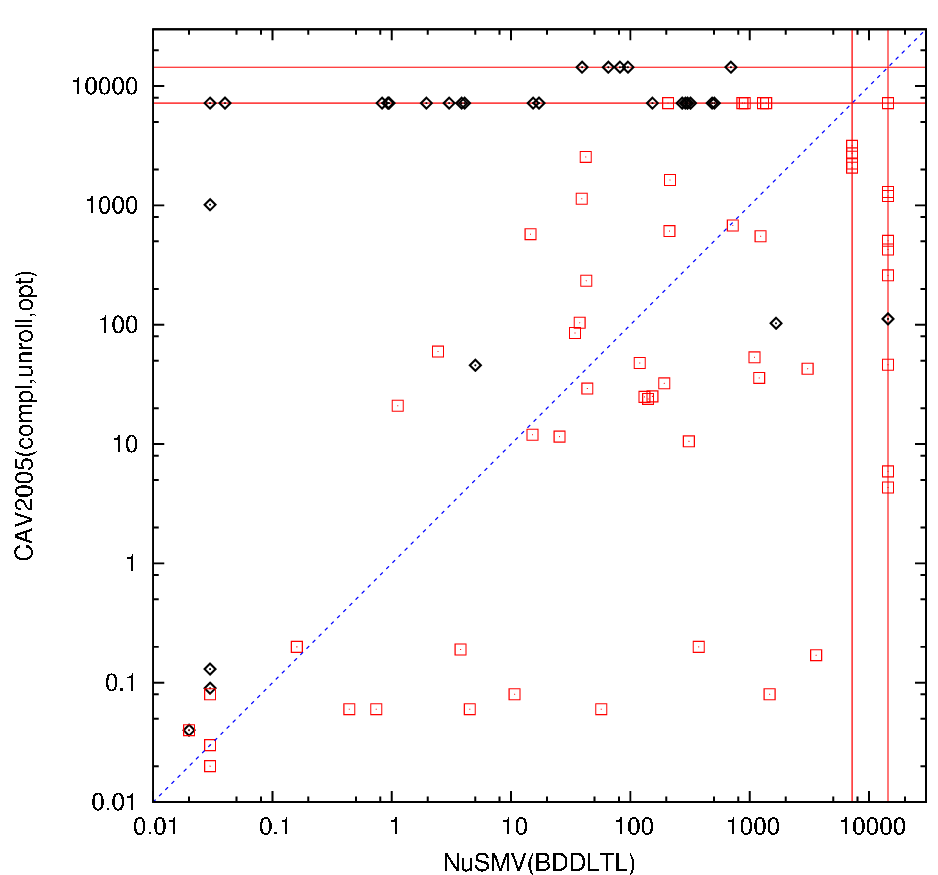}
    &
    \incscatter{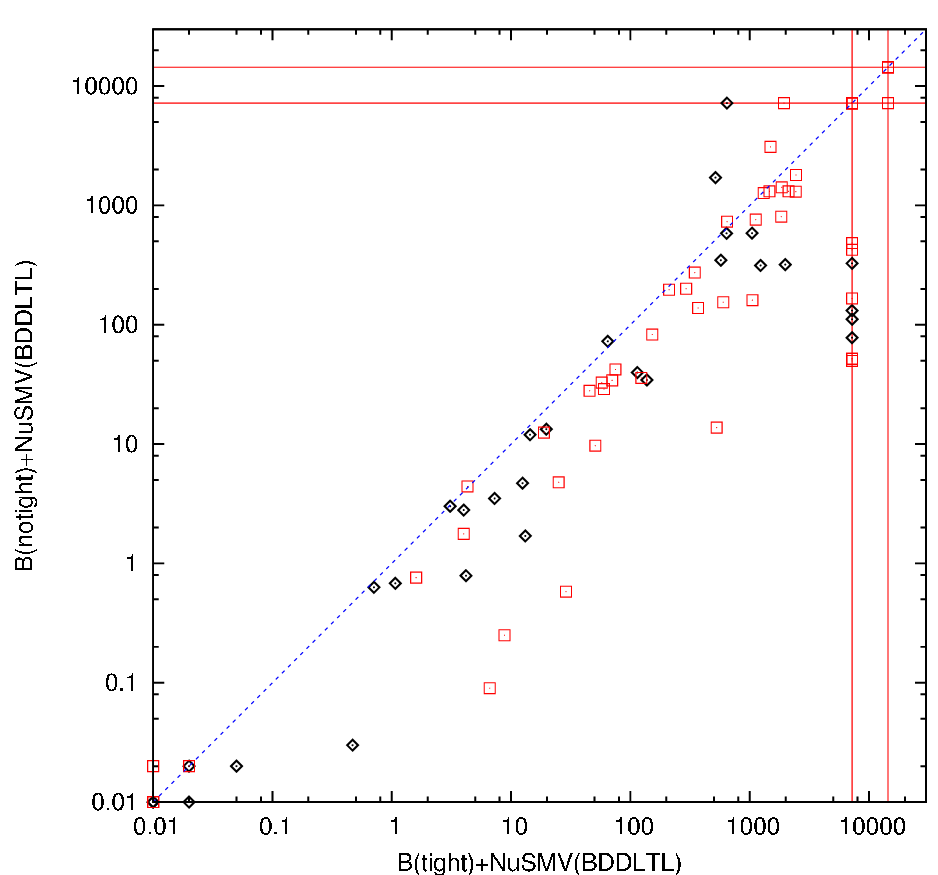}
    \\
    (g) & (h) & (i) \\
    \incscatter{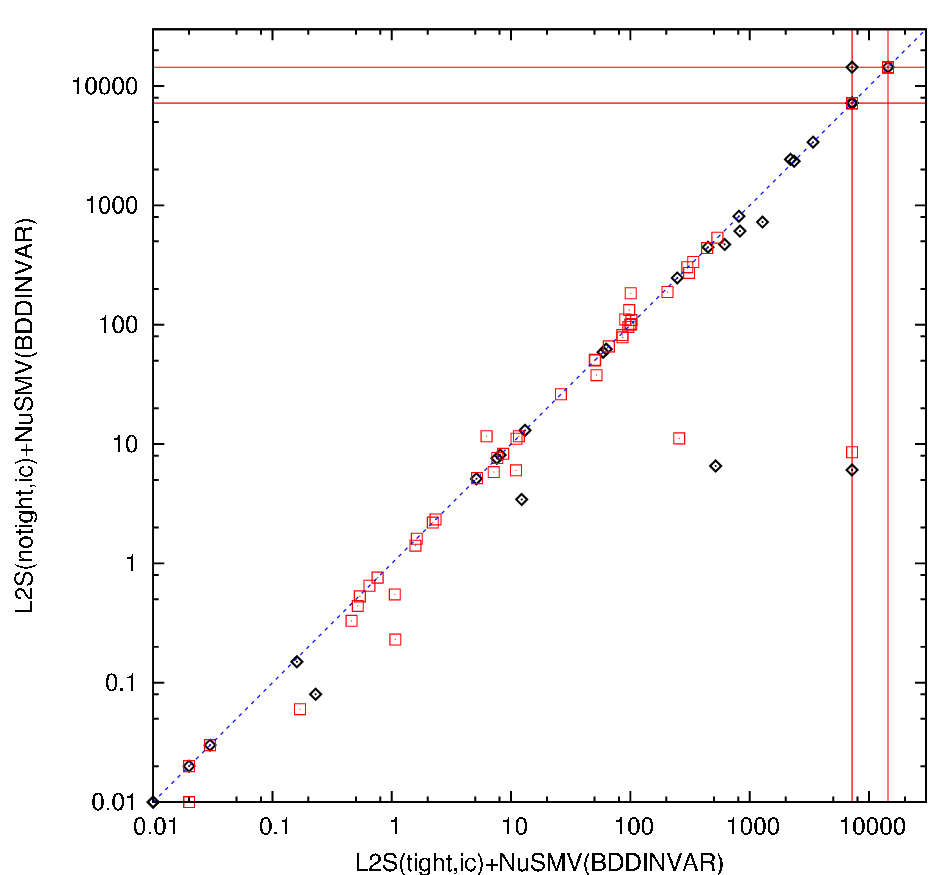}
    &
    \incscatter{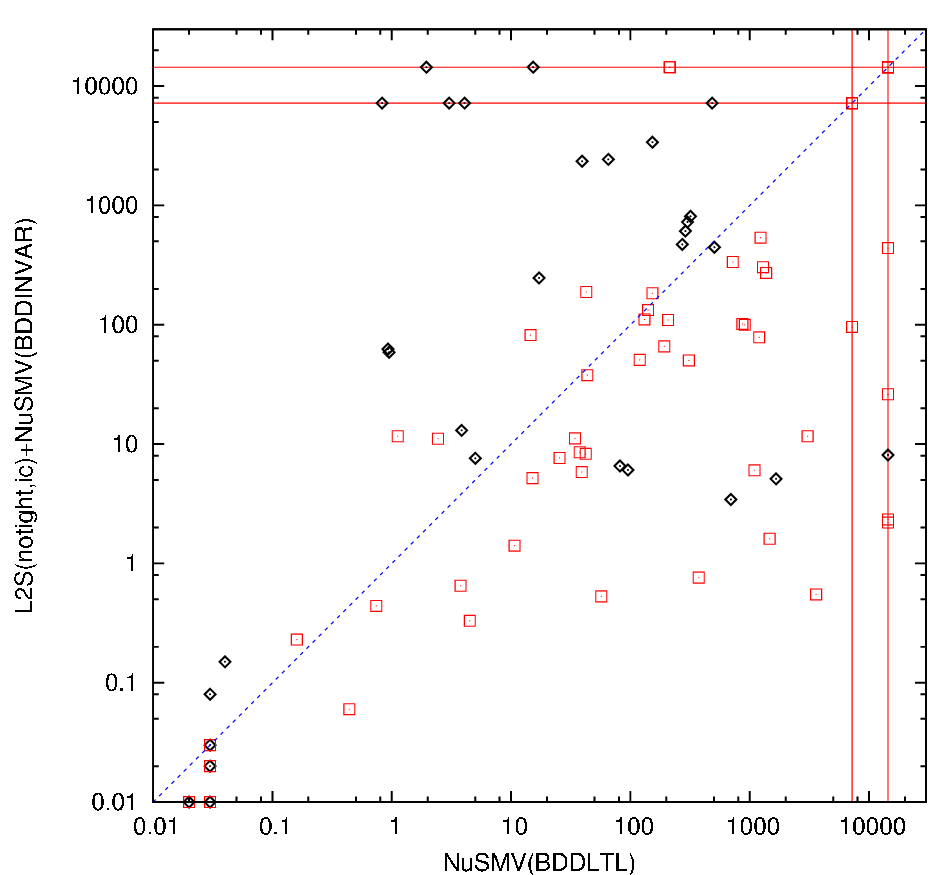}
    &
    \incscatter{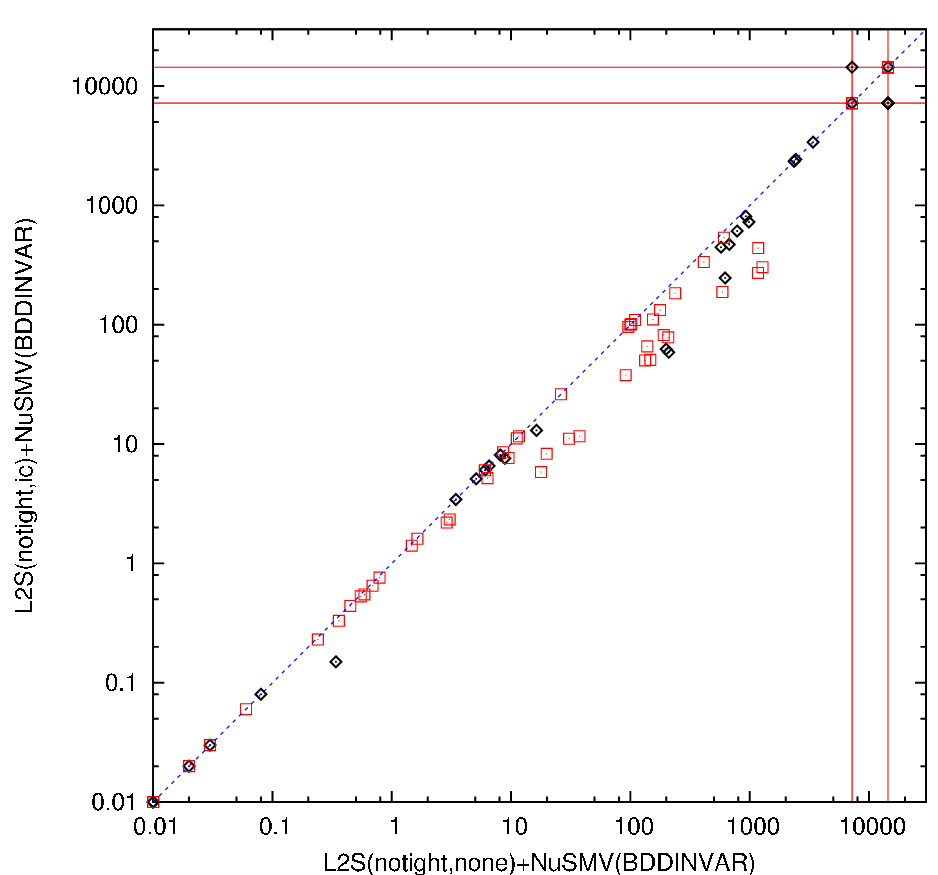}
    \\
    (j) & (k) & (l) \\
  \end{tabular}
  \caption{Scatter plots comparing the running times (in seconds) of different approaches}
  \label{exp:scatters}
\end{figure}

\iffalse
\begin{figure}
  \centering
  \begin{tabular}{cc}
    \includegraphics[width=7cm]{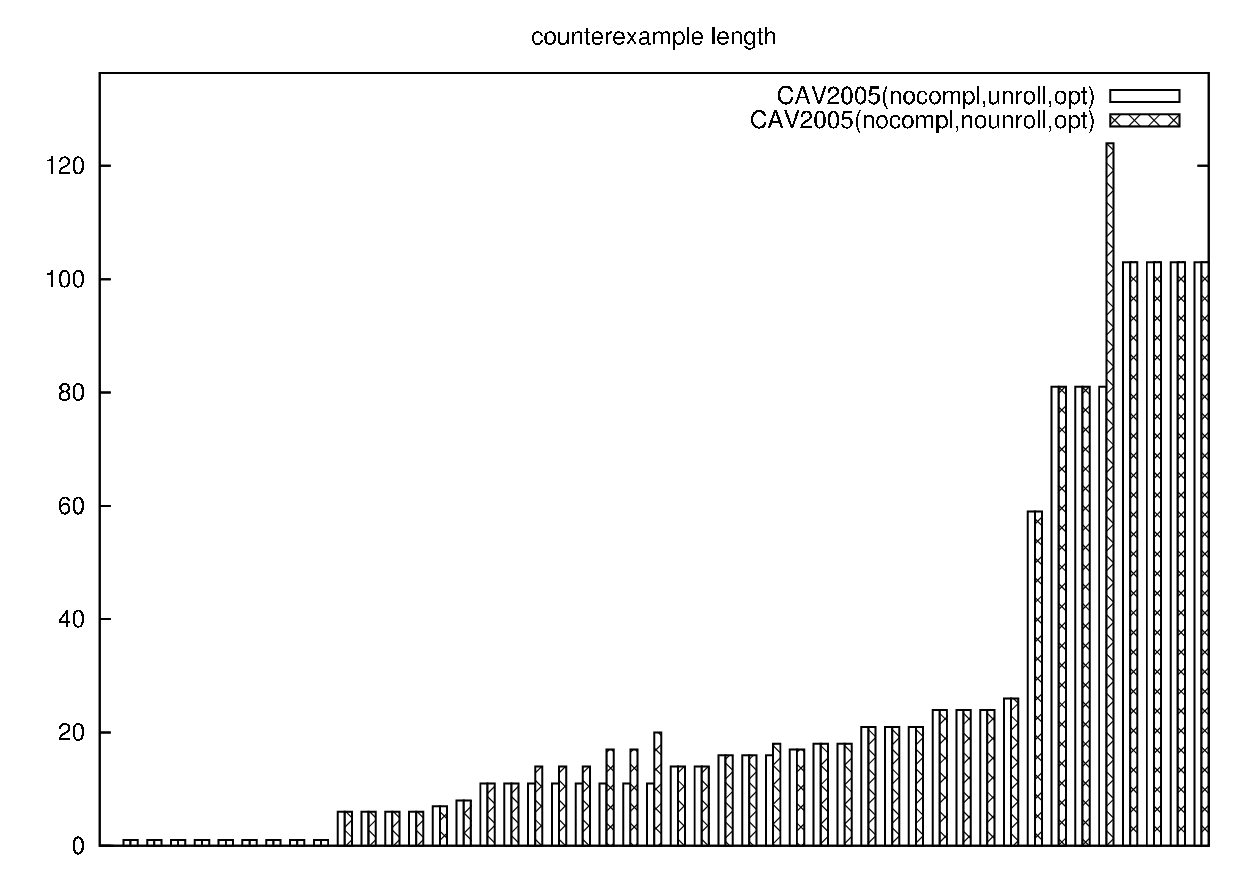}
    &
    \includegraphics[width=7cm]{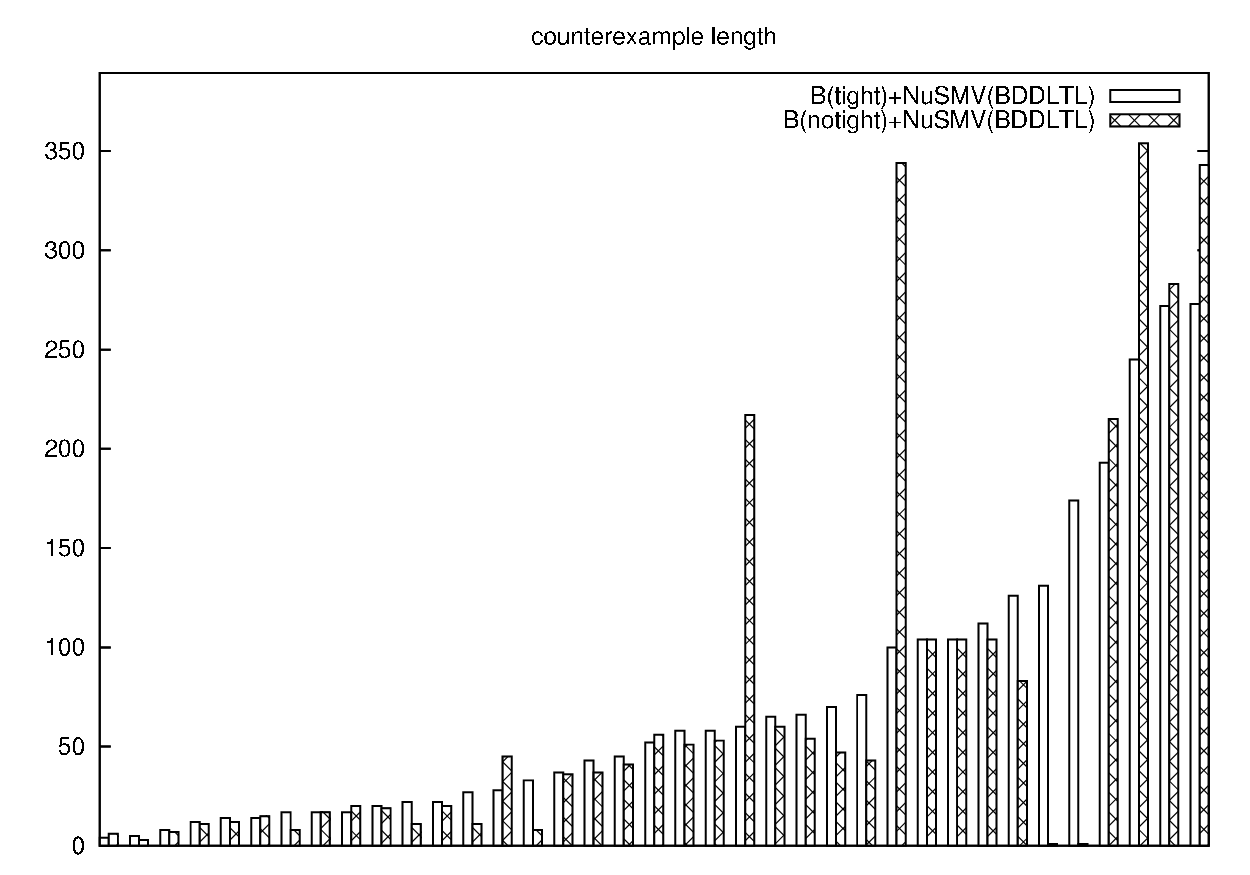}
    \\
    (a) & (b) \\
  \end{tabular}
  \begin{tabular}{c}
    \includegraphics[width=7cm]{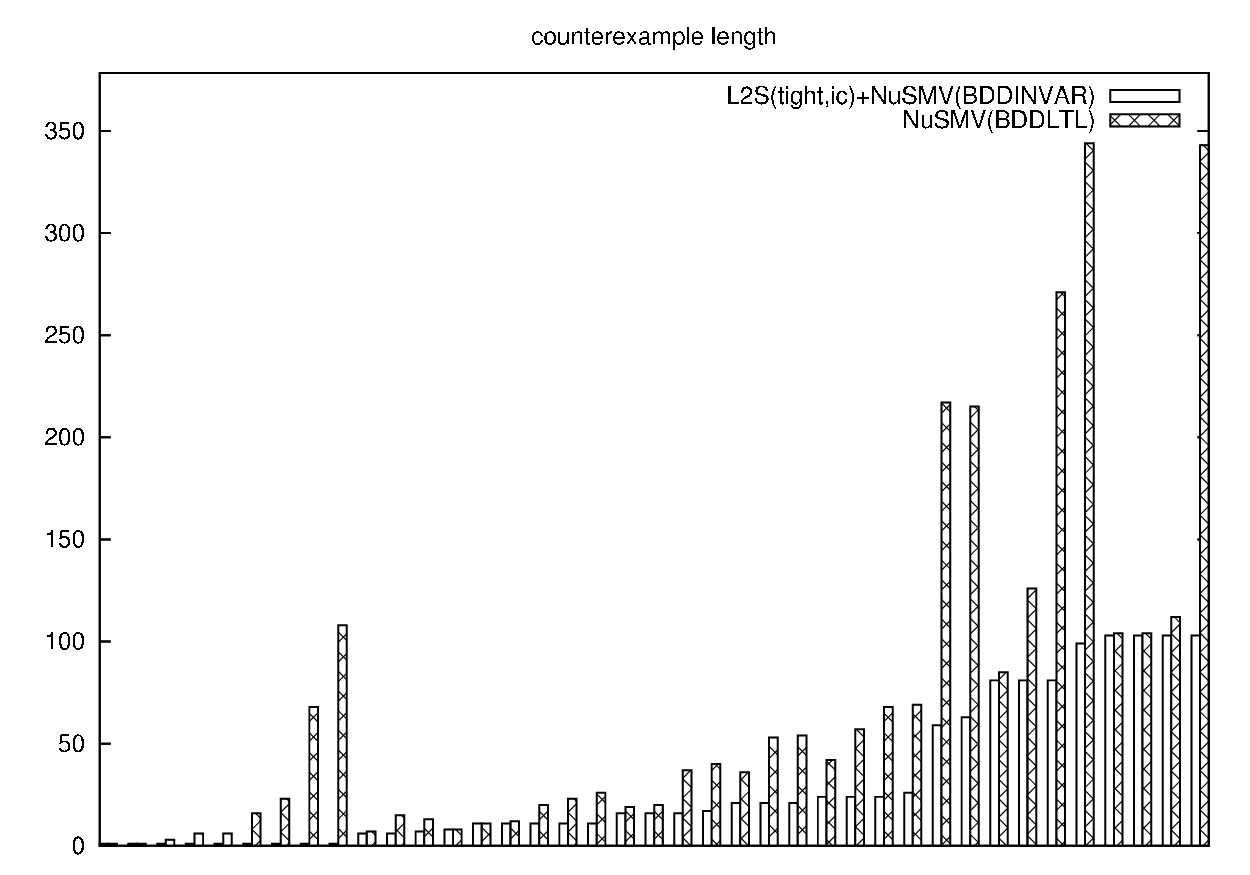}
    \\
    (c)\\
  \end{tabular}
  \caption{Counterexample length comparisons}
  \label{exp:boundplots}
\end{figure}
\else
\newcommand{\cexlengthfigureline}[2]{%
\begin{minipage}{0.1\linewidth}({#1})\end{minipage}&%
\begin{minipage}{0.7\linewidth}%
\includegraphics[width=9.25cm]{#2}%
\end{minipage}%
\\%
\\[0ex]%
}
\begin{figure}
  \centering
  \begin{tabular}{cc}
    \cexlengthfigureline{a}{bound_plots/b_CAV2005_nocompl_unroll_opt_vs_CAV2005_nocompl_nounroll_opt}
    \cexlengthfigureline{b}{bound_plots/b_ltl_tight_none_NuSMV_2_2_3_BDDs_f_vs_ltl_notight_none_NuSMV_2_2_3_BDDs_f}
    \cexlengthfigureline{c}{bound_plots/b_l2sbmc_tight_ic_NuSMV_2_2_3_BDDs_vs_NuSMV_2_2_3_BDDs_f}
  \end{tabular}
  \caption{Counterexample length comparisons}
  \label{exp:boundplots}
\end{figure}
\fi

\section{Discussion and Conclusions} \label{sec:conclusions}
When comparing BMC approaches, the linear sized dedicated BMC
encodings for PLTL offer better performance than alternative
approaches based either on symbolic B\"uchi automata using the
liveness-to-safety transformation
or the original BMC encodings. 
The main advantage of the dedicated encodings over approaches using
symbolic B{\"u}chi automata with fair loop detection
is the ability of the dedicated encodings to also detect no-loop counterexamples.
Adapted to
incremental SAT solving techniques, BMC based on our encodings offers
an efficient method for finding bugs. Virtual unrolling proved a
useful technique to obtain both BMC encodings and B{\"u}chi automata that
accept shortest counterexamples. The BMC  experiments also show that
the shorter counterexamples often lead to shorter times needed to find
counterexamples to PLTL properties.

Using the liveness-to-safety translation with BDD-based invariant
checking represents a competitive way to produce shortest
counterexamples. For both SAT- and BDD-based approaches that find
minimal length counterexamples there are problem instances that are
solved by one approach but not by the other. Thus neither approach
dominates the other.

When it comes to proving complex properties, the BMC approach presented here cannot yet
compete with BDD-based methods. However, there are cases where our BMC approach 
is faster than the BDD-based approaches. Improving the capability to prove properties
with BMC is therefore an important research direction.

There are at least two complementary research directions on proving properties 
of larger systems with BMC. One direction is based on generating stronger invariants than the current
completeness formula. This can be done by adding invariants to formula states
such as $\Trans{\TSF}^d_{k+1}$ to bind variables that are free. The invariants can be
deduced from PLTL semantics. Another approach for generating invariants is formulating
invariants based on the system's behaviour~\cite{MRS03,DBLP:journals/entcs/ArmoniFFHPV05}. 
The capability to prove properties can also be greatly improved if the $\Trans{SimplePath}_k$-predicate 
would have to include fewer state bits. A cone-of-influence reduction \cite{CGP99} tailored for full PLTL
or implementing the variable optimisations mentioned in Sect.~\ref{sec:l2svaropt} also for BMC could
make this possible.
Some insights might be gained by understanding why the combination of $k$-induction
and the liveness-to-safety transformation performs so poorly for proving properties.
We would also like to investigate methods based on Craig interpolants~\cite{McM03}
to better understand the implementation techniques needed and performance obtainable from
that method.

In this work we have concentrated on BMC encodings of PLTL properties. There are also other
places where BMC can be improved.
For example, \cite{DBLP:conf/sat/Sheridan04} discusses
methods to improve CNF generation employed inside a prototype NuSMV variant used in~\cite{CiRoSh:FMCAD04},
an area we have not covered in our BMC implementations.  Using SAT preprocessors, such as \cite{EB05}, 
to simplify the CNF after generation, is an alternative.
Usually bounded model checking papers take the system transition relation $T(s,s')$ as given
and do not try to exploit any special properties it might have. By more careful
encoding of $T(s,s')$ significant performance gains can be obtained, at least for special
classes of systems such as asynchronous
systems~\cite{Heljanko:Concur2001,HelNie03,ttj:otfj,HUT-TCS-A97}.

Although PLTL is exponentially more succinct than LTL, it cannot express all 
$\omega$-regular properties unlike some industry standard specification languages
such as Accellera's PSL~\cite{accellera,ieee1850}.
There are some encouraging initial results on bounded model checking of
$\omega$-regular properties very recently published~\cite{HelJunKeiLanLat:CAV06},
building on top of the work presented here.
For an alternative approach to handling $\omega$-regular properties,
see~\cite{BloCimPilRovSem:CIAA06}.

\subsubsection*{Acknowledgements}
The authors would like to thank the anonymous referees of Logical Methods in Computer Science
for valuable comments that helped us to improve this paper. This work has been
financially supported by the Academy of Finland (projects 109539, 112016, 211025, 213113)
and the Emil Aaltonen Foundation.

\bibliographystyle{alpha}
\bibliography{journal-bmc}

\end{document}